\documentclass[useAMS,usenatbib]{mn2e}
\usepackage{graphicx}

\usepackage{times}

\newcommand{\eref}[1]{(\ref{#1})}
\newcommand{\ion}[2]{\mbox{#1$\,${\sc #2}}}
\newcommand{\kms}{~km~s$^{-1}$}

\title[GAMA: end of survey report and DR2]{Galaxy And Mass Assembly (GAMA): end of survey report and data release 2}
\author[J.~Liske et al.]{J.~Liske,$^{1}$\thanks{E-mail: jliske@eso.org}
I.~K.~Baldry,$^{2}$
S.~P.~Driver,$^{3,4}$
R.~J.~Tuffs,$^{5}$
M.~Alpaslan,$^{6}$
E.~Andrae,$^{5}$\newauthor
S.~Brough,$^{7}$
M.~E.~Cluver,$^{8}$
M.~W.~Grootes,$^{5}$
M.~L.~P.~Gunawardhana,$^{9}$
L.~S.~Kelvin,$^{10}$\newauthor
J.~Loveday,$^{11}$
A.~S.~G.~Robotham,$^{4}$
E.~N.~Taylor,$^{12}$
S.~P.~Bamford,$^{13}$
J.~Bland-Hawthorn,$^{14}$\newauthor
M.~J.~I.~Brown,$^{15}$
M.~J.~Drinkwater,$^{16}$
A.~M.~Hopkins,$^{7}$
M.~J.~Meyer,$^{4}$
P.~Norberg,$^{9}$\newauthor
J.~A.~Peacock,$^{17}$
N.~K.~Agius,$^{18}$
S.~K.~Andrews,$^{4}$
A.~E.~Bauer,$^{7}$
J.~H.~Y.~Ching,$^{14}$\newauthor
M.~Colless,$^{19}$
C.~J.~Conselice,$^{13}$
S.~M.~Croom,$^{14}$
L.~J.~M.~Davies,$^{4}$
R.~De~Propris,$^{20}$\newauthor
L.~Dunne,$^{17,21}$
E.~M.~Eardley,$^{17}$
S.~Ellis,$^{7}$
C.~Foster,$^{7}$
C.~S.~Frenk,$^{9}$
B.~H{\"a}u{\ss}ler,$^{22,23}$\newauthor
B.~W.~Holwerda,$^{24}$
C.~Howlett,$^{25,11}$
H.~Ibarra,$^{26}$
M.~J.~Jarvis,$^{22,27}$
D.~H.~Jones,$^{15,28}$\newauthor
P.~R.~Kafle,$^{4}$
C.~G.~Lacey,$^{9}$
R.~Lange,$^{4}$
M.~A.~Lara-L{\'o}pez,$^{29,7}$
{\'A}.~R.~L{\'o}pez-S{\'a}nchez,$^{7,28}$\newauthor
S.~Maddox,$^{17,21}$
B.~F.~Madore,$^{30}$
T.~McNaught-Roberts,$^{9}$
A.~J.~Moffett,$^{4}$
R.~C.~Nichol,$^{25}$\newauthor
M.~S.~Owers,$^{7}$
D.~Palamara,$^{15}$
S.~J.~Penny,$^{15}$
S.~Phillipps,$^{31}$
K.~A.~Pimbblet,$^{15,32}$\newauthor
C.~C.~Popescu,$^{18,33,5}$
M.~Prescott,$^{27}$
R.~Proctor,$^{34}$
E.~M.~Sadler,$^{14}$
A.~E.~Sansom,$^{18}$\newauthor
M.~Seibert,$^{30}$
R.~Sharp,$^{19}$
W.~Sutherland,$^{35}$
J.~A.~V{\'a}zquez-Mata,$^{11}$
E.~van~Kampen,$^{1}$\newauthor
S.~M.~Wilkins,$^{11}$
R.~Williams$^{2}$
and A.~H.~Wright$^{4}$\\
{\vspace{-0.8mm}\footnotesize $^{1}$European Southern Observatory, Karl-Schwarzschild-Str.~2, 85748 Garching, Germany}\\
{\vspace{-0.8mm}\footnotesize $^{2}$Astrophysics Research Institute, Liverpool John Moores University, IC2, Liverpool Science Park, 146 Brownlow Hill, Liverpool L3 5RF, UK}\\
{\vspace{-0.8mm}\footnotesize $^{3}$Scottish Universities Physics Alliance, School of Physics and Astronomy, University of St Andrews, North Haugh, St Andrews KY16 9SS, UK}\\
{\vspace{-0.8mm}\footnotesize $^{4}$International Centre for Radio Astronomy Research, University of Western Australia, 35 Stirling Highway, Crawley, WA 6009, Australia}\\
{\vspace{-0.8mm}\footnotesize $^{5}$Max-Planck-Institut f{\"u}r Kernphysik, Saupfercheckweg 1, 69117 Heidelberg, Germany}\\
{\vspace{-0.8mm}\footnotesize $^{6}$NASA Ames Research Center, MS 232, Moffett Field, CA 94035, USA}\\
{\vspace{-0.8mm}\footnotesize $^{7}$Australian Astronomical Observatory, PO Box 915, North Ryde, NSW 1670, Australia}\\
{\vspace{-0.8mm}\footnotesize $^{8}$University of the Western Cape, Robert Sobukwe Road, Bellville, 7535, South Africa}\\
{\vspace{-0.8mm}\footnotesize $^{9}$Institute for Computational Cosmology, Department of Physics, Durham University, South Road, Durham DH1 3LE, UK}\\
{\vspace{-0.8mm}\footnotesize $^{10}$Institut f{\"u}r Astro- und Teilchenphysik, Universit{\"a}t Innsbruck, Technikerstraße 25, 6020 Innsbruck, Austria}\\
{\vspace{-0.8mm}\footnotesize $^{11}$Astronomy Centre, Department of Physics and Astronomy, University of Sussex, Falmer, Brighton BN1 9QH, UK}\\
{\vspace{-0.8mm}\footnotesize $^{12}$School of Physics, University of Melbourne, Parkville, VIC 3010, Australia}\\
{\vspace{-0.8mm}\footnotesize $^{13}$School of Physics \& Astronomy, University of Nottingham, University Park, Nottingham NG7 2RD, UK}\\
{\vspace{-0.8mm}\footnotesize $^{14}$Sydney Institute for Astronomy, School of Physics, University of Sydney, NSW 2006, Australia}\\
{\vspace{-0.8mm}\footnotesize $^{15}$School of Physics and Monash Centre for Astrophysics, Monash University, Clayton, VIC 3800, Australia}\\
{\vspace{-0.8mm}\footnotesize $^{16}$Department of Physics, University of Queensland, Brisbane, QLD 4072, Australia}\\
{\vspace{-0.8mm}\footnotesize $^{17}$Institute for Astronomy, University of Edinburgh, Royal Observatory, Blackford Hill, Edinburgh EH9 3HJ, UK}\\
{\vspace{-0.8mm}\footnotesize $^{18}$Jeremiah Horrocks Institute, University of Central Lancashire, Preston PR1 2HE, UK}\\
{\vspace{-0.8mm}\footnotesize $^{19}$Research School of Astronomy and Astrophysics, Australian National University, Canberra, ACT 2611, Australia}\\
{\vspace{-0.8mm}\footnotesize $^{20}$Finnish Centre for Astronomy with ESO, University of Turku, V{\"a}is{\"a}l{\"a}ntie 20, Piikki{\"o}, 21500 Finland}\\
{\vspace{-0.8mm}\footnotesize $^{21}$Department of Physics and Astronomy, University of Canterbury, Private Bag 4800, Christchurch 8140, New Zealand}\\
{\vspace{-0.8mm}\footnotesize $^{22}$Department of Physics, University of Oxford, Keble Road, Oxford OX1 3RH, UK}\\
{\vspace{-0.8mm}\footnotesize $^{23}$University of Hertfordshire, Hatfield, Hertfordshire AL10 9AB, UK}\\
{\vspace{-0.8mm}\footnotesize $^{24}$Leiden Observatory, University of Leiden, Niels Bohrweg 2, 2333 CA Leiden, The Netherlands}\\
{\vspace{-0.8mm}\footnotesize $^{25}$Institute of Cosmology and Gravitation, University of Portsmouth, Dennis Sciama Building, Burnaby Road, Portsmouth PO1 3FX, UK}\\
{\vspace{-0.8mm}\footnotesize $^{26}$National Institute of Astrophysics, Optics and Electronics. Luis Enrique Erro \# 1, Santa Mar{\'i}a Tonatzintla, PC 72840, Puebla, Mexico}\\
{\vspace{-0.8mm}\footnotesize $^{27}$Department of Physics, University of the Western Cape, Private Bag X17, Bellville 7535, South Africa}\\
{\vspace{-0.8mm}\footnotesize $^{28}$Department of Physics and Astronomy, Macquarie University, Sydney, NSW 2109, Australia}\\
{\vspace{-0.8mm}\footnotesize $^{29}$Instituto de Astronom{\'i}a, Universidad Nacional Aut{\'o}noma de M{\'e}xico, A.P.\ 70-264, 04510 M{\'e}xico D.F., Mexico}\\
{\vspace{-0.8mm}\footnotesize $^{30}$Observatories of the Carnegie Institution for Science, 813 Santa Barbara Street, Pasadena, CA 91101, USA}\\
{\vspace{-0.8mm}\footnotesize $^{31}$School of Physics, University of Bristol, Bristol BS8 1TL, UK}\\
{\vspace{-0.8mm}\footnotesize $^{32}$Department of Physics and Mathematics, University of Hull, Cottingham Road, Kingston-upon-Hull HU6 7RX, UK}\\
{\vspace{-0.8mm}\footnotesize $^{33}$Astronomical Institute of the Romanian Academy, Str.\ Cutitul de Argint 5, Bucharest, Romania}\\
{\vspace{-0.8mm}\footnotesize $^{34}$Observat{\'o}rio Nacional, Rua Gal.\ Jos{\'e} Cristino 77, 20921-400, Rio de Janeiro, Brazil}\\
{\vspace{-0.8mm}\footnotesize $^{35}$School of Physics and Astronomy, Queen Mary University of London, Mile End Road, London E1 4NS, UK}
}

\begin{document}

\date{Accepted 2015 January 1. Received 2015 January 1; in original form 2015 January 1}

\pagerange{\pageref{firstpage}--\pageref{lastpage}} \pubyear{2015}

\maketitle

\label{firstpage}

\clearpage

\begin{abstract}
The Galaxy And Mass Assembly (GAMA) survey is one of the largest
contemporary spectroscopic surveys of low-redshift galaxies. Covering
an area of $\sim$$286$~deg$^2$ (split among five survey regions) down
to a limiting magnitude of $r < 19.8$~mag, we have collected spectra
and reliable redshifts for $238\,000$ objects using the AAOmega
spectrograph on the Anglo-Australian Telescope. In addition, we have
assembled imaging data from a number of independent surveys in order
to generate photometry spanning the wavelength range
$1$~nm~--~$1$~m. Here we report on the recently completed spectroscopic
survey and present a series of diagnostics to assess its final state
and the quality of the redshift data. We also describe a number of
survey aspects and procedures, or updates thereof, including changes
to the input catalogue, redshifting and re-redshifting, and the
derivation of ultraviolet, optical and near-infrared photometry.
Finally, we present the second public release of GAMA data. In this
release we provide input catalogue and targeting information, spectra,
redshifts, ultraviolet, optical and near-infrared photometry,
single-component S{\'e}rsic fits, stellar masses, H$\alpha$-derived
star formation rates, environment information, and group properties
for all galaxies with $r < 19.0$~mag in two of our survey regions, and
for all galaxies with $r < 19.4$~mag in a third region ($72\,225$
objects in total). The database serving these data is available at
{\tt http://www.gama-survey.org/}.
\end{abstract}


\section{Introduction}
\label{introduction}

Large galaxy surveys, in particular those with a spectroscopic
component, have undoubtedly played a major role in driving our
understanding of both cosmology and galaxy evolution over the last
decade or so. For example, in cosmology the 2dF Galaxy Redshift Survey
(2dFGRS; \citealp{Colless01,Colless03}) and the Sloan Digital Sky
Survey (SDSS; \citealp{York00,Alam15}) demonstrated convincingly that
the description of large-scale structure formation provided by the
Cold Dark Matter (CDM) paradigm is remarkably accurate
\citep[e.g.][]{Peacock01,Percival01,Tegmark04,Cole05,Eisenstein05,Percival07}.
As a result of this success, large spectroscopic galaxy surveys are
now a well-established tool in cosmology, as evidenced by the large
number of completed, ongoing and planned projects that are seeking to
further explore the cosmological information encoded in the
large-scale distribution of galaxies, such as e.g.\ the WiggleZ Dark
Energy Survey \citep{Drinkwater10}, the Baryon Oscillation
Spectroscopic Survey \citep[BOSS;][]{Dawson13} or the Hobby-Eberly
Telescope Dark Energy Experiment \citep[HETDEX;][]{Hill08}, to name
but a few.

Large galaxy surveys have also been a driving force in the field of
galaxy evolution, although for a different reason. While large-scale
structure formation appears well understood, our theoretical
understanding of the growth of structure on galaxy scales is less
mature. The enormous range of mass and time scales involved in
capturing the gas physics, and the complex interplay between dark
matter, stars, gas, dust and active galactic nuclei (AGN) preclude the
development of a fundamental, comprehensive understanding of galaxy
formation and evolution based on first principles. Instead, we must
resort to approximate models that capture this complexity only to some
level. Although much progress has been achieved in the physical
modelling of galaxy evolution using both semi-analytic techniques
\citep[e.g.][]{Bower06,Guo11} and, most recently, full hydrodynamical
simulations \citep{Vogelsberger14,Schaye15}, advances in this field
are to a significant extent driven by observationally exploring the
physical properties of galaxies, their inter-dependencies and their
evolution with time \citep[e.g.][]{Blanton09}. Large surveys allow us
to systematically study galaxies at different cosmological epochs as a
function of key parameters, such as dynamical, stellar and gas mass,
environment, present and past star formation, stellar and gas-phase
metallicity, size and other structural parameters, morphology,
dynamical state, nuclear activity, dust content, etc. Past results
have shown that much of this information is indeed required in order
to identify and disentangle the various processes responsible for the
evolution of galaxies \citep[e.g.][]{Mo10}.

The Galaxy And Mass Assembly (GAMA)\footnote{\tt
  http://www.gama-survey.org/} survey aims to test the CDM model of
structure formation and to study galaxy evolution by exploiting the
latest generation of ground-based and space-borne, wide-field survey
facilities. GAMA is bringing together data from eight ground-based
facilities and four space missions in order to comprehensively survey
the low-redshift galaxy population. At the heart of this project lies
the GAMA spectroscopic survey of $\sim$$300\,000$ galaxies to $r <
19.8$~mag over $\sim$$286$~deg${^2}$ (split between five survey
regions), mainly conducted with the 2dF/AAOmega facility
\citep{Saunders04p,Smith04,Sharp06} on the 3.9-m Anglo-Australian
Telescope (AAT). In addition, we have coordinated with, and/or
negotiated data sharing agreements with a number of independent
imaging survey teams, conducted our own observing campaigns, and
processed data from publicly available sources (see
Table~\ref{gamasurveys} below) in order to construct a unique
multi-wavelength dataset covering all major galaxy constituents
(i.e.\ young and old stellar populations, ionized and neutral
interstellar medium, AGN and dust).

The main scientific goals that specifically motivated the GAMA
spectroscopic survey include:

(i) The measurement of the dark matter halo mass function (HMF) down
to $10^{12}$~M$_\odot$: since the HMF depends solely on the
cosmological parameters, the nature of gravity, and the dark matter
particle mass, with negligible dependence on baryonic physics, it
represents a fundamental, robust and precise prediction of the CDM
model \citep[e.g.][]{Springel05b}. A measurement of the HMF thus
amounts to a clear-cut, reliable test of the CDM structure formation
model in the {\em non-linear} regime.

(ii) Probing star formation efficiency and feedback: the properties of
the galaxy population within a dark matter halo depend not only on the
halo's mass but also on baryonic processes. Most galaxy formation
models incorporate feedback in order to account for the known
variation of star formation efficiency as a function of halo mass
\citep[e.g.][]{Bower06,DeLucia06}. By invoking different kinds of
feedback for low-mass and high-mass haloes (typically supernova and
AGN feedback, respectively) these models predict a peak in the
stellar-to-halo mass ratio at approximately the mass of the Local
Group. GAMA will characterise this peak and thereby improve our
understanding of feedback mechanisms.

(iii) A comprehensive measurement of the recent galaxy merger rate:
the hierarchical assembly of massive structures is a key feature of
the CDM structure formation paradigm
\citep[e.g.][]{White78,White91}. The build-up of dark matter haloes
through repeated mergers of smaller units is one of the principal
modes of growth in this model \citep[e.g.][]{Fakhouri10a}, and dark
matter halo merger rates are accurately predicted by simulations
\citep[e.g.][]{Fakhouri10b}. Although {\em galaxy} merger rate
predictions are less accurate \citep[e.g.][]{Hopkins10}, limiting the
testability of the CDM structure formation model by merger rate
observations, GAMA merger rate measurements will also be used to constrain
the extent to which mergers are driving various aspects of galaxy
evolution: the build-up of stellar mass in galaxies, in particular in
today's giant elliptical galaxies \citep[e.g.][]{DeLucia06},
morphological transformations
\citep[e.g.][]{Toomre77,Cox06,Hopkins09}, triggering
\citep[e.g.][]{Hopkins13b,Patton13} and truncating
\citep[e.g.][]{Hopkins08} star formation, fuelling central
supermassive black holes \citep[e.g.][]{DiMatteo05,Ellison11}, and
structural and size evolution \citep[e.g.][]{Naab09}.

GAMA stands in the tradition of the SDSS, the 2dFGRS and their
predecessors. Whereas the 2dFGRS essentially `only' provided redshifts
and fluxes in two (photographic) bands, the SDSS added high-quality
spectroscopic data and 5-band optical CCD imaging and photometry,
drastically increasing the available information for each galaxy, and
resulting in a wealth of physical insights into the low-redshift
galaxy population. GAMA builds on this by adding: (i)~$2$~mag in the
depth of the spectroscopic survey, thus probing solidly into the dwarf
regime and allowing a much more robust determination of a galaxy's
environment; (ii)~much higher spectroscopic completeness for pairs,
groups and clusters of galaxies, important for halo mass and merger
rate determinations; (iii)~a factor of $\sim$$2$ higher resolution in
the optical and near-infrared (NIR) imaging (from VST and VISTA), thus
giving reliable access to the internal structure of galaxies;
(iv)~photometric measurements over the wavelength range
$1$~nm~--~$1$~m. As mentioned above, the latter two points are
achieved in cooperation with a number of other independent imaging
surveys (see Table~\ref{gamasurveys}).

The motivation and science case of GAMA was explained in more detail
by \citet{Driver09}. The input catalogue and target selection,
including survey masks, star-galaxy separation, and target
prioritisation was presented by \citet{Baldry10}, while the tiling
algorithm was described by \citet{Robotham10}. The data reduction and
spectroscopic analysis was presented by \citet{Hopkins13a}, and
\citet{Driver11} provided a description of survey procedures and of
the first three years of GAMA data. Aperture-matched optical and NIR
photometry of GAMA galaxies based on processed SDSS and UKIDSS LAS
imaging data was introduced by \citet{Hill11}, while \citet{Kelvin12}
performed two-dimensional single-component S{\'e}rsic model fits to
the surface brightness distributions of GAMA galaxies using the same
data \citep[see also][]{Haussler13}. \citet{Taylor11} used these
photometric measurements, in particular the aperture-matched optical
photometry, to derive stellar masses. \citet{Cluver14} obtained
mid-infrared photometry for GAMA galaxies from reprocessed
\textit{WISE} data. Finally, the environment of GAMA galaxies was
characterised by \citet{Brough13} using galaxy number surface density,
while \citet{Robotham11b} presented the GAMA Galaxy Group Catalogue
\citep[G$^3$C; see also][]{Alpaslan12}.

\begin{table*}
\begin{minipage}{15cm}
\caption{Overview of the GAMA survey regions. The southern G02 and G23
  regions were not part of GAMA\,I. The last column provides the
  magnitude limits of the second data release described in
  Section~\ref{dr2}.}
\label{gamaregions}
\centerline{\begin{tabular}{lr@{\hspace{2mm}--\hspace{2mm}}lr@{\hspace{2mm}--\hspace{2mm}}lr@{\hspace{2mm}--\hspace{2mm}}l@{\hspace{7mm}}c@{\hspace{7mm}}ccc}
\hline
Survey region & \multicolumn{2}{c}{RA range (J2000)} & \multicolumn{4}{c}{Dec.\ range (J2000)} & Area & \multicolumn{3}{c}{$r$-band limits}\\
& \multicolumn{2}{c}{(deg)} & \multicolumn{4}{c}{(deg)} & (deg$^2$) & \multicolumn{3}{c}{(mag)}\\
& \multicolumn{2}{c}{} & \multicolumn{2}{c}{GAMA\,I} & \multicolumn{2}{c}{GAMA\,II} & GAMA\,II & GAMA\,I & GAMA\,II & DR2\\
\hline 
G02 & $30.2$  & $38.8$  & \multicolumn{2}{c}{-} & $-10.25^a$ & $-3.72$ & $55.71$ & - & $19.8$ & -\\
G09 & $129.0$ & $141.0$ & $-1.0$  & $+3.0$ & $-2.0$ & $+3.0$ & $59.98$ & $19.4$ & $19.8$ & $19.0$\\
G12 & $174.0$ & $186.0$ & $-2.0$  & $+2.0$ & $-3.0$ & $+2.0$ & $59.98$ & $19.8$ & $19.8$ & $19.0$\\
G15 & $211.5$ & $223.5$ & $-2.0$  & $+2.0$ & $-2.0$ & $+3.0$ & $59.98$ & $19.4$ & $19.8$ & $19.4$\\
G23 & $339.0$ & $351.0^b$ & \multicolumn{2}{c}{-} & $-35.0$ & $-30.0$ & $50.59$ & - & $i < 19.2^c$ & -\\
\hline
\end{tabular}}
$^a$From 2013 onward the observations focused on the high-priority 
sub-region north of $-6.0$~deg.\\
$^b$The original RA range of the G23 region was $338.1$--$351.9$~deg
but this was revised in 2014.\\
$^c$Originally the magnitude limit of the G23 region was the same as
for the other regions but it was changed from an $r$-band limit of
$19.8$~mag to an $i$-band limit of $19.2$~mag in 2014.
\end{minipage}
\end{table*}

The above have been used, {\em inter alia}, to derive the broad-band
\citep{Loveday12,Loveday15} and H$\alpha$ \citep{Gunawardhana13}
luminosity and stellar mass \citep{Baldry12,Gunawardhana15} functions,
to consider the luminosity and stellar mass functions split by Hubble
type \citep{Kelvin14a,Kelvin14b} and in different environments
\citep{McNaughtRoberts14,Eardley15}, to determine the effect of
mergers on the stellar mass function \citep{Robotham14}, to study
variations and dependencies of the galaxy initial mass function
\citep{Gunawardhana11} and of the star formation rate
\citep[SFR;][]{Wijesinghe12}, and to investigate satellite galaxies
\citep{Prescott11,Schneider13}, the effect of the local environment on
$L^*$ galaxies \citep{Robotham13}, the relations between stellar mass,
metallicity and (specific) SFR \citep{Foster12,LaraLopez13,Bauer13},
and the cosmic spectral energy distribution \citep{Driver12}.

In addition, GAMA provides the basis for numerous follow-up projects
(in particular of group galaxies), and even serves as a stepping stone
for other large, independent survey projects such as the SAMI Galaxy
Survey \citep{Bryant14,Allen14}.

Although it is generally considered good practice to pursue only a
single purpose with any given paper, the intention of the present
paper is in fact fourfold. Accordingly, it consists of four main
sections, each of which may be read somewhat independently of the
other three. First, in Section~\ref{updates} we supplement the earlier
technical papers on the GAMA spectroscopic survey cited above by
reporting on updates to various survey procedures and methods, and by
describing some procedures not yet covered at all by previous GAMA
publications. This includes descriptions of the updated input
catalogue and of the procedures we use to measure redshifts.

Second, in Section~\ref{progress_qc} we report on the recent
completion of the GAMA spectroscopic survey and present its end
product. We describe the progression of the survey, evaluate and
discuss its observing efficiency, and present various diagnostics that
characterise the final dataset, with a particular view towards its
redshift completeness and the quality of the redshifts.

In Section~\ref{photometry} we then move on from the spectroscopic to
the photometric side of GAMA. In this section we provide an updated
description of our procedure for deriving aperture-matched optical and
NIR photometry from processed SDSS and UKIDSS LAS imaging data of the
GAMA survey regions, and we describe for the first time our method of
measuring ultraviolet (UV) fluxes from \textit{GALEX} imaging
data. This section, too, thus represents a supplement to the previous
technical GAMA papers on the subject cited above.

Finally, following the first public data release described by
\citet{Driver11}, we present the second public release of GAMA data
(DR2) in Section~\ref{dr2}, which comprises a large fraction of the
spectroscopic data from the first three years of observations as well
as a wealth of ancillary data. We end with a summary in
Section~\ref{summary}.

\section{Spectroscopic survey procedure updates}
\label{updates}

In this section we report on various aspects, procedures and methods
of the GAMA spectroscopic survey that have either changed
significantly since they were first described, or that have not yet
been described at all in previous GAMA publications. This comprises a
description of the updated input catalogue and target selection
(Section~\ref{inputcat}), our procedures for deriving redshifts
(Sections~\ref{redshifting}--\ref{agnz}), an update of our procedure
to incorporate data from previous spectroscopic surveys into the GAMA
survey (Section~\ref{previous}), and a description of additional
observations of a small number of very bright targets using the
Liverpool Telescope (Section~\ref{LTObs}).

\subsection{Input catalogue and target selection}
\label{inputcat}

Following the first three years of survey operations (2008 February --
2010 May, see \citealp{Driver11}) the GAMA spectroscopic survey on the
AAT was substantially expanded, resulting in a number of significant
changes to the GAMA input catalogue (IC) and target selection. Here
(and in other GAMA publications) the term `GAMA\,I' refers to the data
collected during these first three years, and to all data products
that can be traced back to the original version of the IC (called {\tt
  InputCatAv05}). In contrast, the term `GAMA\,II' refers to the {\em
  entire} GAMA dataset, including all GAMA\,I and all subsequently
collected data, and all data products that can be traced back to the
revised version of the IC ({\tt InputCatAv06} for the equatorial
survey regions, see below).

The GAMA\,I survey extended over three equatorial survey regions of
$48$~deg$^2$ each (called G09, G12 and G15) and down to
extinction-corrected Petrosian magnitude limits of $r < 19.4$~mag in
G09 and G15, and $r < 19.8$~mag in G12, as well as $z < 18.2$~mag and
$K_{\rm AB} < 17.6$~mag, selected from SDSS DR6 \citep{AdelmanMcCarthy08}
and UKIDSS LAS data. The NIR photometry was also used to improve on
the standard SDSS star-galaxy separation. See \citet{Baldry10} for the
full details of the GAMA\,I IC and target selection.

For GAMA\,II we implemented the following main changes to the IC and
the target selection: (i)~the three existing equatorial survey regions
were enlarged from $12 \times 4$ to $12 \times 5$~deg$^2$; (ii)~two
new survey regions were added in the south (called G02 and G23);
(iii)~in the equatorial survey regions the target selection switched
from using SDSS DR6 to DR7 \citep{Abazajian09} photometry, and we
created new input catalogues for the G02 and G23 regions from SDSS DR8
\citep{Aihara11}, CFHTLenS \citep{Heymans12}, KiDS \citep{deJong13}
and VIKING \citep{Edge13} photometry, respectively; (iv)~the $r$-band
Petrosian magnitude limit was set to $19.8$~mag for all survey
regions; in G23 this was later revised to an $i$-band limit of
$19.2$~mag; (v)~the $z$ and $K$-band selections were abandoned. In
addition, the NIR photometry required for the improved star-galaxy
separation mentioned above was only partially available for the
extensions of the equatorial regions, and not at all for G02. Despite
these changes, all objects identified as targets in GAMA\,I (in the
original survey regions) were retained as targets in GAMA\,II for
consistency.

Table~\ref{gamaregions} provides an overview of the main changes. More
details about these changes and the input catalogues used for
selecting targets in the new southern regions G02 and G23 will be
presented by Robotham et al.\ (in preparation) and Moffett et al.\ (in
preparation).

In addition to the changes to the main survey, we have also changed
the selection of `filler' targets (cf.\ section~3.7 of
\citealp{Baldry10}). The purpose of the filler targets was to maximise
the scientific output of the survey by providing useful targets even
in cases where an AAOmega fibre could not be assigned to a main survey
target, either due to physical fibre placement restrictions, or,
towards the end of the survey, due to the scarcity of unobserved main
survey targets. Various samples of filler targets have been defined
over the course of the survey, including radio, optical, far-infrared
and X-ray selected samples, as well as targets randomly selected for
duplicate observations. The latter sample will be used extensively
when assessing the quality of our redshift data in Section~\ref{zqc},
the others will be detailed in future data releases.

\subsection{Tiling, observing and data reduction}

Our tiling, fibre placement, observing and data reduction procedures
have not changed significantly compared to the descriptions provided
by \citet{Robotham10}, \citet{Driver11} and \citet{Hopkins13a}. The
only differences are that we began using dark frames in 2010 November,
and that we are now using the latest version (v5.35) of the data
reduction software {\sc 2dfdr} \citep{Croom04a,Sharp10} provided by
the AAO. Note that, in order to ensure the consistency of the data
reduction, we re-reduce the {\em entire} GAMA\,II dataset whenever a
new version of {\sc 2dfdr} is released.

\subsection{Redshifting and re-redshifting using \textsc{runz}}
\label{redshifting}

In this section we describe the procedure by which we measure the
redshift, $z$, of a given spectrum using the code {\sc runz}. A
summary of this process was already provided by \citet{Driver11} but
here we describe the procedure in full.

As we will see below, {\sc runz} has a number of undesirable features
which motivated the development of a new and improved redshifting
code, {\sc Autoz} (\citealp{Baldry14}; see also Section~\ref{autoz}).
{\sc Autoz} proved to be superior to {\sc runz} in every way (see
Section~\ref{zqc}), and so the {\sc Autoz} redshifts were adopted as
the default for GAMA\,II in 2013. However, DR2 and many of the GAMA
publications to date are based on the {\sc runz} redshifts, and so we
feel it is still important to fully document our {\sc runz}
procedures.

\subsubsection{Initial redshifting}

All GAMA spectra obtained at the AAT (excluding sky spectra) were
`redshifted' by one of the observers at the telescope either on the
same night they were observed or the next day or night. As described by
\citet{Driver11} and \citet{Hopkins13a}, the observations of a
2dF/AAOmega field and the subsequent data reduction process result in
a file containing all of the fully reduced, sky-subtracted and
telluric absorption-corrected spectra of that field ($346$ spectra on
average). The process of redshifting an observation involves running
the program {\sc runz} (developed by Will Sutherland, Will Saunders,
Russell Cannon and Scott Croom; see also \citealp*{Saunders04}) on this
file, meaning that all spectra of a given field are redshifted by the
same person.

For each spectrum {\sc runz} attempts to identify a redshift in two
different ways: (i)~by cross-correlating it with a range of template
spectra, including spectra of star-forming, E+A and quiescent galaxies
(five templates), QSOs (five templates), and A, K and M stars (four
templates); and (ii)~by fitting Gaussians to emission lines (after
having interpolated over strong sky lines) and searching for
multi-line matches, adopting the best-guess single line redshift if no
multi-line match is found. Having thus identified a number of possible
redshifts, a best redshift is automatically chosen from among these
based on the strengths of the cross-correlation peaks and the number
and significance of any identified emission lines.  Except for the
most extreme emission line galaxies this procedure usually results in
the best cross-correlation redshift being chosen as the overall best
redshift.

{\sc runz} then proceeds by presenting the operator with a plot of the
spectrum being redshifted (along with various diagnostic plots),
marking the positions of common nebular emission and stellar
absorption lines at the best automatic redshift. This redshift is then
checked visually by the operator. This check is unfortunately
necessary because the cross-correlation redshift is frequently led
astray by imperfections in the data reduction. If the redshift is
deemed incorrect, the operator may interactively use a number of
methods to try to find the correct one. These methods include checking
the redshifts obtained from the cross-correlations with the various
template spectra, checking all possible emission line redshifts, and
roughly identifying a redshift visually, marking it crudely, and then
fitting absorption and emission lines at the corresponding
positions. A free-format comment can also be attached to the spectrum.

Once satisfied, the operator concludes this process by assigning a
(subjective) quality, $Q$, in the range $0$--$4$ to the final
redshift, where $Q= 4$ signifies a certainly correct redshift, $3$ a
probably correct redshift, and $2$ a possibly correct redshift
requiring independent confirmation. $Q = 1$ indicates that no redshift
could be identified at all, while a value of $0$ flags spectra that are
seriously flawed, in the sense of a complete data reduction failure. A
pure noise spectrum, without any continuum or emission lines, but not
displaying any obvious data reduction issues, is assigned $Q = 1$, not
$0$. By assigning $Q \ge 3$ the operator consents to having this
redshift included in scientific analyses, and thus the distinction
between $Q=2$ and $3$ is clearly the most important one. Note that for
$Q \le 1$ the value of the recorded redshift is meaningless. Note
further that for values $\ge 2$, $Q$ refers to the (subjective)
quality of the {\em redshift}, not of the {\em spectrum}. In
particular, it is sometimes possible to confidently identify a
redshift even in a (partially) damaged spectrum (usually from multiple
strong emission lines). In these cases, too, $Q$ refers to the
confidence in the redshift.

Once $Q$ has been assigned, {\sc runz} moves on to the next spectrum,
and the above process is repeated until all spectra of the field being
processed have been redshifted.

Among the final sample of GAMA\,II spectra we find the fractions of
spectra receiving $Q = 4$--$0$ to be $62$, $20$, $11$, $8$ and
$0.1$~per~cent, respectively.

\subsubsection{Re-redshifting}
\label{re-redshifting}

From the above description it is clear that the process of redshifting
with {\sc runz} is not fully automated, instead involving significant
human interaction, in particular in the assignment of a redshift
quality. In total, no fewer than $56$ GAMA team members have
contributed to the redshifting during observations. These redshifters
have a wide range of experience and differ in their abilities to
(i)~verify the correctness of a given redshift; (ii)~find a
difficult-to-spot redshift; (iii)~not be fooled by spectral features
of non-galaxian origin. Most importantly, the quality assigned to a
redshift is quite subjective and depends strongly on the experience of
the redshifter. These are clearly undesirable features and a fully
automated process for determining the redshifts and their reliability,
as e.g.\ implemented by the SDSS, would be preferable. This motivated
the development of the aforementioned code {\sc Autoz} (see
Section~\ref{autoz}). Until this code became available in 2013,
however, we had to resort to an elaborate double-checking process of
our {\sc runz} redshifts in order to mitigate the effects described
above.

In an effort to confirm or reject redshifts initially classified as
`probable' or `possible', to weed out mistakes and, most importantly,
to homogenise the quality scale of our redshifts, a significant
fraction of our sample has thus been `re-redshifted' independently.
Re-redshifting has been carried out `off-line' (i.e.\ not at the
telescope during observing runs) by volunteering team members in three
separate re-redshifting campaigns. Each of these campaigns
approximately covered the data collected during the year prior to its
launch (2009 August, 2011 April, 2012 February). Thus almost all of
the data collected up to 2011 May (i.e.\ $3.5$ years of observations)
have been subjected to re-redshifting. We now describe this process in
detail.

\paragraph{Selection of spectra for re-redshifting}
\label{re_z_selection}

First of all, we only consider spectra of main survey targets for
re-redshifting. Since the spectra of filler targets
(cf.\ Section~\ref{inputcat}) are generally more difficult to redshift
than those of main survey targets, and since we are interested in
optimising our procedures for the main survey only, spectra of filler
targets (or of flux calibration stars) are not included in the
re-redshifting.

All spectra of main survey targets for which the redshifts were
initially assigned a $Q$ value of $1$, $2$ or $3$ are selected for
re-redshifting. In addition, {\em in each 2dF/AAOmega field} we select
a random sample of $5$~per~cent of spectra with initial $Q=4$
redshifts. Since an entire field is initially redshifted by a single
person, this selection ensures that not only a {\em global} fraction
of $5$~per~cent of spectra with $Q=4$ redshifts are re-redshifted, but
$5$~per~cent of {\em each initial redshifter's} $Q=4$
redshifts. Finally, for each redshifter involved in a given
re-redshifting campaign (including both initial redshifters and
redshifters from previous re-redshifting campaigns) and for each $Q$
value $\ge 1$ we select a random sample of $20$ spectra for
self-checks where possible.

\paragraph{Assignment of spectra to re-redshifters}

On average, $28$ volunteers participated in each of the re-redshifting
campaigns (including team members that had not previously observed and
had hence not done any initial redshifting). The spectra selected for
re-redshifting in a given campaign are assigned to the available
volunteers in the following way.

All spectra with $Q = 1$ or $2$ redshifts are assigned to {\em two}
re-redshifters, so that these spectra are redshifted three times in
total. Spectra with $Q = 3$ or $4$ redshifts are assigned to one
re-redshifter. For $Q = 1$ or $4$ re-redshifters are selected
randomly, but for $Q=2$ or $3$ we attempt to pick re-redshifters that
are `better' than the original redshifter, the idea being that there
is little value of having an inexperienced person re-examine a
spectrum that an experienced redshifter was doubtful about. A strict
implementation of this idea, however, would have placed an
unmanageable burden on experienced redshifters. Thus we chose a scheme
where the probability of a given spectrum being assigned to a
particular re-redshifter depends on the relative `quality' of the
initial redshifter and the re-redshifter: this probability is equal
for re-redshifters that are `better' than the initial redshifter but
decreases linearly for re-redshifters that are `worse'. The metric
used in comparing redshifters is the probability of correctly
identifying redshifts (see below) as derived from earlier
re-redshifting campaigns.\footnote{For the first re-redshifting
  campaign we used a much simpler scheme: we simply ensured that a
  spectrum initially redshifted by an experienced redshifter was not
  assigned to an inexperienced redshifter.}

Spectra selected for self-checks are obviously assigned to the
initial redshifter for re-redshifting.

On average, $\sim$$1100$ spectra were assigned in this way to each
volunteer in each past re-redshifting campaign.

\paragraph{Execution}

Re-redshifters are asked to provide an {\em independent} estimate of
the redshift and of its quality of each of the spectra assigned to
them. To this end they are only given the information {\em which}
spectra have been assigned to them, but not {\em why} these spectra
were selected for re-redshifting or what the original redshift and $Q$
were.

The actual redshifting is done using the same code and procedures as
for the initial redshifting described above, except of course that
{\sc runz} is now run on individual spectra (retrieved directly from
the GAMA team database) and not on an entire field.

From the above it should be clear that the results provided by the
re-redshifting are `independent' of the initial redshifting results
only in a very limited sense. The same data (modulo any changes to the
data reduction that may have been applied in the meantime) and the
same code are being used, hence the `independence' of the results
solely refers to that aspect of the redshift measurement process that
requires human interaction.

\subsubsection{Analysis of the (re-)redshifting data}
\label{re_z_analysis}

Upon completion of a re-redshifting exercise the new data are combined
with all existing redshift data, i.e.\ with those from the original
redshifting as well as with those from any previous re-redshifting
exercises. We now ask how this combined dataset should be used in
order to achieve our goals. In the most na{\"i}ve approach we could
simply assume that any redshift that is confirmed by the
re-redshifting must be correct. If two different redshifts are found
for the same spectrum, however, then we would have no way of knowing
which of these, if either, is correct and hence would be forced to
discard both (although a third `opinion'\footnote{In the following we
  will refer to the combination of $z$ and $Q$ found by a
  (re-)redshifter for a given spectrum as that redshifter's `opinion'
  of that spectrum. The range of possible opinions explicitly includes
  $Q = 1$, i.e.\ that no redshift can be found in this spectrum (in
  which case the value of $z$ is of course meaningless). Note that
  every spectrum has at least one opinion associated with it (from the
  initial redshifting).} might help in deciding). In this simplistic
approach it is also not clear how to use the additional information
encoded in the $Q$ values attached to the redshifts, how to account
for the different levels of ability and experience of the many
redshifters, or how to create a homogenised quality scale. Clearly,
this approach does not use all of the available information.

Instead, we now develop a probabilistic approach which enables us to
quantify the probability of a given redshift being correct. The
underlying assumption of our approach is that the process of
redshifting a given spectrum can be viewed as a multinomial process in
which the redshifter attempts to identify the correct redshift from a
set of possible redshifts. The idea is then to use the re-redshifting
data to measure the probability of correctly identifying a redshift as
a function of redshifter and $Q$. From these probabilities we can then
calculate the probability that a given redshift is correct, taking
into account all of the available opinions as well as the
reliabilities of those who offered them. For example, if two
redshifters, $i$ and $j$, independently find the same redshift for a
given spectrum, we can calculate the increased probability (compared
to having either only $i$'s or only $j$'s opinion) of this redshift
being correct from $i$, $j$, $Q_i$ and $Q_j$. Similarly, if their
redshifts disagree this lowers the probabilities of either $i$'s or
$j$'s value being correct. This can be generalised to an arbitrary
number of agreeing or differing opinions. Furthermore, this method
allows us to unambiguously identify the `best' (i.e.\ most probably
correct) redshift (or else that no redshift can be determined) for
{\em every} spectrum. This allows us to statistically treat all
spectra in the same way, even those that have not been re-redshifted
at all. Finally, for every best redshift our method provides us with a
homogeneous measure of confidence which we can use to decide whether
to accept this redshift for scientific analyses or not.

We emphasise that in this context we use the word `correct' in a very
narrow sense. The re-redshifting data do not allow us to determine the
probability of a redshift being correct in any absolute sense
(although this can be achieved by referring to duplicate observations
of the same object, see Section~\ref{zqc}). We can only determine the
probability that other people with similar training, given the same
data and code, will identify the same redshift. A `correct' redshift
in this sense is simply the most popular one.

While having to make this distinction is of course in general an
undesirable feature, it does have one advantage: it allows us to
ignore the (small) complication that would otherwise be introduced by
targets that are in fact two unresolved objects at different
redshifts. While the spectra of these targets may well display two
real redshifts, we will nevertheless be able to assume in the
following that there is only one `correct' (i.e.\ most popular)
redshift.

We now describe our method of using the re-redshifting data to measure
the probability of `correctly' identifying a redshift (in the above
sense) as a function of redshifter and $Q$, i.e.\ $p(i,Q)$ where $i =
1 \dots N_{\rm RS}$ is an index identifying redshifters, of which
there are $N_{\rm RS}$. The general idea is to consider the `agreement
fractions' among pairs of redshifters, i.e.\ the fraction of spectra
for which the opinions of two redshifters agree. We will model these
fractions in terms of the sought-after parameters $p(i,Q)$, and then
fit this model to the observed values of the agreement
fractions. Readers not interested in the details of this process may
wish to skip ahead to the results, as shown in Figs.~\ref{p_rs_Q} and
\ref{prms_rs_Q} and discussed in the accompanying text.

We begin by considering all spectra with at least two $Q \ge 2$
opinions.\footnote{In the following we will disregard all opinions
  with $Q \ge 2$ and $z > 0.9$. Almost all of these opinions are of
  spectra showing broad emission lines. These spectra are obviously
  very different from those of our main survey targets for which we
  wish to optimise our procedures, and hence the high-redshift
  opinions are excluded.} For all of these spectra we identify all
pairs of opinions of the same spectrum where both opinions have $Q \ge
2$, and sort these into groups according to the originators ($i$ and
$j$) and $Q$ values ($Q_i$ and $Q_j$) of these opinions, such that
each group is uniquely identified by the tuple $(i,Q_i,j,Q_j)$.
Following the last re-redshifting campaign we have $5824$ such groups,
containing a total of $54\,733$ opinion pairs which involve $92\,902$
individual opinions of $43\,765$ unique spectra from $N_{\rm RS} = 55$
redshifters. The large number of these groups is of course a
consequence of the way in which we assign spectra to re-redshifters
(see above), which guarantees a high degree of `intermixing' of
redshifters. Indeed, of the $3N_{\rm RS} = 165$ 
possible ($i,Q$) combinations ($Q$ can take on $3$ different values
here), $155$ are cross-checked by more than $10$ other redshifters,
and the median number of cross-checking redshifters for each
combination is $32$. However, in many cases the number of opinion
pairs in each group is of course quite small. It ranges from $1$ to
$251$, but the distribution is strongly skewed towards small values,
with a mean and median of $9.4$ and $4$ pairs, respectively.

In each group we then determine the fraction of opinion pairs where
the redshifts agree with one another. Whether two redshifts agree or
not is determined using {\em all} available $Q \ge 2$ opinions of that
spectrum and a one-dimensional friends-of-friends method with a
generously large linking length of $\Delta z = 0.0035$, chosen after
inspection of the full $\Delta z$ distribution of all $Q \ge 3$
opinion pairs. The redshift agreement fractions $f_z$ are then
modelled by:
\begin{eqnarray}
\label{pmodel}
\lefteqn{f_z(i, Q_i, j, Q_j) =} \nonumber \\
& & p(i,Q_i) \; p(j,Q_j) + [1 - p(i,Q_i)] \; [1-p(j,Q_j)] \; p_{\rm a}.
\end{eqnarray}
The second term on the right-hand side accounts for the possibility of
both $i$ and $j$ being `wrong' and yet identifying the same (`wrong')
redshift, where the parameter $p_{\rm a}$ denotes the probability of
such `accidental' agreement. The value of $p_{\rm a}$ is not
negligibly small because in practice there is only a finite number of
plausible redshifts to choose from. For reasons described below we
somewhat arbitrarily set $p_{\rm a} = 0.2$ but note that the exact
value of this parameter has little effect on the final results.

So far we have only considered $Q \ge 2$ opinions because of the
qualitative difference between the meanings of $Q$ values $2$ and
greater (in which case at least some redshift has been identified) and
values of $1$ and $0$ (in which case no redshift could be identified
and the reported value of the redshift is entirely meaningless).
Adopting a procedure similar to the one described above we can ask
what the binomial probability of a given redshifter is to `correctly'
identify a spectrum as not yielding any redshift at all (where we
again use the word `correct' in the sense described above).

\begin{figure}
\includegraphics[width=\columnwidth]{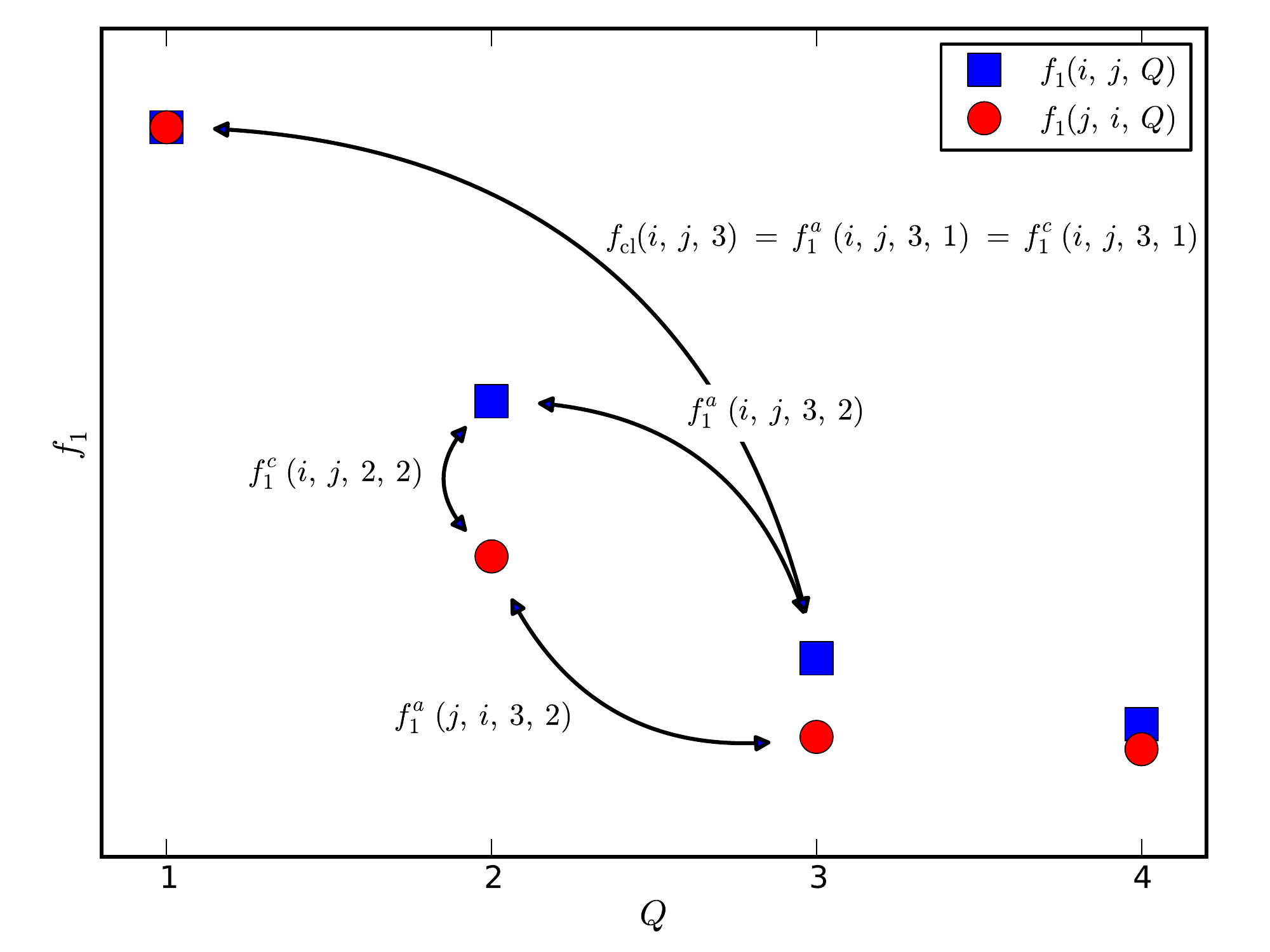}
\caption{Sketch illustrating various ratios discussed in the text. The
  blue squares and red circles show $f_1(i,j,Q)$ and $f_1(j,i,Q)$,
  respectively, as a function of $Q$. $f_1(i,j,Q)$ is the fraction of
  $i$'s $Q_i = 1$ opinions among $j$'s $Q_j = Q$ opinions, relative to
  $i$'s total fraction of $Q_i = 1$ opinions among all $i,j$ opinion
  pairs. By construction, we have $f_1(i,j,1) = f_1(j,i,1)$. The $f_1$
  values are observables that could, in principle, be used to
  constrain the parameters we are after, i.e.\ all $p(i,1)$. However,
  our model for $f_1$ [equations~\eref{p2}--\eref{f1}] contains an
  inconvenient proportionality factor. It turns out, though, that this
  factor only depends on the pair of redshifters $i,j$ (or $j,i$), and
  is thus the same for all points shown in the figure. By taking
  ratios of these quantities (indicated by arrows) we thus eliminate
  the inconvenient constant. We label ratios of same-coloured points
  as `auto'-ratios, those of differently coloured points as
  `cross'-ratios. In total there are $27$ such ratios, of which only
  six are independent. We choose to use those six ratios that have
  $f_1(i,j,1) = f_1(j,i,1)$ as the denominator, and label these
  $f_{\rm cl}$.}
\label{f1_sketch}
\end{figure}

\begin{figure}
\includegraphics[angle=-90,width=\columnwidth]{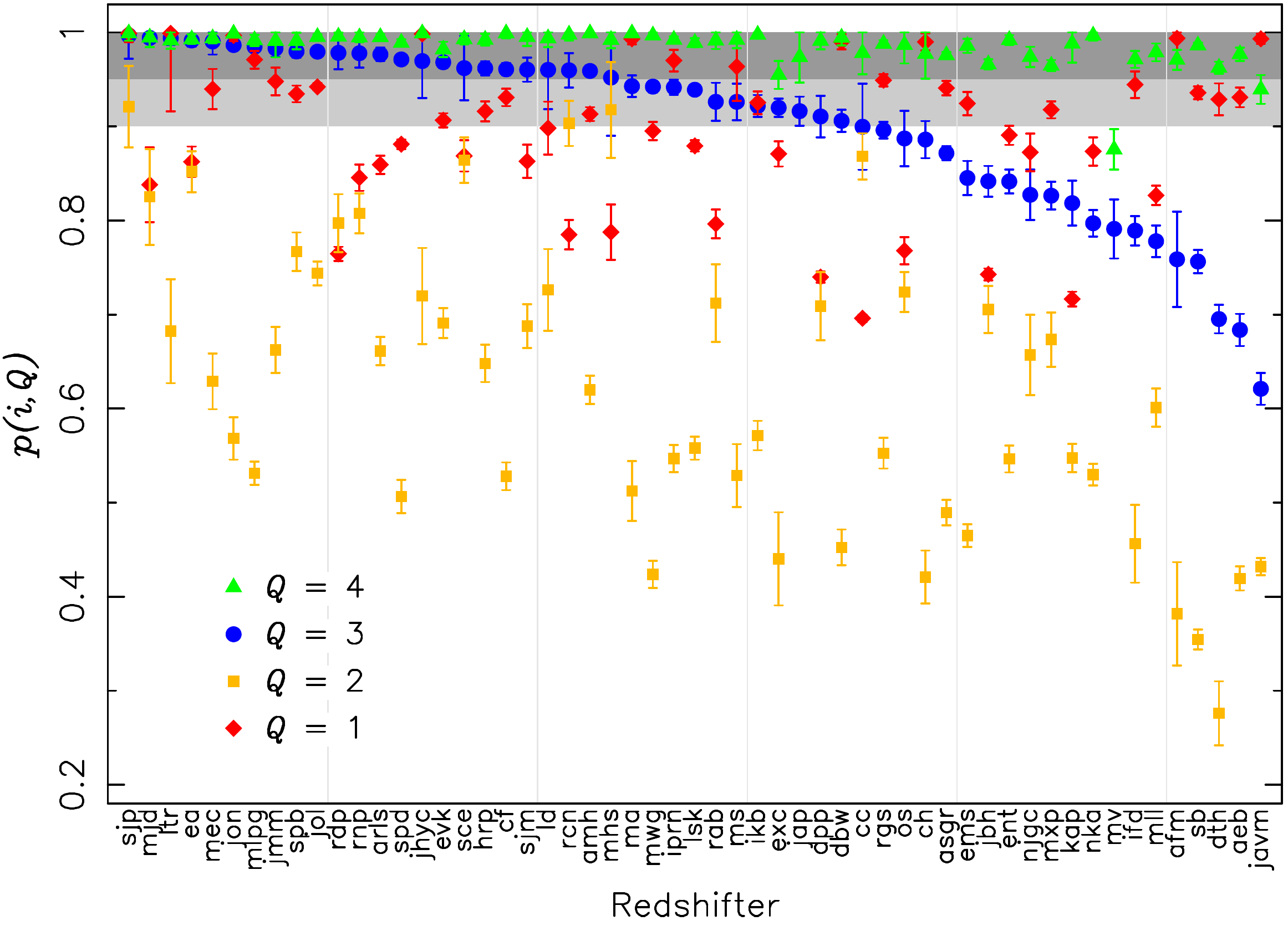}
\caption{Probability to `correctly' identify a redshift, or to
  `correctly' identify a spectrum as not yielding a redshift, as a
  function of redshifter and $Q$. These values are the result of
  fitting the model of equations~\eref{pmodel} and \eref{p1model} to
  the observed redshift agreement and normalised class disagreement
  fractions. Redshifters are identified by their initials along the
  abscissa, and are arbitrarily ordered by their $p(i,3)$
  values. Green triangles, blue points, orange squares and red
  diamonds are for $Q = 4$, $3$, $2$ and $1$, respectively, as
  indicated. The dark and light shaded regions mark the $p(z)$ ranges
  to which we assign $nQ = 4$ and $3$, respectively (see Section
  \ref{final_z}).}
\label{p_rs_Q}
\end{figure}

\begin{figure}
\includegraphics[angle=-90,width=\columnwidth]{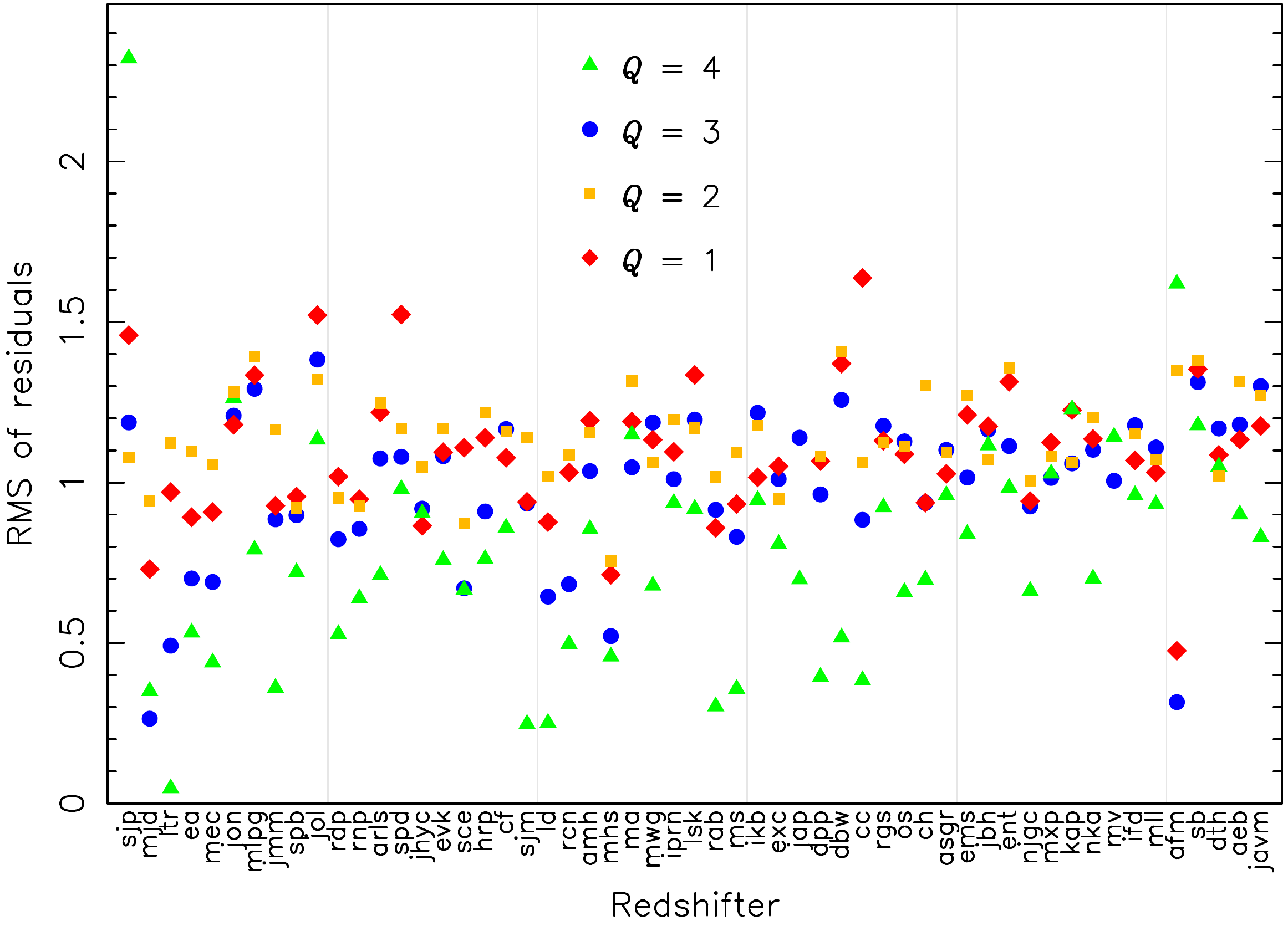}
\caption{Rms of the residuals between the fitted model of
  equations~\eref{pmodel} and \eref{p1model}, and the observed
  redshift agreement and class disagreement fractions as a function of
  redshifter and $Q$. Redshifters are identified by their initials
  along the $x$-axis and are ordered in the same way as in
  Fig.~\ref{p_rs_Q}. Green triangles, blue points, orange squares and
  red diamonds are for $Q = 4$, $3$, $2$ and $1$, respectively.}
\label{prms_rs_Q}
\end{figure}

To derive these probabilities, $p(i,Q=1)$, from the re-redshifting
data, consider all opinion pairs of redshifters $i$ and $j$ (where
both opinions of a given pair of course refer to the same
spectrum). Let us denote the number of such pairs by $n_{ij}$. Now
further consider that sub-set of $i$, $j$ pairs where $j$ assigned
$Q_j = Q$, and let us denote the number of these pairs by $n_{ij}(Q_j
= Q)$. If $j$ assigned $Q$ values completely randomly, we would expect
the fraction of these pairs in which $i$ assigned $Q_i=1$ to be
independent of $Q$ and equal to the total fraction of pairs in which
$i$ assigned $Q_i=1$:
\begin{equation}
\label{nij}
\frac{n_{ij}(Q_i = 1, Q_j = Q)}{n_{ij}(Q_j = Q)} \; = \; \frac{n_{ij}(Q_i =
    1)}{n_{ij}}.
\end{equation}
However, $j$ does not assign $Q$ values randomly of course, and we
expect the left-hand side of the above equation to be smaller for
larger $Q$. So how do we modify the right-hand side to reflect this
dependence on $Q$?  Clearly, if $i$ assigns $Q_i = 1$ and $j$ assigns
$Q_j = Q \ge 2$ either $i$ or $j$ or both of them must be `wrong' (in
the sense discussed above). If $j$ assigns $Q_j = Q = 1$ then either
both are `right' or both are `wrong'. The modulation factor to be
applied to the right-hand side of equation~\eref{nij} above must
therefore be proportional to
\begin{eqnarray}
\label{p2}
\lefteqn{P(i,j,Q)} \nonumber \\
& \equiv & p(i,1) \; [1-p(j,Q)] \; + \; [1-p(i,1)] \; p(j,Q) \nonumber \\
& & \mbox{} + [1-p(i,1)] \; [1-p(j,Q)] \nonumber \\
& = & 1 - p(i,1) \; p(j,Q)
\end{eqnarray}
when $Q\ge2$, and
\begin{eqnarray}
\label{p1}
& \equiv & p(i,1)\;p(j,1) \; + \; [1-p(i,1)] \; [1-p(j,1)]
\end{eqnarray}
when $Q = 1$. Rearranging equation~\eref{nij} to define
\begin{eqnarray}
f_1(i,j,Q) &\equiv& \frac{n_{ij}(Q_i = 1, Q_j = Q) \; / \; n_{ij}(Q_j = Q)}{n_{ij}(Q_i = 1) \; / \; n_{ij}}
\end{eqnarray}
we thus find
\begin{eqnarray}
\label{f1}
f_1(i,j,Q) &=& c_{ij} \; P(i,j,Q),
\end{eqnarray}
where $c_{ij}$ is a proportionality constant. $f_1(i,j,Q)$ is simply
the fraction of $i$'s $Q_i = 1$ opinions among $j$'s $Q_j = Q$
opinions, relative to $i$'s total fraction of $Q_i = 1$ opinions. In
Fig.~\ref{f1_sketch} we sketch the behaviour of $f_1$ as a function of
$Q$.

While the constant $c_{ij}$ could in principle be determined from the
`integral constraint':
\begin{equation}
\label{intconst}
\sum_{Q=1}^{4} n_{ij}(Q_i=1, Q_j=Q) \; = \; n_{ij}(Q_i=1),
\end{equation}
leading to
\begin{equation}
c_{ij} \; = \; \frac{n_{ij}}{\sum_{Q=1}^{4} P(i,j,Q) \; n_{ij}(Q_j = Q)},
\end{equation}
its presence in the model is clearly an inconvenience. Note, however,
that for $Q=1$ the left-hand side of equation~\eref{f1} above is
symmetric in $i$ and $j$, and that $P(i,j,1)$ is also symmetric. This
means that $c_{ij}$ must also be symmetric in $i$ and $j$. $c_{ij}$
thus cancels out in all `auto'-ratios of the form
\begin{equation}
f_1^{\rm a}(i,j,Q_1,Q_2) \; \equiv \; \frac{f_1(i,j,Q_1)}{f_1(i,j,Q_2)} 
\; = \; \frac{P(i,j,Q_1)}{P(i,j,Q_2)} 
\end{equation}
as well as in all `cross'-ratios of the form
\begin{equation}
f_1^{\rm c}(i,j,Q_1,Q_2) \; \equiv \; \frac{f_1(i,j,Q_1)}{f_1(j,i,Q_2)}
\; = \; \frac{P(i,j,Q_1)}{P(j,i,Q_2)}
\end{equation}
(see Fig.~\ref{f1_sketch} for a visualisation of these ratios). In
other words, the above ratios only depend on $p(i,Q)$ and $p(j,Q)$,
including the sought-after parameters $p(i,1)$ and $p(j,1)$. We can
therefore use the observed $f_1^{\rm a}$ and $f_1^{\rm c}$ ratios to
constrain the $p(i,1)$ values.

Note that each redshifter pair yields (at most) six independent data
points: although we can compute up to four $f_1(i,j,Q)$ and four
$f_1(j,i,Q)$ values, two of these are identical by construction
[$f_1(i,j,1) = f_1(j,i,1)$], and given any six values the seventh can
be determined from the integral constraint of equation~\eref{intconst}
above. Therefore, of the $27$ possible $f_1^{\rm a,c}$ ratios
(cf.\ Fig.~\ref{f1_sketch}) only six are independent. Any
appropriately chosen group of six should yield the same results. For
simplicity we choose to use the three $f_1^{\rm a}(i,j,Q,1)$ and the
three $f_1^{\rm a}(j,i,Q,1)$ ratios (where $Q \ge 2$). Dubbing these
the `normalised class\footnote{The term `class' refers to the
  distinction between $Q < 2$ and $Q \ge 2$.}  disagreement
fractions', $f_{\rm cl}$ (see Fig.~\ref{f1_sketch}), we find:
\begin{eqnarray}
\lefteqn{f_{\rm cl}(i,j,Q) \; \equiv \; f_1^{\rm a}(i,j,Q,1) 
\; = \; \frac{f_1(i,j,Q)}{f_1(i,j,1)}} \nonumber \\
& = &\frac{n_{ij}(Q_i = 1, Q_j = Q) \; / \; n_{ij}(Q_j = Q)}
{n_{ij}(Q_i = 1, Q_j = 1) \; / \; n_{ij}(Q_j = 1)}.
\end{eqnarray}
As described above, these are modelled by:
\begin{equation}
\label{p1model}
f_{\rm cl}(i,j,Q) = \frac{1 - p(i,1) \; p(j,Q)}{p(i,1)\;p(j,1) 
\; + \; [1-p(i,1)] \; [1-p(j,1)]}.
\end{equation}

To summarise, we use the (re-)redshifting data to derive the
sought-after probabilities $p(i,Q)$ to `correctly' identify a redshift
($Q \ge 2$) or to `correctly' identify a spectrum as not yielding a
redshift ($Q < 2$) by fitting the model of equations~\eref{pmodel} and
\eref{p1model} to all of the observed $f_z$ and $f_{\rm cl}$
simultaneously.

Before we can perform the fit, however, we need to estimate errors for
the various fractions above. We use Bayes' theorem with a uniform
prior to estimate $68$ percentile confidence intervals from the
posterior distributions which, in general, are asymmetric around the
measured values. These errors are robust even when the fractions are
based on small number statistics and/or are close to $0$ or $1$, as is
frequently the case. The asymmetry of the errors is taken into account
during the fit.

The $p(i,Q)$ values that result from the fit are shown in
Fig.~\ref{p_rs_Q} as a function of redshifter and $Q$. Note that we
arbitrarily chose to order the redshifters along the abscissa
according to their $p(i,3)$ values, which causes the apparently
regular behaviour of these values as a function of redshifter.  The
redshift data used for this fit comprise all currently available data,
i.e.\ from the original redshifting and from all three re-redshifting
exercises carried out so far. Fig.~\ref{p_rs_Q} clearly reveals the
different abilities and/or different meanings the various redshifters
have attached to the $Q$ values (note that it is generally not
possible to distinguish the two), underlining the importance of the
re-redshifting process.  Gratifyingly, almost all $p(i,4)$ values lie
at $\ge 0.95$. For several redshifters their $p(i,3)$ values are the
same as their $p(i,4)$ values to within the errors, meaning that these
redshifters essentially did not distinguish between $Q=3$ and
$4$. Others clearly made a distinction while still maintaining high
$p(i,3)$ values. However, there are also some redshifters whose
$p(i,3)$ values are clearly inconsistent with the definition of $Q=3$
as a `probably' correct redshift to be accepted for scientific
analyses. On the other hand, almost all redshifters {\em did} make a
very clear distinction between $Q=3$ and $2$, reflecting the important
distinction between the definitions of these values (i.e.\ whether the
redshift is to be accepted for scientific analyses or not).

Note that for several redshifters we find $p(i,2) < 0.5$. This does
{\em not} necessarily indicate worse-than-random performance because
for $Q \ge2$, $p$ represents a {\em multi}nomial probability. The
assumption that {\em all} redshifters perform better than random even
for $Q=2$ led us to adopt $p_{\rm a} = 0.2$ above. In contrast, for
$Q=1$, $p$ represents a {\em bi}nomial probability, and gratifyingly
we find that all $p(i,1)$ values lie well above $0.5$.

There is also a clear anti-correlation between $p(i,2)$ and
$p(i,1)$. This can be understood by considering the extremes of the
redshifters' behaviours when confronted with a spectrum where the
`correct' redshift is difficult to identify. A particularly ambitious
or conscientious redshifter will always attempt to find a redshift,
and will too often assign $Q=2$, while reserving $Q=1$ only for the
very worst spectra. More balanced redshifters will hence almost always
agree with the latter assignments, resulting in a high $p(i,1)$ value
relative to others, but less frequently with the former, resulting in
a relatively low $p(i,2)$ value. Vice versa, a redshifter at the
other extreme will too frequently assign $Q=1$ while reserving $Q=2$
for comparatively `easy' cases. Again, more balanced redshifters will
thus often agree with the latter assignments but not with the former,
resulting in relatively high $p(i,2)$ and low $p(i,1)$ values.

We point out that the model of equations~\eref{pmodel} and
\eref{p1model} with its $4 N_{\rm RS} = 220$ free parameters (where
$4$ is the number of values that $Q$ can take on) does not
in fact provide a formally acceptable fit to the $5824$ $f_z$ and
$4381$ $f_{\rm cl}$ observed data points: we find a minimum $\chi^2$
per degree of freedom of $1.23$ [$P(>\chi^2) \approx 0$]. We attribute
this to shortcomings of the model itself (see below) as well as to the
inability of our use of asymmetric errorbars in the fit to fully
capture the extreme asymmetry and non-Gaussianity of the error
distribution of $f_z$ near values of $1$ and $0$. Despite this poor
formal fit quality, the inspection of the residuals between the data
and the best-fit model inspires confidence that the fit is
nevertheless meaningful, and we find an rms of the residuals of
$1.09$. In Fig.~\ref{prms_rs_Q} we show the rms of the residuals as a
function of redshifter and $Q$. While the $Q \le 3$ values all scatter
around a value of $1.1$, the $Q=4$ values are clearly lower on
average. This offset is explained by the high $p(i,4)$ values since
the underlying assumption of a probabilistic process breaks down for
$p \approx 1$.

Fig.~\ref{prms_rs_Q} is also a useful diagnostic to detect individual
redshifters whose data cannot be fit by our model, which could be
caused, e.g., by inconsistent $Q$ assignments as a function of
time. Only one redshifter stands out (afm), with three of the four rms
values being outliers. These are explained, however, by small number
statistics, as this redshifter's results have been checked by only one
other person (sjp). Similarly, sjp's $Q=4$ value is also unusually
high, but again, this value is based on a comparison with only two
other redshifters (one of whom is afm). We thus conclude that overall
our model provides an acceptable fit to the observed redshift
agreement and normalised class disagreement fractions.

A shortcoming of our approach is that we have to measure a given
redshifter's $p$ as a function of the discrete parameter $Q$. Not only
is this parameter discrete, it will also be `fuzzy' at least to some
extent, in the sense that no redshifter can be expected to be entirely
consistent in assigning $Q$ values in borderline cases. Ideally, we
would like to measure $p$ as a function of some continuous, completely
reproducible measure of a spectrum's propensity to having its redshift
correctly identified, even if the scale of this measure varied from
redshifter to redshifter.\footnote{Note that this measure would be
  related to, but would not be synonymous with easily quantifiable
  measures of `data quality'. For example, even a spectrum with low
  continuum signal-to-noise ratio may still yield a secure redshift if
  multiple strong emission lines are present.} The difficulty of
defining such a measure, however, is the very reason why redshifters
have to assign a redshift quality in the first place. We thus have to
use $Q$ as a proxy and accept that we are unable to capture any
variation of $p(i,Q)$ within $Q$.

Similarly, we do not capture any possible variations of $p(i,Q)$ as
a function of time, which could be caused, e.g., by a redshifter
gaining more experience with the redshifting process over time. We
have attempted to eliminate this particular cause by subjecting all
redshifters new to the process to an extensive training programme
before they begin redshifting in earnest.

Finally, we note that the redshifting results of nine initial
redshifters have not yet been subjected to re-redshifting. For these
redshifters we therefore have no information regarding their $p(i,Q)$
values. Since we will need $p$ values for all redshifters in the
following, we choose to assign values of $0.9$, $0.6$, $0.9$ and
$0.95$ for $Q = 1$ to $4$, respectively. The first two are the
averages of the corresponding values in Fig.~\ref{p_rs_Q}, while the
latter two are conservatively chosen as the lowest $p$ values that
will result in redshifts marked as $Q=3$ or $4$ by these redshifters
being assigned $nQ=3$ or $4$, respectively (see Section \ref{final_z}
below).

\subsubsection{Assignment of final redshifts and qualities}
\label{final_z}

With $p(i,Q)$ values for all redshifters in hand, we can now proceed
to evaluate, for each spectrum, the relative merit of all offered
opinions for this spectrum by computing the probability that they are
`correct' (in the sense described in the previous section). For those
spectra with multiple opinions this will allow us to identify the
`best' redshift for each spectrum (i.e.\ the one most likely to be
`correct'), and to provide a homogenised measure of confidence for all
redshifts.

\begin{figure}
\includegraphics[width=\columnwidth]{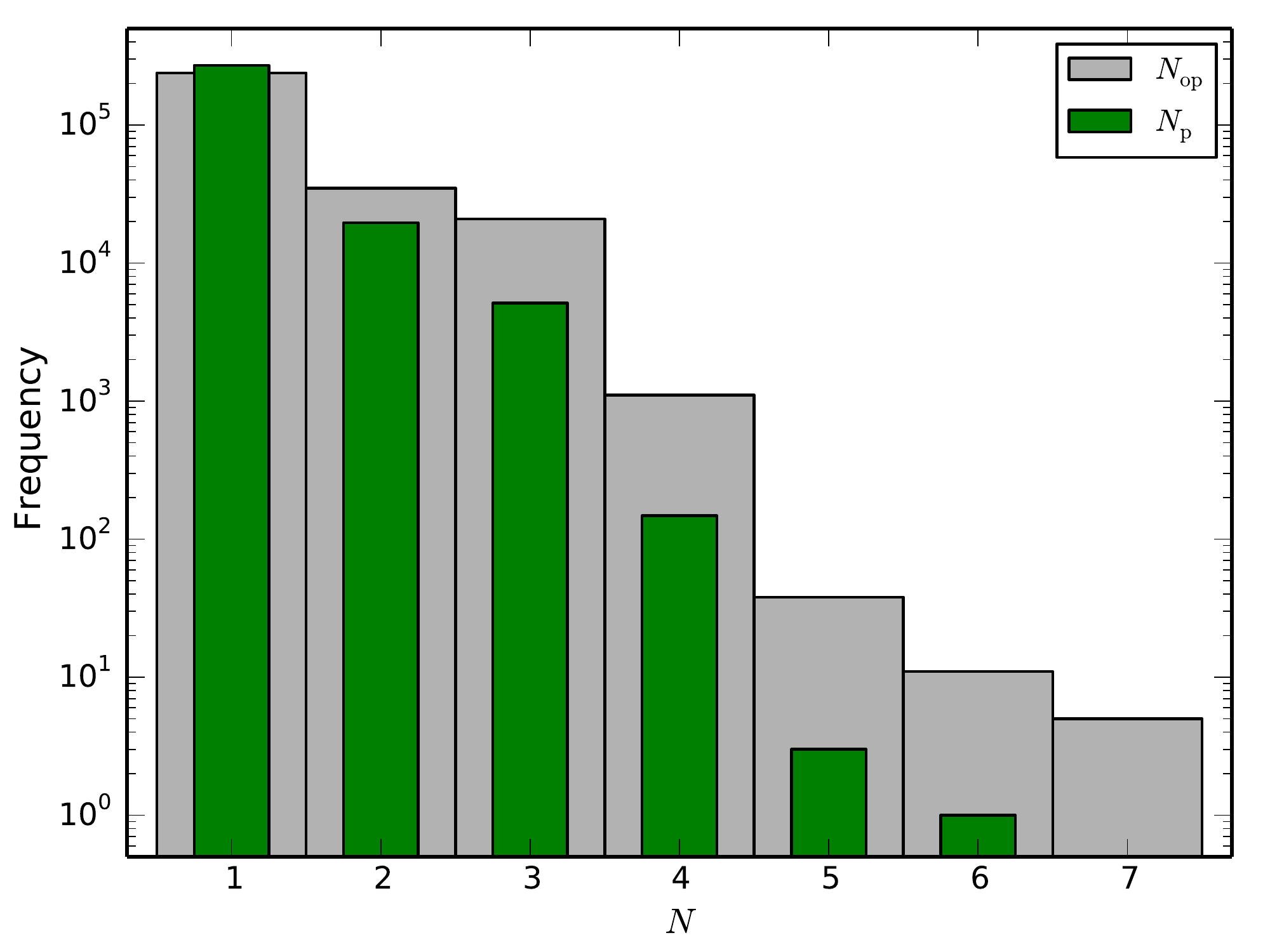}
\caption{The grey histogram shows the distribution of the number of
  offered opinions per spectrum, $N_{\rm op}$, for all GAMA\,II
  spectra of main survey targets. The green histogram shows the
  distribution of the number of possibilities per spectrum, $N_{\rm
    p}$ (see text for details).}
\label{rez_ns}
\end{figure}

Consider the general case of a spectrum for which $N_{\rm op} = N_{Q2}
+ N_{Q1}$ opinions have been offered, where $N_{Q2}$ and $N_{Q1}$ are
the number of $Q \ge 2$ and of $Q < 2$ opinions, respectively. We
begin by collating these opinions into a set of $N_{\rm p}$ distinct
`possibilities', where each possibility $x_i$ is supported by the set
of opinions $S_i$ of size $N_i$. These possibilities consist of all of
the {\em different} redshifts, $z_i$, among the offered $Q\ge2$
opinions, as well as the possibility that the spectrum does not yield
a redshift. This last possibility is of course only included if at
least one $Q=1$ opinion was offered (i.e.\ if $N_{Q1} > 0$). Thus we
have $N_{\rm p} = N_z + \delta_{N_{Q1}}$, where we define
$\delta_{N_{Q1}} = 0$ ($1$) if $N_{Q1} = 0$ ($N_{Q1} > 0$), and where
$N_z$ is the number of {\em different} redshifts found for this
spectrum ($0 \le N_z \le N_{Q2}$). To obtain these different redshifts
we sort the offered $Q \ge 2$ opinions into sets $S_i$ by identifying
groups of similar redshifts (where a `group' often consists of only a
single member, i.e.\ $N_i = 1$) using the same one-dimensional
friends-of-friends method as in Section~\ref{re_z_analysis} above,
with the same generous linking length of $\Delta z = 0.0035$. The
$z_i$ are then simply taken as the average redshifts of these groups.

For each possibility $x_i$ we now compute its probability of being
`correct' as:
\begin{equation}
\label{p_x}
p(x_i) = \frac{q(x_i)}{\sum_{j=1}^{N_{\rm p}}q(x_j) + q(c)},
\end{equation}
where
\begin{eqnarray}
\label{q_x}
q(x_i) & = & \prod_{j \in S_i} p(r_j, Q_j) \; \prod_{j \notin S_i} [1 - p(r_j,Q_j)] \nonumber\\
& &  \mbox{} \times f_i(p_{\rm a}, N_{Q2}, N_z, \{N_j\}),
\end{eqnarray}
and where $q(c)$ is the (unnormalised) probability of the complement
of all offered possibilities being `correct' (i.e.\ of the possibility
that all offered possibilities are `incorrect'). The first product in
the above equation runs over all $N_i$ opinions supporting $x_i$, and
the second product over all other (disagreeing) opinions. $r_j$ and
$Q_j$ refer to the originator and $Q$ value of opinion $j$. Note that
if $x_i$ is `correct' then all agreements on (other) redshifts must be
`accidental' [see equation~\eref{pmodel}]. $f_i$ represents the
probability of these accidental agreements, which depends on $p_a$,
$N_{Q2}$, $N_z$ and the distribution of the $N_{Q2}$ opinions among
the $N_z$ different redshifts.

Finally, we identify the `best' possibility, $x_b$, as the one with
the highest probability of being `correct'.\footnote{For spectra with
  $N_{\rm op} = 1$ this step is obviously trivial, but the procedure
  nevertheless holds.} If this `best' possibility corresponds to a
redshift, $z_b$, then this is adopted as the final redshift of the
spectrum. If, on the other hand, $x_b$ corresponds to the possibility
that the spectrum does not yield a redshift then of course the
redshift of the spectrum is undefined.

\begin{figure}
\includegraphics[width=\columnwidth]{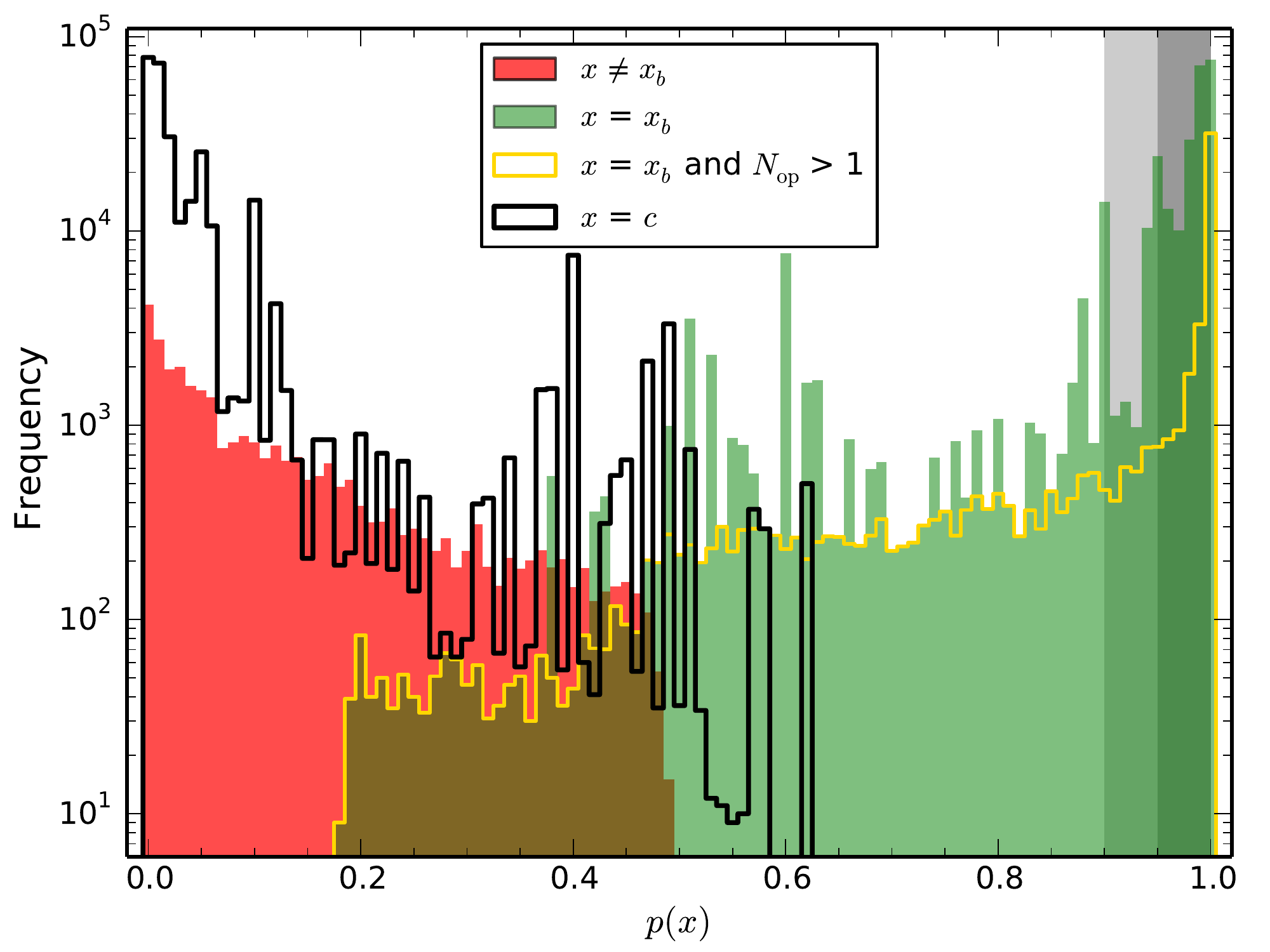}
\caption{The green histogram shows the distribution of the probability
  of the `best' possibility to be `correct' for all GAMA\,II spectra
  of main survey targets. The yellow line shows the same but only for
  those spectra that have been re-redshifted (i.e.\ those with $N_{\rm
    op} > 1$). The red histogram shows the `correctness' probability
  distribution for all other (i.e.\ `non-best') possibilities (which,
  by construction, always refer to spectra with $N_{\rm op} >
  1$). Note that this histogram cuts off at $p(x) = 0.5$, as it
  must. The black line shows the distribution of the probability of
  the complement to be correct, i.e.\ that none of the offered
  possibilities are correct, for all spectra. The many more or less
  isolated peaks in the green histogram relative to the yellow line
  are caused by spectra with $N_{\rm op} = 1$, and the peaks
  correspond to the $p(i,Q)$ values in Fig.~\ref{p_rs_Q}, except for
  those at $0.6$, $0.9$ and $0.95$. These three peaks are largely
  artificial, as they are caused by setting the $p(i,Q)$ of the nine
  untested initial redshifters to these values (see end of
  Section~\ref{re_z_analysis}). Note that all peaks are mirrored in
  the black $p(c)$ distribution. The dark and light grey shaded
  regions mark the $p(x)$ ranges to which we assign $nQ = 4$ and $3$,
  respectively (only if $x$ corresponds to a redshift).}
\label{rez_probs}
\end{figure}

In Fig.~\ref{rez_ns} we show the distributions of the numbers of
opinions and possibilities, $N_{\rm op}$ and $N_{\rm p}$, for all
GAMA\,II spectra (taken from {\tt SpecCatv27}). As mentioned in
Section~\ref{re-redshifting}, the re-redshifting campaigns have so far
only covered the data collected up to 2011 May. This resulted in
$56\,989$ spectra ($19$ per cent of the total) having $N_{\rm op} >
1$. For $24\,898$ of these spectra ($44$ per cent, $8$ per cent of the
total) there was at least some disagreement among the multiple
opinions, leading to $N_{\rm p} > 1$.

In Fig.~\ref{rez_probs} we show as the green filled histogram the
distribution of $p(x_b)$ for all GAMA\,II spectra, while the yellow
histogram shows the same for all spectra with $N_{\rm op} >
1$. Gratifyingly, these distributions are strongly peaked at $p \ga
0.93$, meaning that in general the `best' possibility is
well-distinguished from any other offered possibilities (shown in
red), as well as from the complement (shown in black). Nevertheless,
the $p(x_b)$ distribution of course extends down to quite low
values. Users of the $z_b$ should therefore define an appropriate
threshold value $p_{\rm min}$ and only include those $z_b$ in their
scientific analyses for which $p(z_b) > p_{\rm min}$ (or, more
sophisticatedly, devise a $p(z_b)$-based weighting scheme). To this
end, and to replace the familiar single-redshifter $Q$ parameter, we
have defined a normalised quality parameter $nQ$ thus:
\begin{equation}
\label{nQ}
nQ = \left \{
\begin{array}{lrcccl}
4 \qquad & 0.95 & \le & p(z_b) & \le & 1\\
3 \qquad & 0.9 & \le & p(z_b) & < & 0.95\\
2 \qquad & & & p(z_b) & < & 0.9\\
\end{array}\right .
\end{equation}
and $nQ = 1$ is assigned to those spectra where $x_b$ corresponds to
the possibility that the spectrum does not yield a redshift. The
$p(z_b)$ ranges above are somewhat arbitrary, but they were chosen at
an early stage of the survey on the basis of an earlier version of
Fig.~\ref{p_rs_Q} to roughly reflect the meanings of the
single-redshifter $Q$ values. In particular, the intention was to make
$nQ = 3$ and $2$ the divide between accepting a redshift for
scientific analyses and not accepting it, thus reflecting the
distinction between $Q = 3$ and $2$. In other words, we have set
$p_{\rm min} = 0.9$. This is the value we have used in all of our own
studies using these redshifts to date.  However, we stress that in
principle the choice of $p_{\rm min}$ may depend on the scientific
application at hand, and users may wish to define $p_{\rm min}$
differently or indeed use a more sophisticated scheme than a simple
threshold.

We have thus finally achieved what we set out to do: we have
unambiguously identified, in a probabilistic manner, the `best'
redshift $z_b$ (or else that no redshift can be determined) for {\em
  every} spectrum, and we have computed a homogeneous quality measure
for these redshifts [$p(z_b)$ and $nQ$], free of the idiosyncrasies of
individual redshifters,\footnote{The only caveat to this statement is
  the fact that the $p(i,Q)$ values of nine of the initial redshifters
  have not yet been measured, as mentioned in the previous section.}
thus allowing us to statistically treat all spectra the same,
regardless of the number of opinions available for them.

\subsubsection{Overall effect of re-redshifting}

\begin{figure}
\includegraphics[width=\columnwidth]{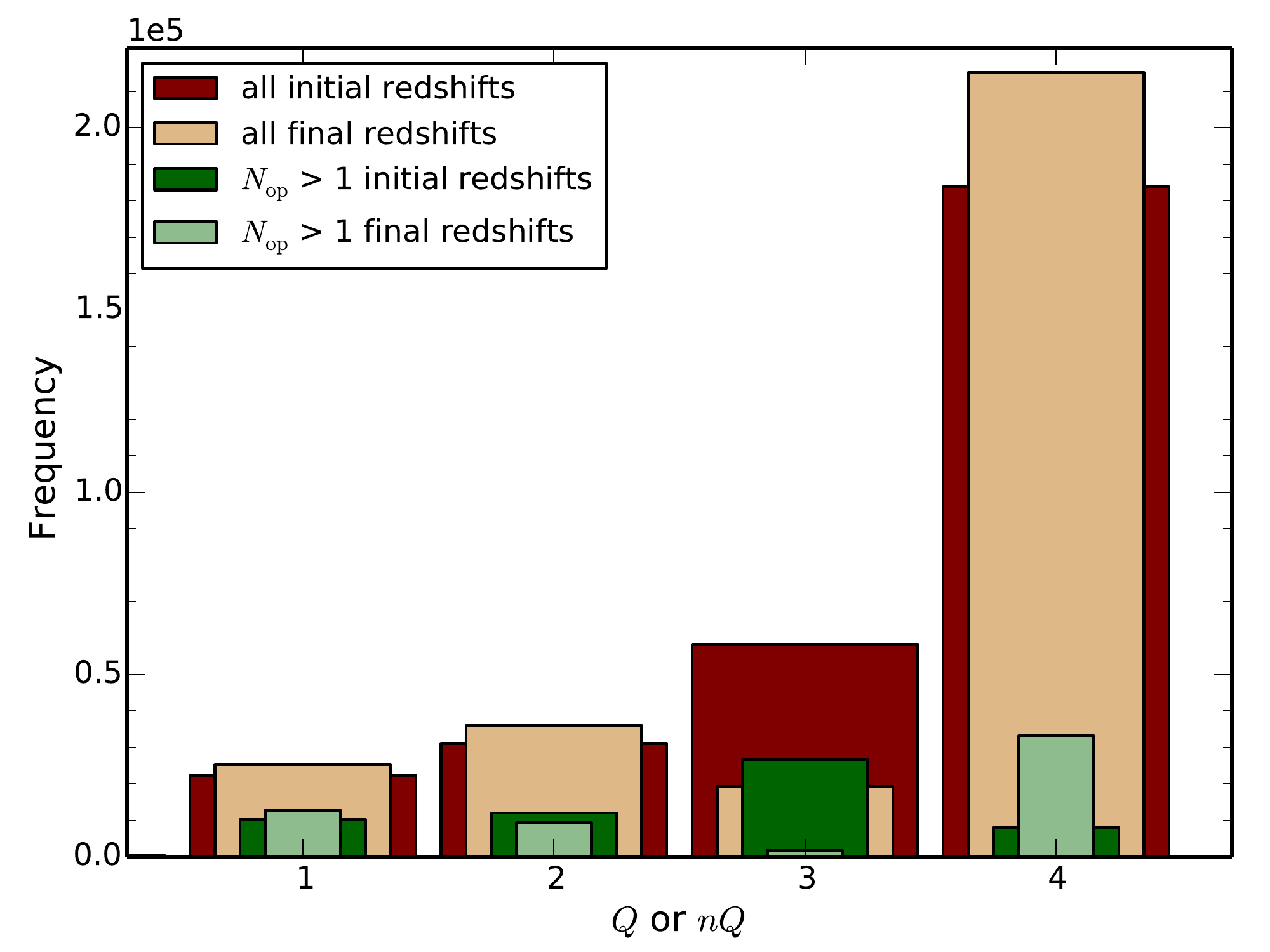}
\caption{The dark green histogram shows the $Q$ distribution of the
  initial redshifts of all GAMA\,II spectra of main survey targets with
  $N_{\rm op} > 1$. The light green histogram shows the $nQ$ histogram
  of the final redshifts for the same spectra. For comparison, the
  brown and beige histograms show the same but now for all GAMA\,II
  spectra.}
\label{rez_qs}
\end{figure}

We now briefly compare the distributions of the initial redshifts and
qualities ($z_{\rm ini}$ and $Q$) to those of the final redshifts and
qualities ($z_b$ and $nQ$) in order to illustrate the overall effect
of the re-redshifting. 

As explained above, one of the purposes of re-redshifting is to
confirm or reject redshifts initially classified as `probable' or
`possible', i.e.\ as $Q=3$ or $Q=2$. Thus we expect the $Q$ and $nQ$
distributions of the initial and final redshifts to differ. In
Fig.~\ref{rez_qs} we plot these distributions as the dark and light
green histograms, respectively, for all spectra with $N_{\rm op} > 1$,
i.e.\ for all spectra that have been re-redshifted. The two
distributions are indeed very different, with the relative heights of
the $Q$,$nQ = 3$,$4$ bars roughly interchanged.\footnote{The
  relatively small number of $Q=4$ initial redshifts in this sample is
  of course due to the way in which we selected spectra for
  re-redshifting (see Section~\ref{re_z_selection}).} The difference
between these distributions is the net result of the initially
uncertain redshifts either being strongly confirmed or clearly
rejected: $75$~per~cent of $Q=3$ redshifts were strongly confirmed,
and thus received $nQ=4$, while $21$~per~cent were not confirmed, and
thus received $nQ=2$ ($17$~per~cent) or $nQ=1$ ($4$~per~cent). Only
$4$~per~cent were confirmed but remained somewhat uncertain, and thus
received $nQ=3$. Similarly, of the $Q=2$ redshifts, $37$~per~cent were
strongly confirmed ($nQ=4$), while $59$~per~cent were not confirmed
(equally split among $nQ=2$ and $1$), and again only $4$~per~cent
received $nQ=3$. We also note in passing that $80$~per~cent of spectra
with $Q=1$ remained at $nQ=1$ (with a further $9$~per~cent receiving
$nQ=2$), and that $96$~per~cent of redshifts with $Q=4$ were clearly
confirmed.

Since only part of the data have been re-redshifted, the effect of the
re-redshifting on the full dataset is not quite as dramatic, as
evidenced by the brown and beige histograms in Fig.~\ref{rez_qs}. The
change in the relative heights of the $Q$,$nQ = 3$,$4$ bars is
nevertheless quite clear.

\begin{figure}
\includegraphics[width=\columnwidth]{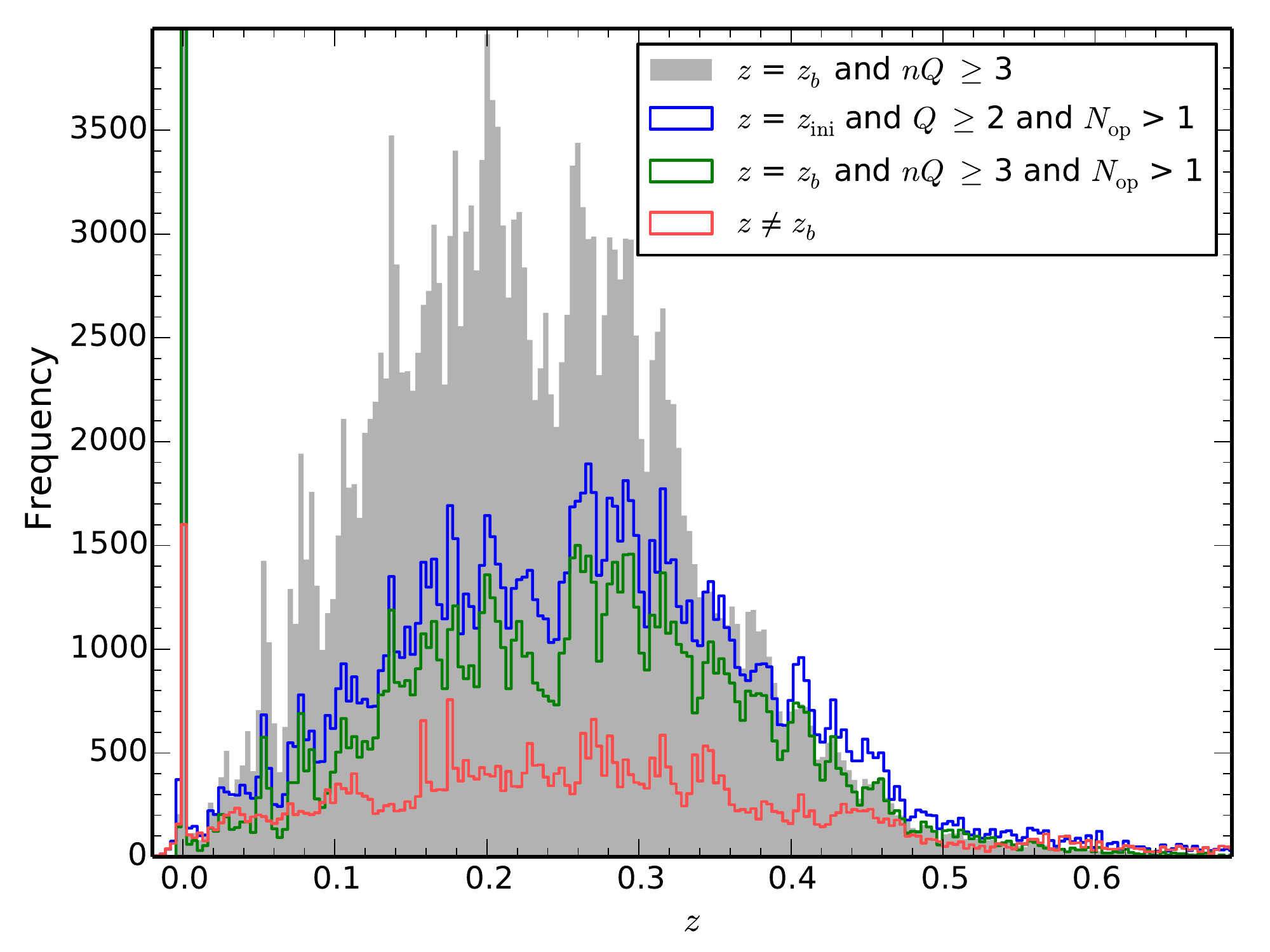}
\caption{The grey shaded histogram shows the distribution of the
  finally assigned (i.e.\ `best') redshifts with $nQ \ge 3$ for all
  GAMA\,II spectra of main survey targets. The green line shows the
  same but only for those spectra that have been re-redshifted
  (i.e.\ those with $N_{\rm op} > 1$). For these same spectra the
  blue line shows the distribution of the initial redshifts
  with $Q \ge 2$. Finally, the red histogram shows the distribution of
  the redshifts that are not the `best'. For clarity, the green and
  blue histograms have been multiplied by a factor of~$3$, while the
  red histogram has been multiplied by~$2$.}
\label{rez_zs}
\end{figure}

In Fig.~\ref{rez_zs} we show as the blue histogram the distribution of
the initial redshifts with $Q\ge2$ for spectra with $N_{\rm op} >
1$. In other words, this is the distribution of the redshifts that
went into the re-redshifting process. We first of all note in passing
that these redshifts are not a random sub-set of the overall redshift
sample, which is shown as the grey filled histogram.\footnote{The
  striking gap in this distribution at $0.225 \la z \la 0.25$ raises
  the question whether some property of our spectra or of the
  redshifting process systematically prevents us from successfully
  identifying redshifts in this range. This is not the case, since the
  gap is only evident in the equatorial survey regions (the data from
  which dominate this distribution), but not in G02 or G23.} Clearly, the
blue distribution is skewed towards higher values, meaning that the
initial redshifters tend to be more uncertain when assigning higher
redshifts.

The main point of Fig.~\ref{rez_zs}, however, is to compare the
redshift distributions before and after re-redshifting. To this end we
show as the green histogram the distribution of the redshifts that
came out of the re-redshifting process, i.e.\ the distribution of the
final redshifts with $nQ \ge 3$ for spectra with $N_{\rm op} > 1$. We
find that the blue and green distributions are reasonably similar,
both on small and large scales, and so we conclude that the
re-redshifting does not alter the redshift distribution dramatically.

Finally, we show as the red histogram the distribution of the
redshifts that were {\em not} identified as the `best' redshift. In
other words, these are redshifts for spectra for which at least one
other, more likely `correct' redshift has been found. This
distribution is clearly quite different from the others, significantly
broader and not reproducing the same peaks on small scales.
Furthermore, the two most pronounced peaks in this distribution,
namely those at $z \approx 0.16$ and $0.175$, are clearly due to the
frequent misidentification of the residuals of certain strong sky
features: at $z = 0.159$, H$\alpha$ is shifted to the blue trough of
the telluric O$_2$ A-band at $7606$~\AA, while at $z = 0.175$ both
H$\alpha$ and the [\ion{Si}{ii}] $\lambda 6731$ line happen to
coincide exactly with two prominent atmospheric OH lines. Based on our
past experience with {\sc runz} and similar AAT data in the context of
the 2dFGRS and MGC surveys, these two peaks were in fact expected. We
thus find that the distribution of the `non-best' redshifts inspires
confidence in our selection of the `best' redshifts. It should be kept
in mind, however, that the redshifting mistakes represented by the red
histogram are still present among the data that have not yet been
re-redshifted (cf.\ also Section~\ref{zqc}).

\subsection{Fully automated redshifts using \textsc{Autoz}}
\label{autoz}

As we already mentioned at the beginning of Section~\ref{redshifting},
in 2013 we completed the development of a new and fully automated
redshifting code called {\sc Autoz}. This new code was fully described
and illustrated by \citet{Baldry14}. In brief, it determines redshifts
using cross-correlation of our survey spectra with galaxy and stellar
templates. Note that no QSO templates are included at present, meaning
that the redshift of a spectrum with broad emission lines cannot be
confidently identified using {\sc Autoz}. The stellar templates were
taken from SDSS DR5\footnote{{\tt
    http://classic.sdss.org/dr5/algorithms/spectemplates/}} (IDs
$0$--$22$), and we created eight galaxy templates from the BOSS galaxy
eigenspectra (\citealp{Bolton12}). Both the template and survey
spectra were robustly high-pass filtered prior to cross-correlation.
In addition, each high-pass filtered spectrum is clipped so that the
deviations lie within plus or minus thirty times the mean absolute
deviation. This reduces the impact on the cross-correlation function
from strong lines or unknown bad data, which could give rise to false
peaks. The aim was to make the code robust to spectrophotometric
uncertainties and artefacts.

The best-estimated redshift for each survey spectrum is taken from the
highest cross-correlation peak, normalised by a root mean square
value, across all the templates. The allowed redshift range for the
galaxy templates is up to $0.9$. For each redshift we estimate a
figure of merit (FOM) primarily by comparing the height of the highest
correlation peak with those of the next three best redshifts (outside
$600$\kms\ from each other). We then derive the redshift confidence,
i.e.\ the probability that the redshift is correct, $p(z)$, from the
redshift's FOM. The relation between these parameters is calibrated
using duplicate observations of the same targets. Finally, analogous
to our procedure used for {\sc runz} (see Section~\ref{final_z}), we
define a quality parameter $nQ$ based on the value of $p(z)$. However,
this time we are slightly more conservative by assigning $nQ=4$ only
to redshifts with $p(z) \ge 0.98$ [cf.\ equation~\eref{nQ}].

In Section~\ref{zqc} we will compare the performance of {\sc Autoz}
with that of {\sc runz}, both in terms of the precision of the
redshifts, and in terms of the fraction of redshifts that are
catastrophically wrong. As we shall see, {\sc Autoz} turns out to be
superior to {\sc runz} in all respects, and therefore we adopted the
{\sc Autoz} redshifts as the default for GAMA\,II in 2013 (although
for quality control purposes we have continued to measure redshifts
with {\sc runz} as well). The {\sc Autoz} redshifts have already been
used in some of the most recent GAMA publications.

\subsection{Improving redshift confidence using combined spectra}
\label{combspec}

As we will describe in more detail in Section~\ref{progress} below, a
main survey target that was unsuccessfully observed, in the sense that
its spectrum did not yield a robust (i.e.\ $nQ\ge3$) redshift,
remained on the target list until a subsequent observation proved
successful. Many targets were thus observed more than once. For some,
however, {\em all} of the spectra obtained are of insufficient quality for
{\sc Autoz} to be able to reliably measure a redshift from these
individually. With the survey now completed, and thus with no further
re-observations forthcoming, the question arises whether we can
nevertheless obtain reliable redshifts for at least some of these
objects by combining their spectra together and using the combined,
higher S/N spectra for the redshift measurements.

For all objects with multiple spectra that do not already have a
high-quality ($nQ = 4$) {\sc Autoz} redshift from one of these we thus
combine their high-pass filtered and clipped spectra and attempt to
measure a redshift from the combined spectrum as described in the
previous section. If the redshift measured from the combined spectrum
has a higher FOM than those measured from the individual spectra then
the redshift from the combined spectrum is used for this object. A
total of $5348$ objects thus receive a `new' redshift with an improved
redshift confidence, increasing the number of main survey objects with
a reliable ($nQ \ge 3$) redshift by $1654$. Note that the `new'
redshift may or may not be different from the redshifts measured from
the individual spectra, but it always has an improved confidence.

\subsection{AGN redshifts}
\label{agnz}

Since {\sc Autoz} does not use any QSO templates and does not consider
redshifts $> 0.9$ (see Section~\ref{autoz}), it often fails to
reliably identify a redshift for AGN spectra. Since these spectra
display prominent emission lines, however, their redshifts are usually
reliably determined by {\sc runz}. For spectra of main survey objects
without any good ($nQ \ge 3$) redshift from either {\sc Autoz} or from
a previous survey (see Section~\ref{previous}) we thus continue to use
their {\sc runz} redshift if $nQ_{\mbox{\small \sc runz}} = 4$ and if
$z_{\mbox{\small \sc runz}} > 0.9$ or the spectrum is flagged as an
AGN by a {\sc runz} redshifter. Thus we `recover' the redshifts of
$283$ main survey objects.

\begin{table*}
\begin{minipage}{14.7cm}
\caption{Details of the publicly available spectroscopic data we have
  merged with the GAMA\,II survey.}
\label{prevspec}
\centerline{\begin{tabular}{llrrrl}
\hline
Survey & Provides data in & No.\ of spectra & No.\ of objects$^a$ & 
No.\ of MS objects$^b$ & Reference\\
\hline
SDSS/BOSS DR10 & G02, G09, G12, G15 & $102\,160$ & $61\,986$ & $25\,625$ & \citet{Ahn14}\\
2dFGRS    & G02, G12, G15, G23 &  $31\,300$ & $26\,836$ & $19\,599$ & \citet{Colless01}\\
MGC$^c$	  & G12, G15	       &  $4551$    &  $4098$ & $2078$ & \citet{Driver05}\\
6dFGS$^d$ & All                &  $1894$    &  $1529$ & $1108$ & \citet{Jones09}\\
2QZ$^e$	  & G12, G15, G23      &  $12\,053$ &  $7620$ & $695$ & \citet{Croom04b}\\
2SLAQ$^f$-LRG& G09, G12, G15   &  $3150$    &  $1735$ & $300$ & \citet{Cannon06}\\
WiggleZ$^g$& G09, G15          & $29\,499$  &  $3258$ & $166$ & \citet{Parkinson12}\\
VVDS$^h$  & G02	               & $12\,481$  &   $177$ & $109$ & \citet{LeFevre13}\\
2SLAQ-QSO & G09, G12, G15      &  $3603$    &  $1012$ & $81$ & \citet{Croom09}\\
UZC$^i$   & G09, G12, G15      &    -       &   $377$ & $269$ & \citet{Falco99}\\
NED$^j$   & G12, G15           &    -       &     $5$ & $5$\\
\hline
Total & & $200\,691$ & $95\,488$ & $41\,747$\\
With $nQ\ge3$ & & & $92\,090$ & $40\,901$\\
\hline
\end{tabular}}
$^a$Number of unique matched GAMA\,II objects (not limited to main survey targets);
the totals account for inter-survey duplications.\\
$^b$Number of unique matched GAMA\,II main survey objects;
the totals account for inter-survey duplications.\\
$^c$Millennium Galaxy Catalogue;
$^d$6dF Galaxy Survey;
$^e$2dF QSO Redshift Survey;
$^f$2dF SDSS LRG and QSO survey;
$^g$WiggleZ Dark Energy Survey;
$^h$VIMOS VLT Deep Survey;
$^i$Updated Zwicky Catalog;
$^j$NASA/IPAC Extragalactic Database; UZC and NED provide only redshifts,
not spectra.\\
\end{minipage}
\end{table*}

\subsection{Spectra and redshifts from other surveys}
\label{previous}

Previous spectroscopic surveys already obtained spectra and redshifts
for a significant number of GAMA\,II main survey objects. These
objects were targeted by GAMA with lower priority (depending on the
quality of the pre-existing redshift) than previously unobserved
objects. The GAMA\,II survey dataset by itself is thus not complete
and it needs to be merged with the data from these previous surveys in
order to obtain a complete sample.

We have thus downloaded all publicly available spectra and redshifts
in the GAMA\,II survey regions as detailed in
Table~\ref{prevspec}. Note that we did not restrict ourselves to data
for main survey targets. We also included all available data within
$0.5$~deg of the nominal GAMA\,II survey regions. These additional
data may be useful in the future for mitigating edge effects when
determining the environments of GAMA main survey objects. We also
included all duplicate observations for completeness. Our current
sample of `external' spectra represents a significant update and
extension of the earlier samples described by \citet{Baldry10} and
\citet{Driver11}.

The external spectra were associated with GAMA objects by positional
matching. To be able to resolve duplications, and thus to merge the
samples from the different surveys with each other and with the
GAMA\,II sample, it was necessary to define a common (preferably
homogeneous) redshift quality parameter. We have thus translated the
various quality parameters of the different surveys to our $nQ$ system
(see Section~\ref{final_z}). This was straightforward for almost all
of the surveys since they used simple quality parameters very similar
to ours. The only exception was the SDSS for which we used the
following $nQ$ definition:
\begin{eqnarray}
\lefteqn{nQ = 1 + (\Delta \chi^2_\nu > 0.001) + (\mbox{\textsc{zwarning}} == 0)} \nonumber\\
& & \mbox{} + (\mbox{\textsc{zwarning}} == 0 \mbox{  AND  } \Delta \chi^2_\nu  > 0.05)\nonumber\\
& & \mbox{} + (\mbox{\textsc{zwarning}} == 0 \mbox{  AND  } \Delta \chi^2_\nu  > 0.2) 
\end{eqnarray}
where each of the terms takes on the value of $1$ if the condition
inside the parentheses is true and $0$ otherwise. $\Delta \chi^2_\nu$
is the difference between the reduced $\chi^2$ of the best and the
second best redshifts as determined by the SDSS, and {\sc zwarning} is
the SDSS redshift warning flag. Note that for SDSS redshifts $nQ$ may
take on a value of $5$ which we do not use for any other survey
including GAMA. This is owed to the exceptional reliability of these
redshifts.

Although main survey objects with a good (i.e.\ $nQ\ge3$) pre-existing
redshift from a previous survey were targeted only with a lower
priority than previously unobserved objects (depending on the value of
$nQ$) there is nevertheless significant overlap between the sample of
external spectra and the GAMA sample for these objects: of the
$40\,901$ main survey objects that have at least one $nQ\ge3$ redshift
from one of the other surveys, $16\,266$ ($40$~per~cent) also have at
least one $nQ\ge3$ redshift from GAMA\,II. This helps in improving the
overall homogeneity of the combined sample, both in terms of the
redshifts as well as in terms of the spectra, especially when
considering that, unlike GAMA and SDSS spectra, the spectra from all
of the other surveys are not flux calibrated.

\subsection{Additional observations of bright targets}
\label{LTObs}

Recall that the GAMA spectroscopic survey was carried out with the
AAOmega multi-fibre spectrograph on the AAT. For such instruments
observations of very bright targets may lead to cross-talk between
adjacent spectra on the detector. To avoid this the GAMA target
selection for AAT observations included a bright magnitude limit
(GAMA\,I: $r_{\rm fib} > 17.0$~mag, \citealp{Baldry10}; GAMA\,II:
$r_{\rm fib} > 16.6$~mag, where $r_{\rm fib}$ is the SDSS $r$-band
fibre magnitude). Most objects brighter than this limit had already
been observed by one of the previous spectroscopic surveys as
discussed in the previous section. Here we briefly describe
observations using the robotic Liverpool Telescope (LT) of $20$
targets that were too bright for the AAT, and which had no
pre-existing data.

All $20$ targets were observed between 2009 November and 2010 June
with FRODOSpec, an integral field spectrograph consisting of a $12
\times 12$ lenslet array coupled to a dual-beam spectrograph using
fibres \citep{MoralesRueda04}. Two consecutive exposures (usually of
$500$~s each) were taken of each target using the $R \approx 2200$
gratings. Unfortunately, the blue spectrograph arm had significantly
reduced throughput at the time and so only the red-arm data
($580$--$940$~nm) was usable. We reduced the data using the pipeline
by \citet*{Barnsley12} to the point where it provides a non-sky
subtracted datacube (later stages of the pipeline were designed with
point sources in mind). Summed-flux images were then used to determine
object and sky apertures. Cosmic rays were rejected before combining
the object and sky spectra across these apertures and finally
producing the integrated, sky-subtracted object spectrum.

To determine the redshifts, the spectra were cross-correlated with the
stellar and galaxy templates (IDs $0$--$15$ and $23$--$27$,
respectively) used by the SDSS. Only a generic telluric correction was
applied to each spectrum so one or two of the strongest telluric
regions were masked to avoid spurious cross-correlations. The redshift
range allowed was $-0.002$ to $0.002$ for the stellar templates and
$0.002$ to $0.2$ for the galaxy templates. For each spectrum, the
best-matching template was selected by comparing the peak of the
cross-correlation function in the allowed range, divided by its rms in
the range $-0.1$ to $0.2$. This parameter was also used in assessing
the quality of the final, selected redshift. Only one of the $20$
redshifts was assigned $Q = 2$, all others received $Q = 3$. Half of
the sample were identified as stars.

\begin{table}
\begin{minipage}{\columnwidth}
\caption{Global statistics of the GAMA\,II spectroscopic survey after
  the completion of all observations. Note that these numbers include
  the data from the full G02 region, not just from its smaller,
  high-priority sub-region.}
\label{survey_stats}
\begin{tabular}{lrl}
\hline
Survey parameter & & Comment\\
\hline
AAT observations:\\
Allocated nights & $209.5$ & \\
Useful$^a$ nights & $130.9$ & $62.5$~per~cent\\
Observed 2dF fields & $930$ & $4.4$ ($7.1$) / night (useful night)\\
On-sky$^b$ fibres & $344\,460$ & $370.4$ / field\\
Unused$^c$ fibres & $646$\\
Broken fibres & $20\,517$ & $6.0$~per~cent of on-sky fibres\\
Object spectra & $321\,465$ & $24.7$ sky spectra / field\\
Spectra of galaxy targets & $318\,550$ & $342.5$ / field, $3.1$ stars$^d$ / field\\
--- \raisebox{-0.5ex}{''} --- with $nQ \ge 3$ & $275\,424$ & $86.5$~per~cent gross $z$ success\\
Spectra of MS$^e$ targets & $295\,853$ & $318.1$ / field\\
--- \raisebox{-0.5ex}{''} --- with $nQ \ge 3$ & $259\,720$ & $87.8$~per~cent gross $z$ success\\
Unique MS targets observed & $245\,424$ & $263.9$ / field\\
--- \raisebox{-0.5ex}{''} --- with $nQ \ge 3$ & $237\,900$ & $96.9$~per~cent net $z$ success\\
\hline
\multicolumn{3}{l}{Including spectra from previous surveys and GAMA LT observations:}\\
Spectra of galaxy targets$^f$ & $517\,979$ &\\
--- \raisebox{-0.5ex}{''} --- with $nQ \ge 3$ & $456\,649$ &\\
Spectra of MS$^e$ targets & $354\,059$ &\\
--- \raisebox{-0.5ex}{''} --- with $nQ \ge 3$ & $318\,256$ &\\
Unique MS targets observed & $270\,710$ &\\
--- \raisebox{-0.5ex}{''} --- with $nQ \ge 3$ & $263\,719$ & $88.1$~per~cent from GAMA\\
\hline
\end{tabular}
$^a$Excluding downtime due to adverse weather and technical problems, as
assessed by the observers.\\
$^b$Excluding guide fibres.\\
$^c$Fibres that could not be allocated to any targets due to
fibre collisions.\\
$^d$Used for flux calibration.\\
$^e$Main survey.\\
$^f$Includes objects outside of the nominal GAMA\,II survey regions,
see Section~\ref{previous}.
\end{minipage}
\end{table}

\section{End of survey report and quality control}
\label{progress_qc}

The $6.5$-year observing campaign for the GAMA\,II spectroscopic survey
using the 2dF/AAOmega facility on the AAT came to an end in 2014.
While the equatorial survey regions (G09, G12 and G15) were completed
as planned, achieving an exceptionally high redshift completeness of
$98.5$~per~cent, the southern regions (G02 and G23) could
unfortunately not be completed within their original scope. 

For G02 it became clear in 2013 that the allocated observing time
would not suffice to complete this region in full, and so from then on
observations focused on what we have termed the `high-priority
sub-region' of G02 north of declination $-6$~deg
(cf.\ Table~\ref{gamaregions}). By this time, however, significant
observing effort had already been spent on the rest of G02, and so we
have continued to consider all of G02 to be part of the GAMA\,II
survey. In the high-priority sub-region the final redshift
completeness is $95.0$~per~cent, while in the full region it is
$54.5$~per~cent. Note that the full G02 sample, despite its low
completeness, is nevertheless very valuable for the identification of
AGN and members of galaxy clusters detected by the XXL survey
of the same region \citep{Pierre11}.

Similarly, in early 2014 it became clear that G23, too, could not be
completed as planned. In contrast to G02 though, this region was not
yet in an advanced state of completion. Hence we descoped G23 both in
terms of its size as well as in terms of its magnitude limit
(cf.\ Table~\ref{gamaregions}), but did so in a way that minimised the
`loss' of already observed objects while still allowing us to complete
the region within the allocated time. In the end, we were able to
achieve a redshift completeness of $94.2$~per~cent in G23.

In the following we will discuss the survey's progression and
observing efficiency in some detail, and we will present a number of
diagnostics that characterise the quality of final GAMA\,II
spectroscopic dataset.

\subsection{Survey progression and efficiency}
\label{progress}

\begin{figure}
\includegraphics[width=\columnwidth]{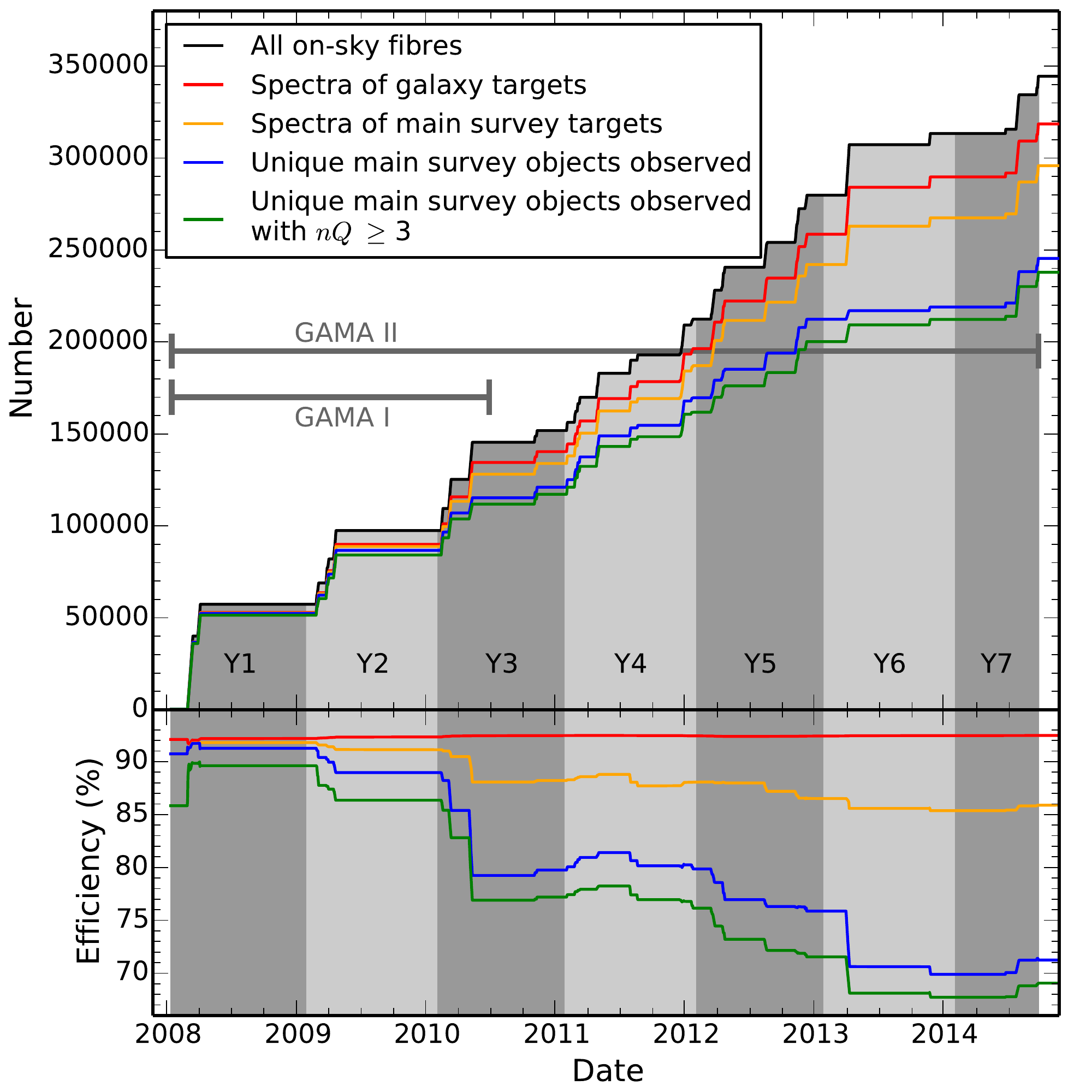}
\caption{Evolution of the GAMA\,II spectroscopic survey on the
  AAT. The upper panel shows the progress of the survey in terms of
  the numbers of on-sky fibres, target spectra and unique objects
  observed, as indicated. The lower panel shows various measures of
  the average past survey efficiency, i.e.\ the numbers of the upper
  panel relative to the number of on-sky fibres. The colour coding is
  the same as in the upper panel. The temporary increase of the
  efficiency in 2010--2011 corresponds to the expansion of the survey
  from GAMA\,I to GAMA\,II (cf.\ Section~\ref{inputcat}).}
\label{survey_progress}
\end{figure}

\begin{figure*}
\includegraphics[angle=270,width=\textwidth]{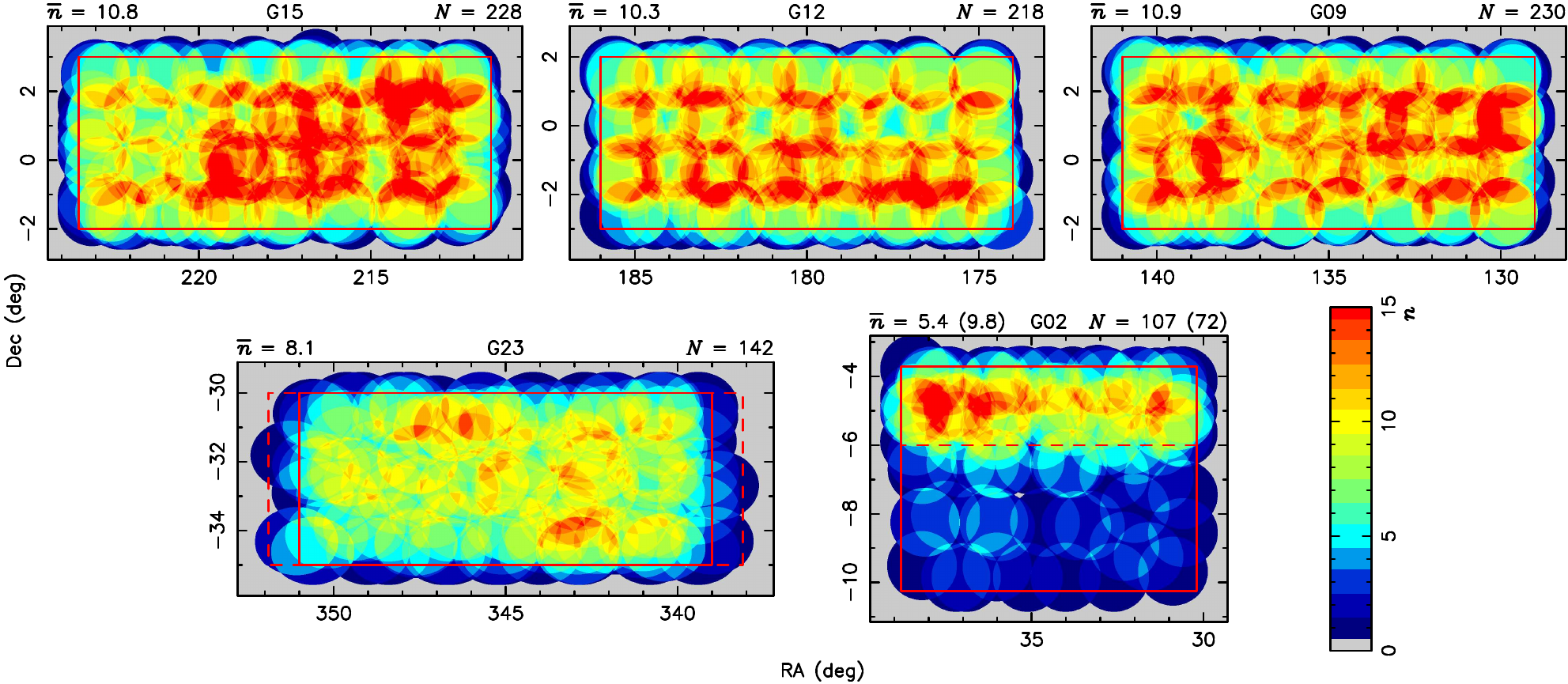}
\caption{Distribution of observed 2dF/AAOmega fields (assumed to be
  circles of $1$~deg radius) in the GAMA\,II survey regions. The
  nominal survey regions are shown as red rectangles. The dashed red
  lines in G23 indicate the original, slightly more extended region,
  while the dashed line in G02 shows the lower declination limit of
  the high-priority sub-region (cf.\ Table~\ref{gamaregions}). The
  colour scale indicates the number of fields, $n$, covering a given
  position. In each panel, the average value of $n$ within the nominal
  survey region, $\bar{n}$, is indicated at the top left, while the
  total number of fields in each region, $N$, is indicated at the top
  right. The numbers in parentheses in the G02 panel refer to the
  high-priority sub-region.}
\label{tdf_coverage}
\end{figure*}

The GAMA\,II spectroscopic survey has been carried out over a total of
$209.5$ nights, spread over $31$ observing runs, in the period 2008
February to 2014 September. Of these, we estimate that only
$63$~per~cent were useful, mostly due to exceptionally bad weather in
the period 2010--2012. During this time we have successfully observed
$930$ 2dF/AAOmega fields, resulting in $295\,853$ spectra of
$245\,424$ unique main survey objects. For $237\,900$
($96.9$~per~cent) of these we have been able to measure a secure
(i.e.\ $nQ \ge 3$) redshift using {\sc Autoz}
(cf.\ Sections~\ref{autoz}--\ref{agnz}). Merging these data with
publicly available spectra from previous surveys in the GAMA\,II
regions (see Section~\ref{previous}) and the GAMA LT spectra
(Section~\ref{LTObs}) increases these numbers to $354\,059$ spectra of
$270\,710$ unique main survey objects, of which $263\,719$ have a
secure redshift. Additional global statistics of the survey are
provided in Table~\ref{survey_stats}, and the evolution with time of
some of these parameters is shown in Fig.~\ref{survey_progress}.

Although the 2dF/AAOmega facility nominally has $400$ fibres, on
average only $342.5$ of these were available for science targets. The
rest ($14.4$~per~cent) were either needed to guide the telescope or to
observe calibration spectra (sky and flux calibration stars), or they
were broken.

Initially, the density of main survey targets exceeded the density of
available fibres by a factor of $\sim$$9.8$. Despite this high value,
not all fibres available for {\em science} targets could be allocated
to {\em main survey} targets. Fibre placement restrictions and a low
density of main survey targets once a survey region neared completion
(which forced us to allocate fibres to filler targets,
cf.\ Section~\ref{inputcat}) reduced the average number of fibres
available for main survey targets to $318.1$.

The first observation of a main survey target led to a secure
(i.e.\ $nQ \ge 3$) redshift only in $90.9$~per~cent of all cases. This
was not high enough to meet our (secondary) survey requirement of
$99$~per~cent redshift completeness at all magnitudes down to the
survey limit \citep{Robotham10}. Since the high initial target density
necessitated multiple visits to every patch of sky in any case,
unsuccessfully observed main survey targets thus remained on the
target list until a robust redshift had been obtained, although with a
progressively decreasing priority. As a result, $42\,241$ main survey
targets were observed more than once. These duplicate observations
raised the fraction of observed unique main survey targets with robust
redshifts to $96.9$~per~cent.\footnote{These numbers are based on the
  {\sc Autoz} redshifts. The equivalent numbers using the {\sc runz}
  redshifts are somewhat lower: $82.8$~per~cent for the redshift
  success of the initial observation, $90.2$~per~cent for the final
  fraction of main survey targets with a robust redshift (both
  including re-redshifting). Note that {\sc Autoz} only became
  available in 2013. Until then, the decision on whether to re-observe
  a given object was obviously based on the {\sc runz} results.} On
average, the survey has thus produced robust redshifts for $1136$
unique main survey objects per allocated night ($1817$ per useful
night).

In view of this number, the question arises whether the survey has
made optimal use of its allocated time. In other words, could the
survey have progressed any faster? To answer this question let us
decompose the progression rate into a product of three factors:
(i)~the number of observed 2dF/AAOmega fields per allocated night;
(ii)~the number of on-sky fibres per field; and (iii)~the number of
main survey targets with a robust redshift per on-sky fibre (for which
we will use the term `efficiency' hereafter). While each of these
factors in turn depends on a number of parameters, for the first two
factors we could control only one of these. First, we maximised the
number of observed fields per night by reducing the exposure time per
field to its smallest sensible value (which is set by the time
required by the 2dF positioner to configure the following observing
plate). Second, to maximise the number of on-sky fibres per field,
all we could do was to ensure that essentially no fibres were left
unused at any time. The question of the survey's optimal progression
thus boils down to its efficiency.

\begin{figure}
\includegraphics[width=\columnwidth]{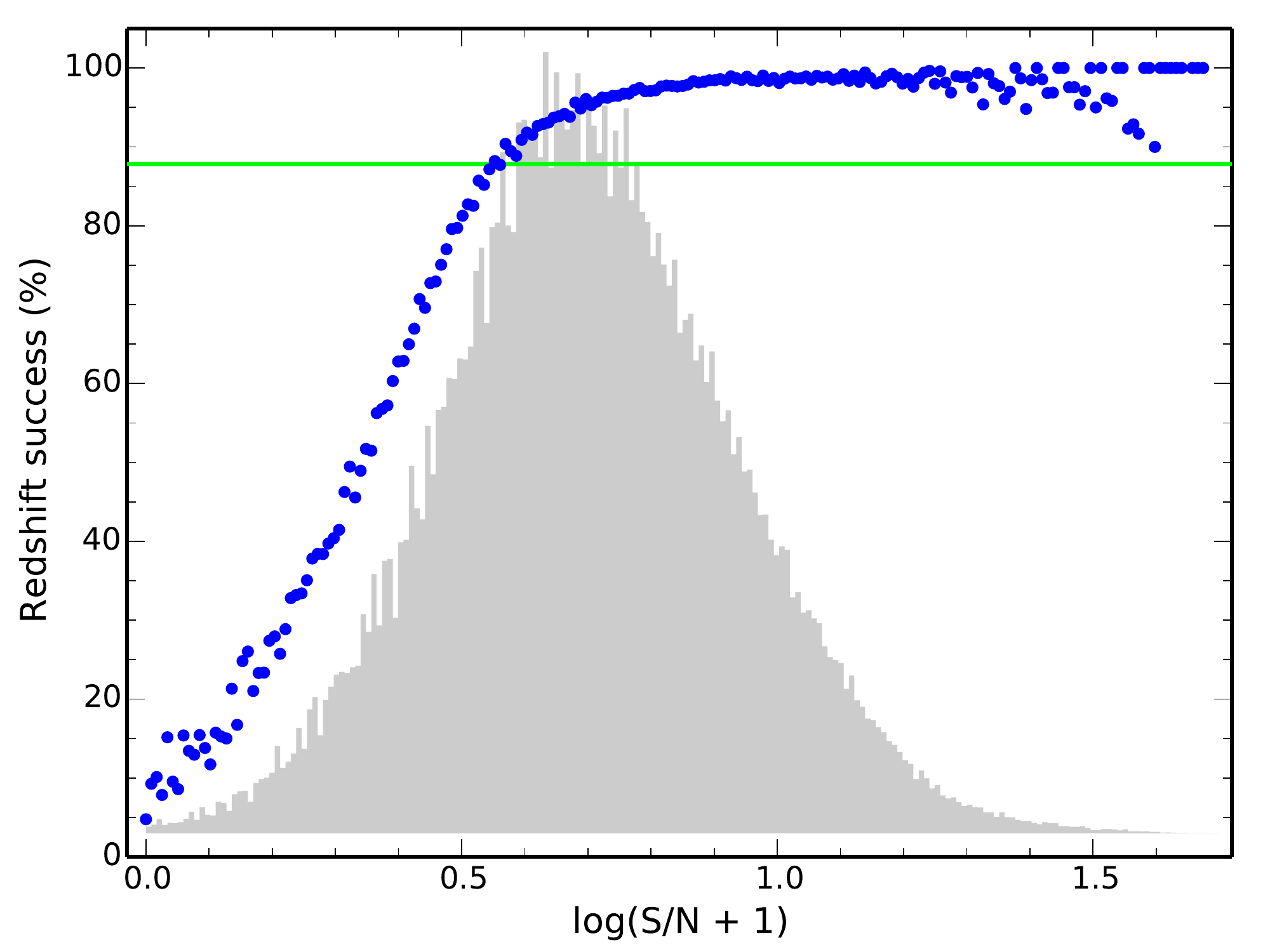}
\caption{The blue points show redshift success, i.e.\ the fraction of
  spectra of main survey targets with secure redshifts, as a function
  of the spectral S/N per pixel, averaged over the full spectrum
  (excluding bad pixels). The horizontal green line shows the survey's
  overall redshift success, while the grey shaded histogram shows the
  S/N distribution of the spectra (on an arbitrary linear scale).}
\label{zsuc_sn}
\end{figure}

The survey's final average efficiency is $69.1$~per~cent. However, as
can be seen from the lower panel of Fig.~\ref{survey_progress}, unlike
the other two factors the survey efficiency is a function of
time. Apart from a small, constant inefficiency required by the
survey's calibration needs (cf.\ the red line in the lower panel of
Fig.~\ref{survey_progress}), the survey's inefficiency is mainly
driven by the duplicate observations (blue line). However, as
described above, these duplicate observations were essential in order
to achieve the survey's high redshift completeness requirement. The
only way to reduce the duplication rate, and hence to increase the
survey's efficiency, would have been to increase the exposure time per
field. However, this dependence is sub-linear. In contrast, the number
of observed fields per night depends linearly on the exposure time, so
that the progression rate would in fact have decreased if the exposure
time had been increased.

The only true inefficiency thus lies in the number of fibres that had
to be allocated to filler targets (orange line in
Fig.~\ref{survey_progress}). As the survey progressed and main survey
targets were removed from the target list, it was impossible to
entirely avoid observing filler targets. A further contributor to this
inefficiency was the decision in 2014 to descope the G23 survey region
as discussed above. This descope had the unavoidable side-effect of
somewhat increasing the inefficiency of the survey because all objects
outside of the new selection limits, including those that had already
been observed, retroactively lost their main survey status and are
thus now considered filler targets by definition. Nevertheless, mainly
due to the large initial main survey target density, the inefficiency
due to the observation of filler targets amounts to only
$6.6$~per~cent for the completed survey.

We thus conclude that the seemingly low final overall survey
efficiency of $69.1$~per~cent was essentially unavoidable given our
high redshift completeness goal.

Finally, in Fig.~\ref{tdf_coverage} we show the distribution of the
observed 2dF/AAOmega fields on the sky. This distribution is the
result of the tiling algorithm described by \citet{Robotham10}, which,
for any given state of the survey in a particular survey region,
chooses the position of the next field to be observed as the one that
most improves the spatial redshift completeness in that region. In the
equatorial survey regions (G09, G12 and G15), which are the most
complete, the average number of fields that cover a given position
ranges from $10.3$ to $10.9$, with more than $99$~per~cent of these
regions covered by $\ge5$ fields. These high covering numbers are to
some extent driven by the inefficiencies discussed above, but they are
nevertheless a key feature of the survey's design, one that provides
us with an important advantage over single-pass surveys: it allows us
to ensure high redshift completeness even for closely packed pairs and
groups of galaxies (see Fig.~\ref{zcomp_nndist} below). Without
multiple visits the redshift completeness of close pairs and groups
would be severely impaired by physical fibre placement restrictions.

\subsection{Redshift success}

The gross redshift success of the survey is defined as the fraction of
spectra of main survey targets for which we were able to measure a
secure redshift ($nQ \ge 3$, using {\sc Autoz}, disregarding redshifts
from combined spectra). In Table~\ref{survey_stats} we have already
seen that the overall redshift success of the survey is
$87.8$~per~cent. In this section we will briefly consider the redshift
success in more detail.

\begin{figure}
\includegraphics[width=\columnwidth]{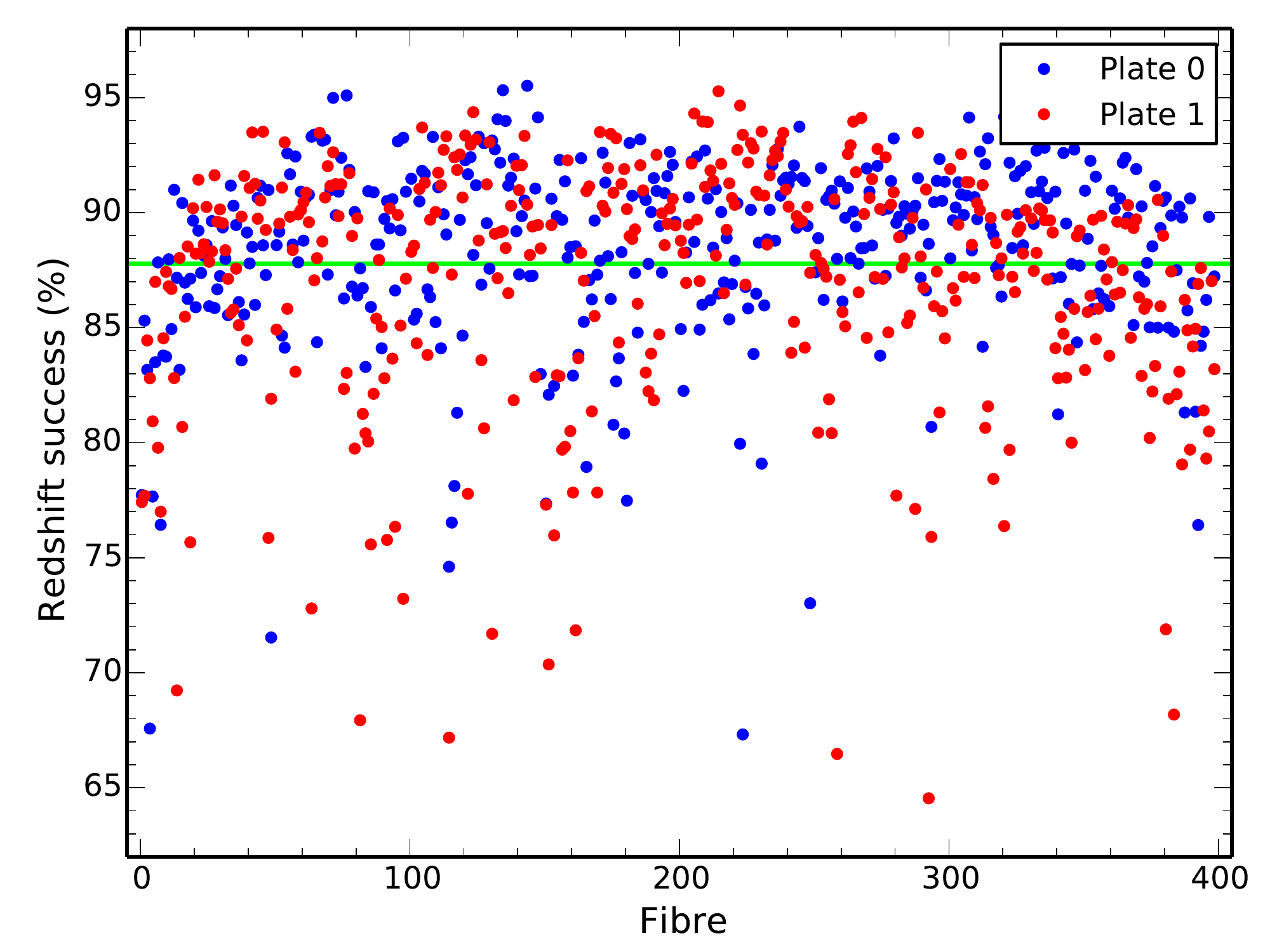}
\caption{Redshift success as a function of fibre number, separated by
  2dF plate as indicated. The horizontal green line shows the survey's
  overall redshift success.}
\label{zsuc_fibre}
\end{figure}

\begin{figure*}
\includegraphics[width=0.49\textwidth]{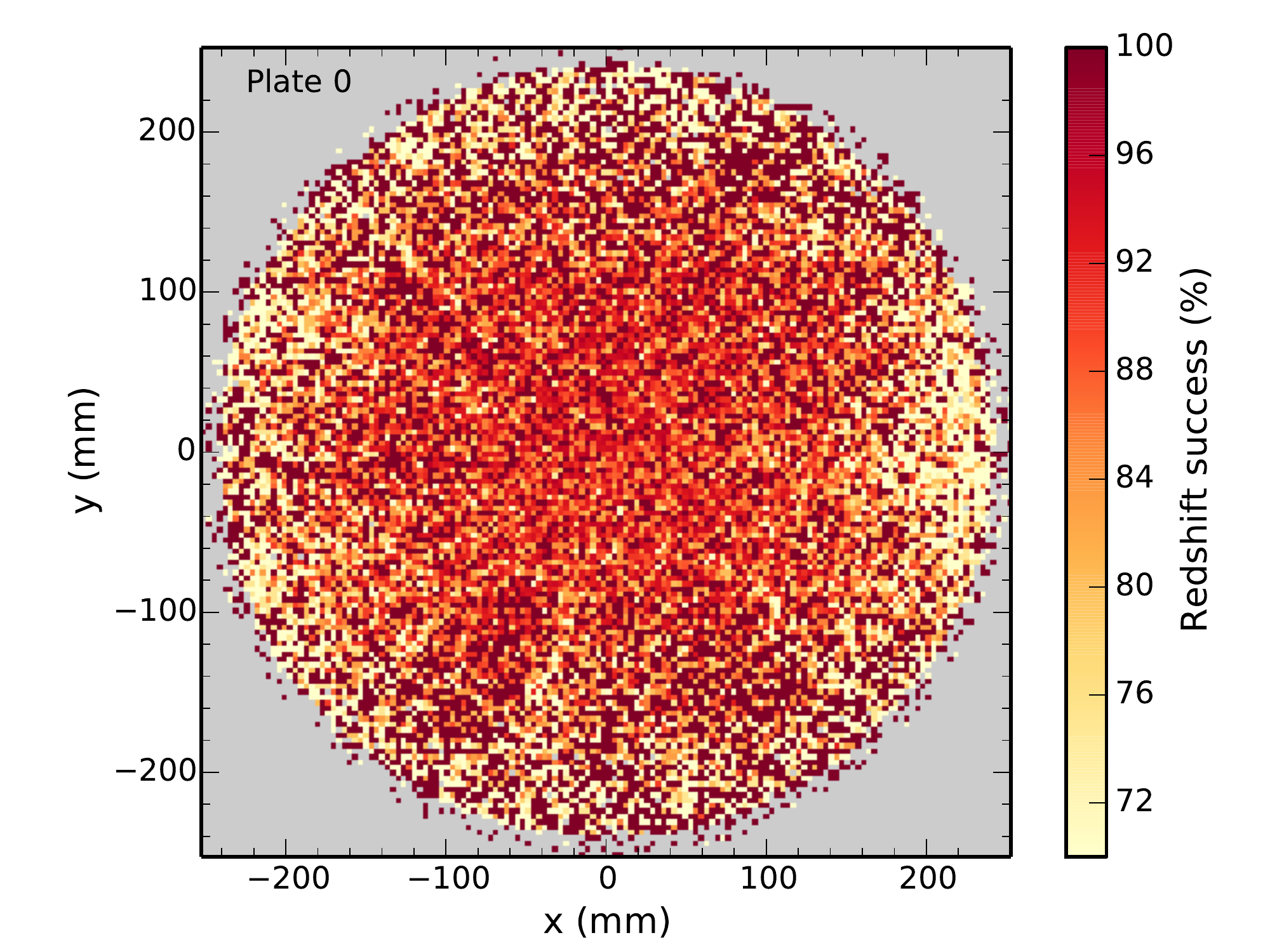}
\includegraphics[width=0.49\textwidth]{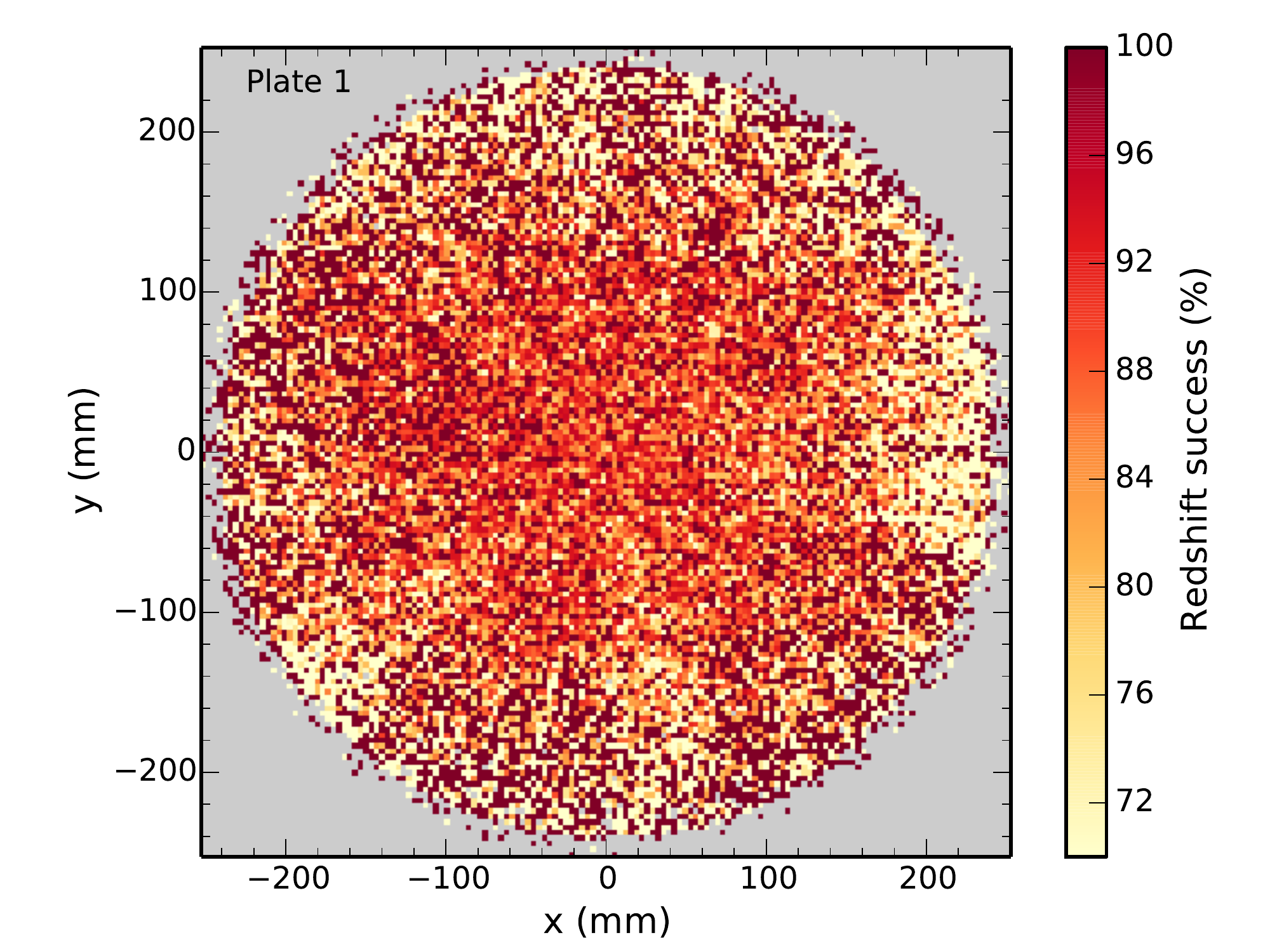}
\caption{Redshift success as a function of the fibre position on the
  plate for each of the two 2dF plates as indicated.}
\label{zsuc_xy}
\end{figure*}

In Fig.~\ref{zsuc_sn} we show how the redshift success varies as a
function of the average spectral signal-to-noise ratio (S/N). While
the redshift success turns down sharply for S/N $< 3$, we note that it
does not quite drop to $0$. The reason is of course that a redshift
can still be measured reliably from emission lines even in the
complete absence of a stellar continuum. The few data points at high
S/N and relatively low redshift success are due to only a very small
number of spectra, as evidenced by the grey histogram in
Fig.~\ref{zsuc_sn}. Most of these spectra are of stars and QSOs (for
which {\sc Autoz} has no templates at present), and many are severely
affected by data reduction issues.

From Fig.~\ref{zsuc_sn} it is clear that S/N is an excellent predictor
of redshift success. Redshift success will therefore also strongly
correlate with the observational parameters and target properties that
determine the S/N, i.e.\ exposure time, sky brightness, airmass and
atmospheric transparency, seeing, and the target's brightness and
light distribution [cf.\ also Fig.~5 of \citet{Hopkins13a}]. In the
following we will briefly ask whether redshift success also depends on
any instrumental parameters.

Fig.~\ref{zsuc_fibre} shows the redshift success as a function of the
fibre through which the spectra were observed, separately for each of
the two 2dF plates. For several fibres the redshift success is clearly
significantly lower than for the bulk of the fibres, in particular on
plate~1. We believe that the most likely explanation for these low
values is that these fibres have significantly lower transmission than
the others \citep[cf.][]{Sharp13}. Fibre transmission variations will
be further investigated in the context of efforts to improve the flux
calibration scheme of the survey (Maier et al., in preparation).

\begin{figure}
\includegraphics[width=\columnwidth]{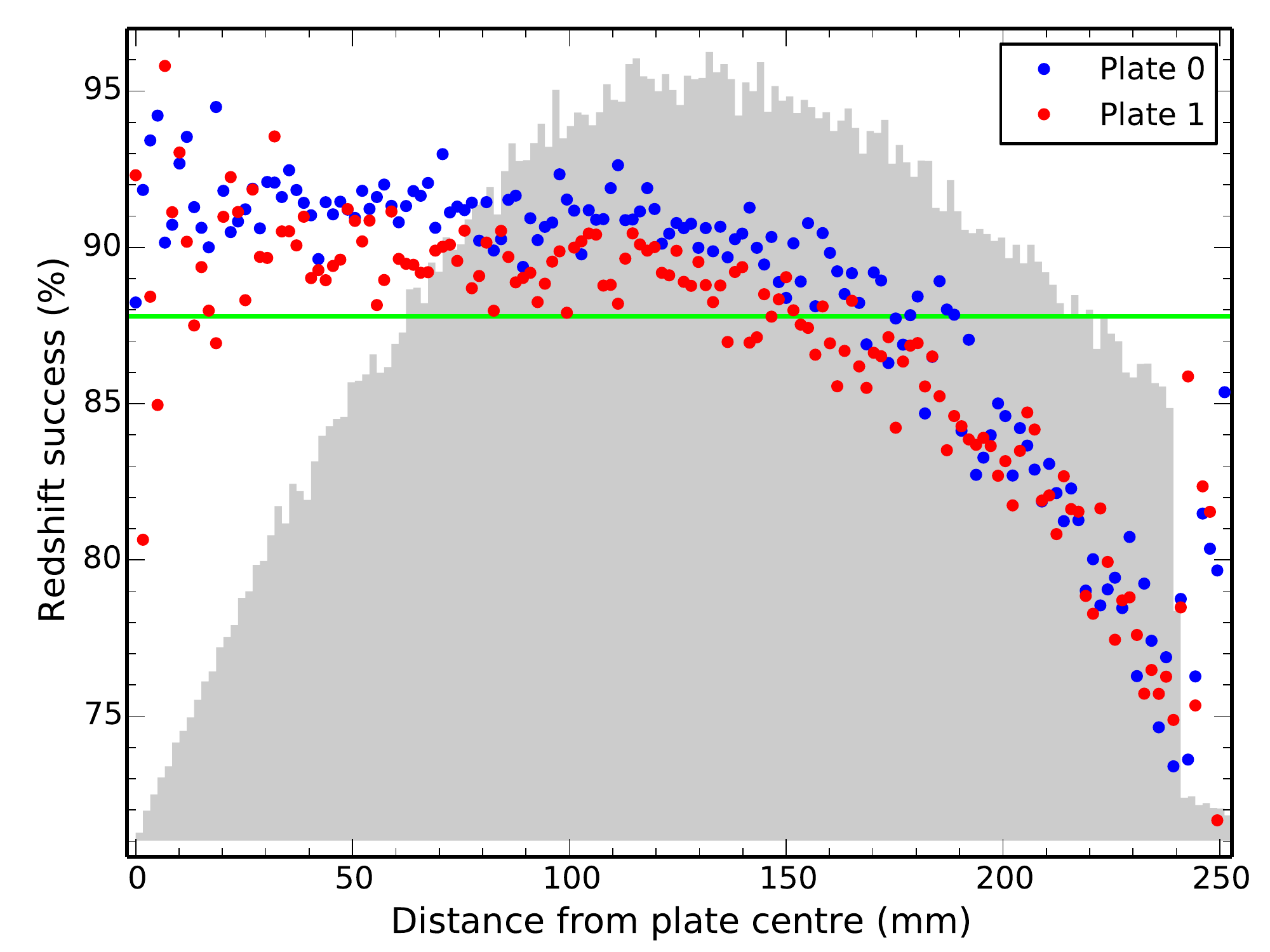}
\caption{Redshift success as a function of distance from the plate
  centre, separated by 2dF plate as indicated. The horizontal green
  line shows the survey's overall redshift success, while the grey
  shaded histogram shows the distribution of distances (on an
  arbitrary linear scale).}
\label{zsuc_radial}
\end{figure}

In Fig.~\ref{zsuc_xy} we show the redshift success as a function of
the fibre position on the plate, separately for each of the two 2dF
plates as indicated. In both cases we can clearly see structure in the
spatial distribution of the redshift success. While there are some
differences between the two plates, many features are shared. The most
obvious feature is the radial dependence. This is shown more clearly
in Fig.~\ref{zsuc_radial} where we plot redshift success as a function
of the distance from the plate centre. This figure bears a remarkable
resemblance to Fig.~18 of \citet{Croom04a}, who already identified
this same effect in the 2QZ survey. As discussed by these authors, the
radial dependence of the redshift success could be caused by a number
of effects, including systematic errors in the astrometry or field
rotation and atmospheric refraction effects. In addition,
\citet{Sharp13} found that the transmission of a given fibre also
depends on the fibre's distance from the plate centre, which they
mainly attributed to radial variations of the apparent fibre diameter,
of focal ratio degradation and of non-telecentricity. Whatever the
cause of the radial dependence of the redshift success may be, the
concern here is of course that the distribution seen in
Fig.~\ref{zsuc_xy} may also be imprinted on the spatial distribution
of the redshift completeness on the sky. As we will see in the next
section, this is not the case, presumably due to the large amount of
overlap among the observed 2dF fields and their irregular positioning
on the sky (cf.\ Fig.~\ref{tdf_coverage}).

\subsection{Redshift completeness}

In this section we turn to the redshift completeness, defined as the
fraction of main survey targets for which we were able to obtain at
least one secure redshift ($nQ \ge 3$, either using {\sc Autoz}, now
including redshifts from combined spectra, or from a previous
survey). The redshift completeness thus includes the effects of
targeting completeness, redshift success, and duplicate observations.

The overall redshift completeness in the equatorial survey regions
(G09, G12 and G15) is $98.48$~per~cent, in the high-priority
sub-region of G02 it is $94.95$~per~cent, and in G23 the completeness
is $94.19$~per~cent. In the equatorial regions almost all of the
incompleteness is due to redshift measurement failures, since only
$158$ main survey targets in these regions ($0.08$~per~cent) remain
unobserved. In G02 and G23 the contribution of the targeting
incompleteness is somewhat larger. Here we have failed to observe
$251$ and $863$ main survey targets ($1.2$ and $1.9$~per~cent),
respectively.

\begin{figure}
\includegraphics[width=\columnwidth]{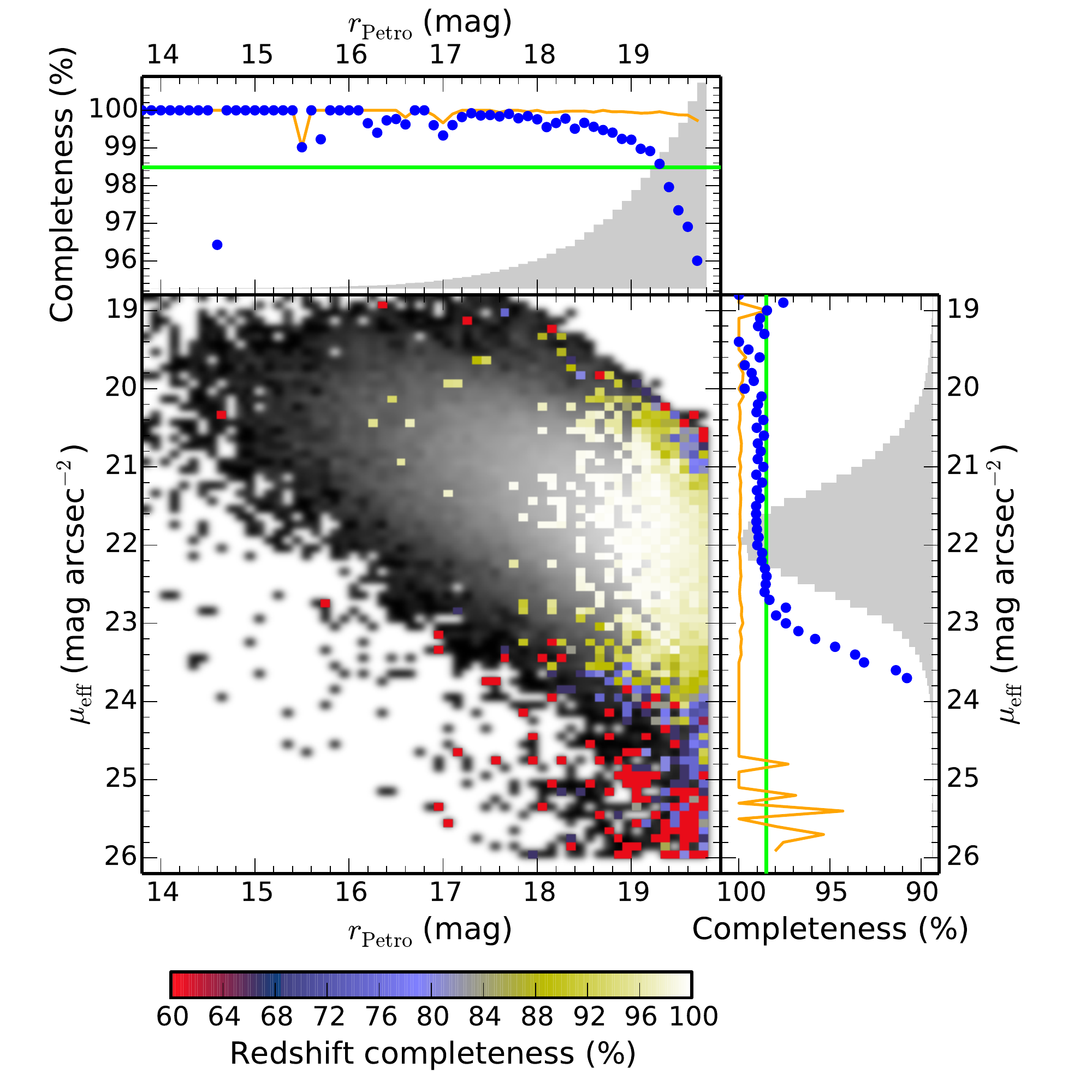}
\caption{The colour image in the main panel shows the redshift
  completeness of the equatorial survey regions as a bivariate
  function of SDSS DR7 $r$-band Petrosian magnitude and effective
  surface brightness, both corrected for Galactic extinction. The
  grey-scale image in the background shows the distribution of main
  survey targets in this plane using an arbitrary logarithmic
  scale. The blue points and orange lines in the side panels show the
  redshift and targeting completeness as a function of just one of
  these parameters, respectively. The green lines in these panels show
  the overall redshift completeness in the equatorial survey
  regions. The grey shaded histograms show the target distributions
  (now using an arbitrary linear scale).}
\label{zcomp_r_mu}
\end{figure}

Since the redshift incompleteness is mostly due to redshifting
failures and not targeting incompleteness, we must expect the
incompleteness to be biased towards faint and low surface brightness
galaxies. As we can see from Fig.~\ref{zcomp_r_mu} this is indeed the
case. In this figure we show the redshift completeness in the
equatorial survey regions as a function of $r$-band magnitude and
surface brightness. For G02 and G23 the plot looks quite similar,
albeit at slightly lower overall completeness levels. The cut-offs of
the data at $r = 19.8$~mag and $\mu_{\rm eff} = 26$~mag~arcsec$^{-2}$
are the explicit selection limits imposed on main survey targets
\citep{Baldry10}.

From Fig.~\ref{zcomp_r_mu} we can see that the completeness is
reasonably uniform across the bulk of the target galaxy
population. Still, there is a small, but nonetheless significant trend:
the completeness drops from $\sim$$99$~per~cent at $r = 19.2$~mag to
$\sim$$96$~per~cent at the faint limit of $19.8$~mag, where of course
the magnitude distribution peaks (cf.\ top panel of
Fig.~\ref{zcomp_r_mu}). There is also a significant trend with surface
brightness (cf.\ right panel of Fig.~\ref{zcomp_r_mu}). The
completeness is roughly constant at $\sim$$99$~per~cent down to
$\mu_{\rm eff} = 22.8$~mag~arcsec$^{-2}$, from where it drops to
$\sim$$92$~per~cent at $23.7$~mag~arcsec$^{-2}$. While the
completeness is thus constant across the peak of the surface
brightness distribution, the drop nevertheless affects a significant
fraction of the target galaxy population. Below
$23.7$~mag~arcsec$^{-2}$ the completeness drops even further, down to
$\sim$$60$~per~cent at $26$~mag~arcsec$^{-2}$ (not shown in the right
panel). However, only a tiny fraction of the target population is
affected by these low completeness levels. 

\begin{figure}
\includegraphics[width=\columnwidth]{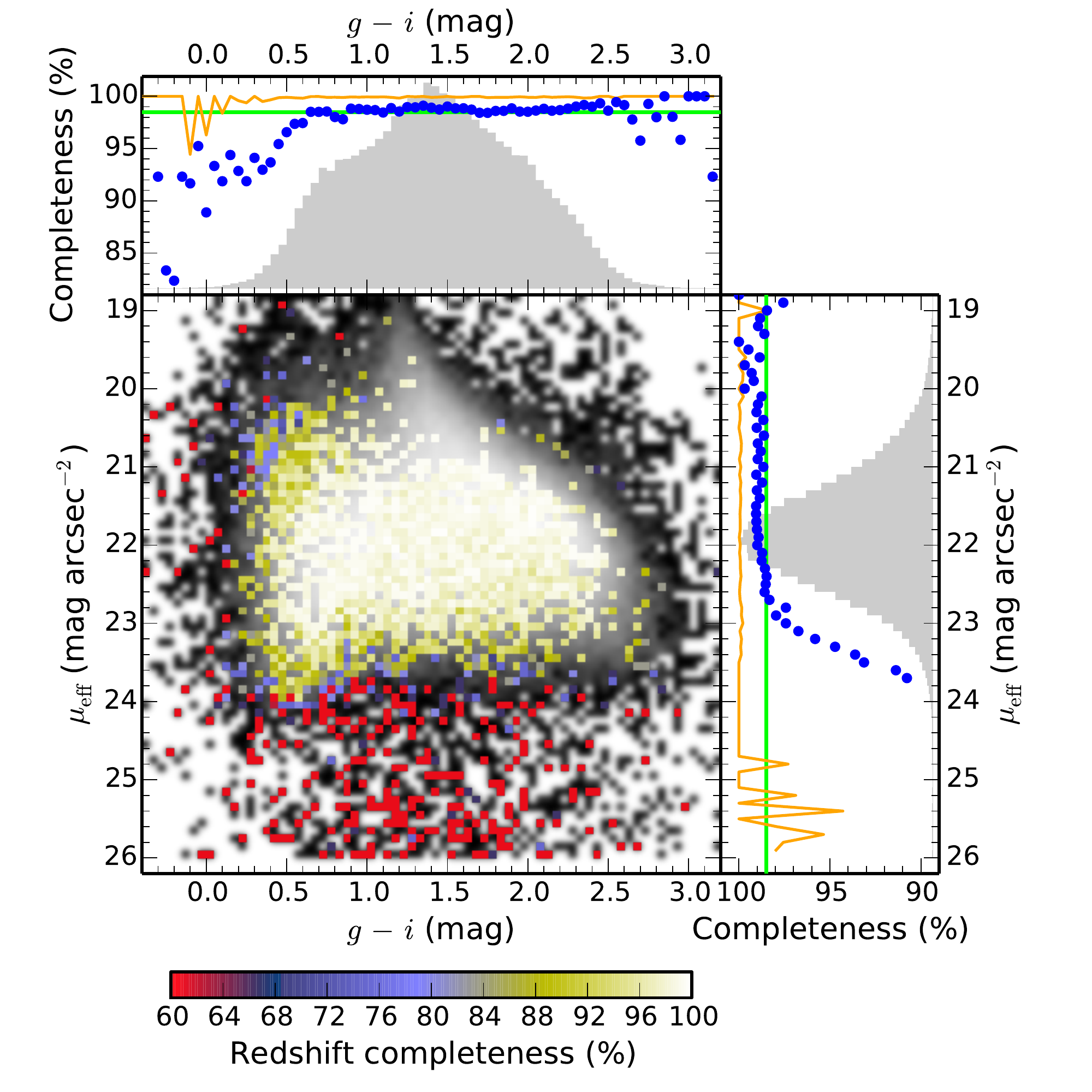}
\caption{As Fig.~\ref{zcomp_r_mu} for observed SDSS DR7 $g - i$ colour
  (using model magnitudes) and $r$-band effective surface brightness.}
\label{zcomp_col_mu}
\end{figure}

In Fig.~\ref{zcomp_r_mu} we can also see a pocket of lower
completeness at faint magnitudes and {\em high} surface
brightness. Having inspected the relevant spectra, we believe that
this pocket is mostly caused by QSOs (cf.\ Section~\ref{agnz}) and
stars (our star-galaxy separation is not perfect). This hypothesis is
further supported by the colour of the incompleteness pocket. Since
QSOs are in general quite blue compared to galaxies, and since the
stellar contamination of our main survey sample is highest at $g-i <
1$~mag (cf.\ Fig.~6 of \citealp{Baldry10}), we expect the high surface
brightness incompleteness to mainly affect the blue end of our
sample. From Fig.~\ref{zcomp_col_mu} we can see that it is indeed
largely confined to $g-i < 0.7$~mag.

Note that these objects alone cannot explain the observed drop in the
completeness from $\sim$$99$~per~cent at $g - i = 0.6$~mag down to
$\sim$$86$~per~cent at $g - i = 0$~mag. The low surface brightness
incompleteness discussed above also contributes to this decline,
consistent with the notion of low surface brightness galaxies being
gas-rich and star-forming, and therefore blue.

Summarising the above, we find that, although redshift completeness
variations are small across the bulk of the target galaxy population,
significant trends with magnitude, surface brightness and colour
nevertheless exist, and should be corrected for when using the
redshift data.

\begin{figure}
\includegraphics[width=\columnwidth]{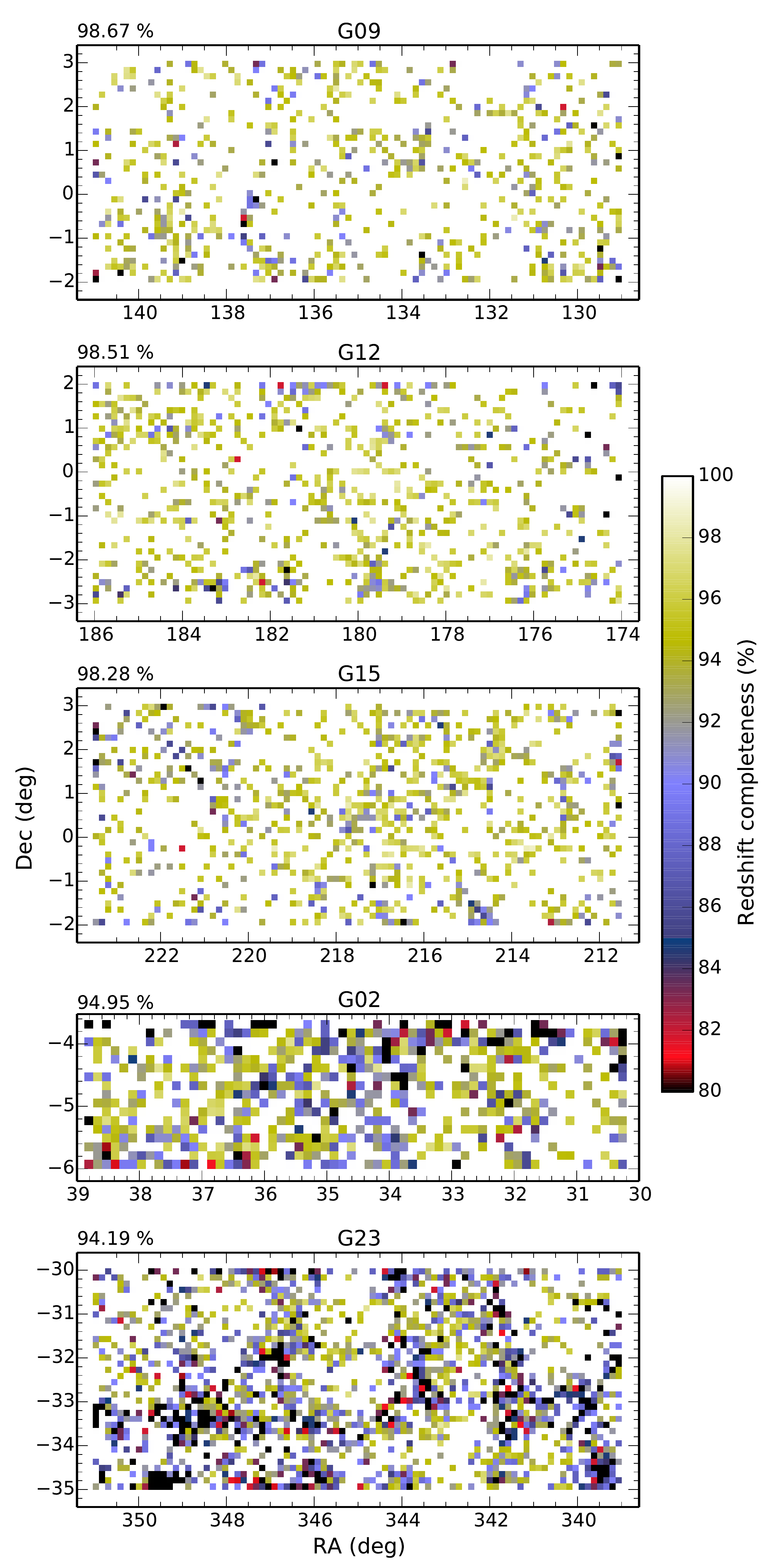}
\caption{Redshift completeness of the five GAMA\,II survey regions, as
  indicated, in bins of $0.14$~deg size. For G02 we only show the
  high-priority sub-region. The average completeness of each region is
  indicated at the top left of each panel.}
\label{zcomp_regions}
\end{figure}

In Fig.~\ref{zcomp_regions} we show the spatial distribution of the
redshift completeness on the sky for each of our survey regions. No
large-scale trends or patterns are evident. The dependence of redshift
success on the distance from the 2dF plate centre seen in
Figs.~\ref{zsuc_xy} and \ref{zsuc_radial} thus appears to have had
little impact on the redshift completeness distribution on the sky.

\begin{figure}
\includegraphics[width=\columnwidth]{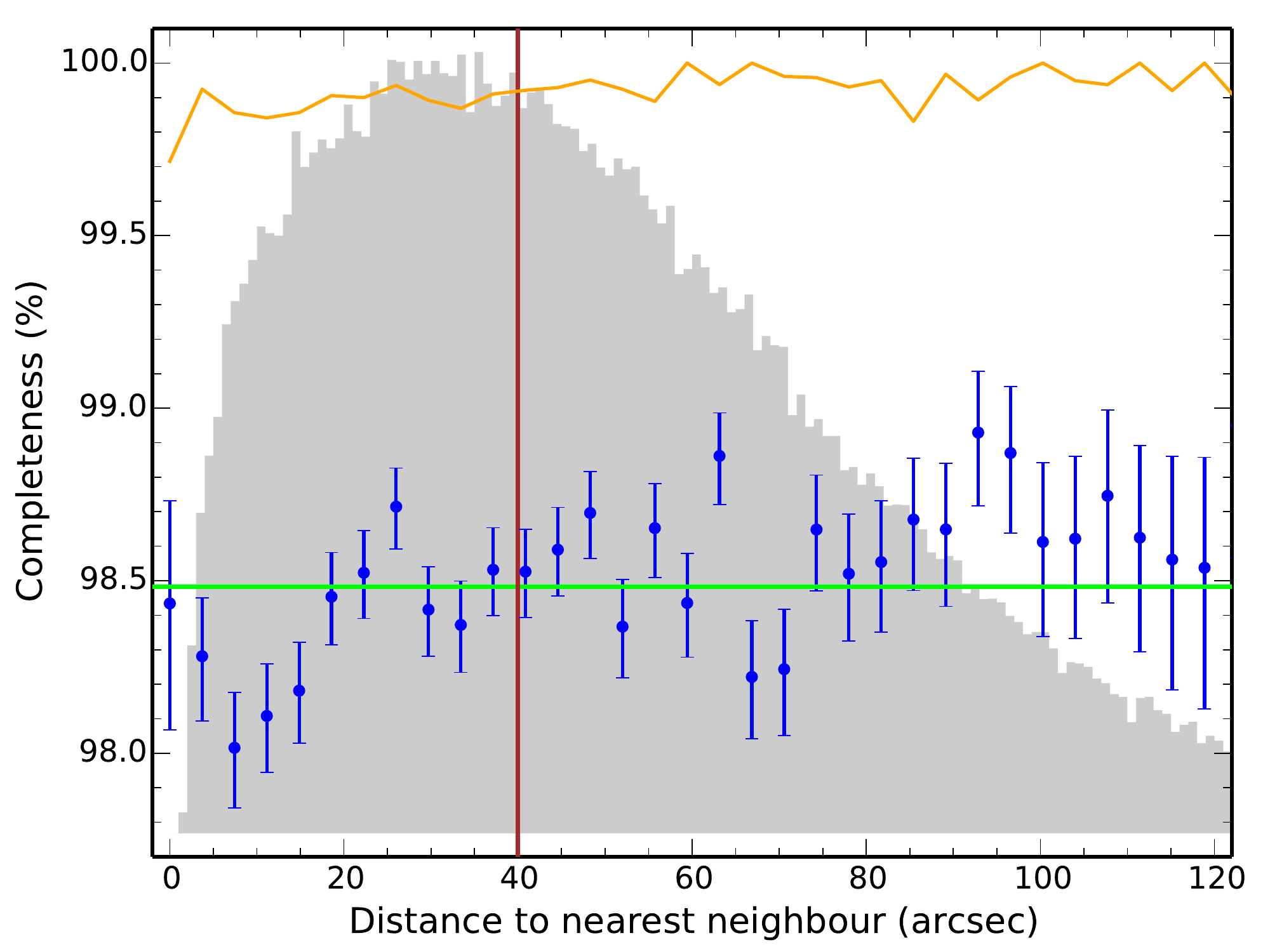}
\caption{The blue points and orange line show the redshift and
  targeting completeness of the equatorial survey regions as a
  function of distance to the nearest neighbour among main survey
  targets, respectively. The horizontal green line shows the overall
  redshift completeness in the equatorial survey regions. The vertical
  brown line marks the distance out to which fibre collisions may
  occur. Two targets separated by less than this distance cannot
  usually both be allocated a fibre in the same configuration. The
  grey shaded histogram shows the distribution of all nearest
  neighbour distances (on an arbitrary linear scale).}
\label{zcomp_nndist}
\end{figure}

Given the importance of galaxy groups and close pairs to the GAMA
survey's main scientific goals we are of course also interested in the
survey's redshift completeness on small angular scales. In
Fig.~\ref{zcomp_nndist} we show the redshift completeness as a
function of distance to the nearest neighbour among main survey
targets. One might expect the completeness to be affected out to a
nearest neighbour distance of $\sim$$40$~arcsec by the fact that two
targets separated by less than this distance cannot in general both be
allocated a fibre in the same configuration due to physical fibre
placement constraints. However, thanks to our fibre placement
strategy, which prioritizes targets with many close neighbours
\citep{Robotham10}, and thanks to the large number of visits to each
patch of sky (cf.\ Fig.~\ref{tdf_coverage}) we find that the redshift
completeness is largely independent of the distance to the nearest
neighbour. The only residual effect is a small, but apparently still
significant reduction of the completeness by $\sim$$0.5$ percentage
points in the nearest neighbour distance range $4$--$20$~arcsec.

\begin{figure}
\includegraphics[width=\columnwidth]{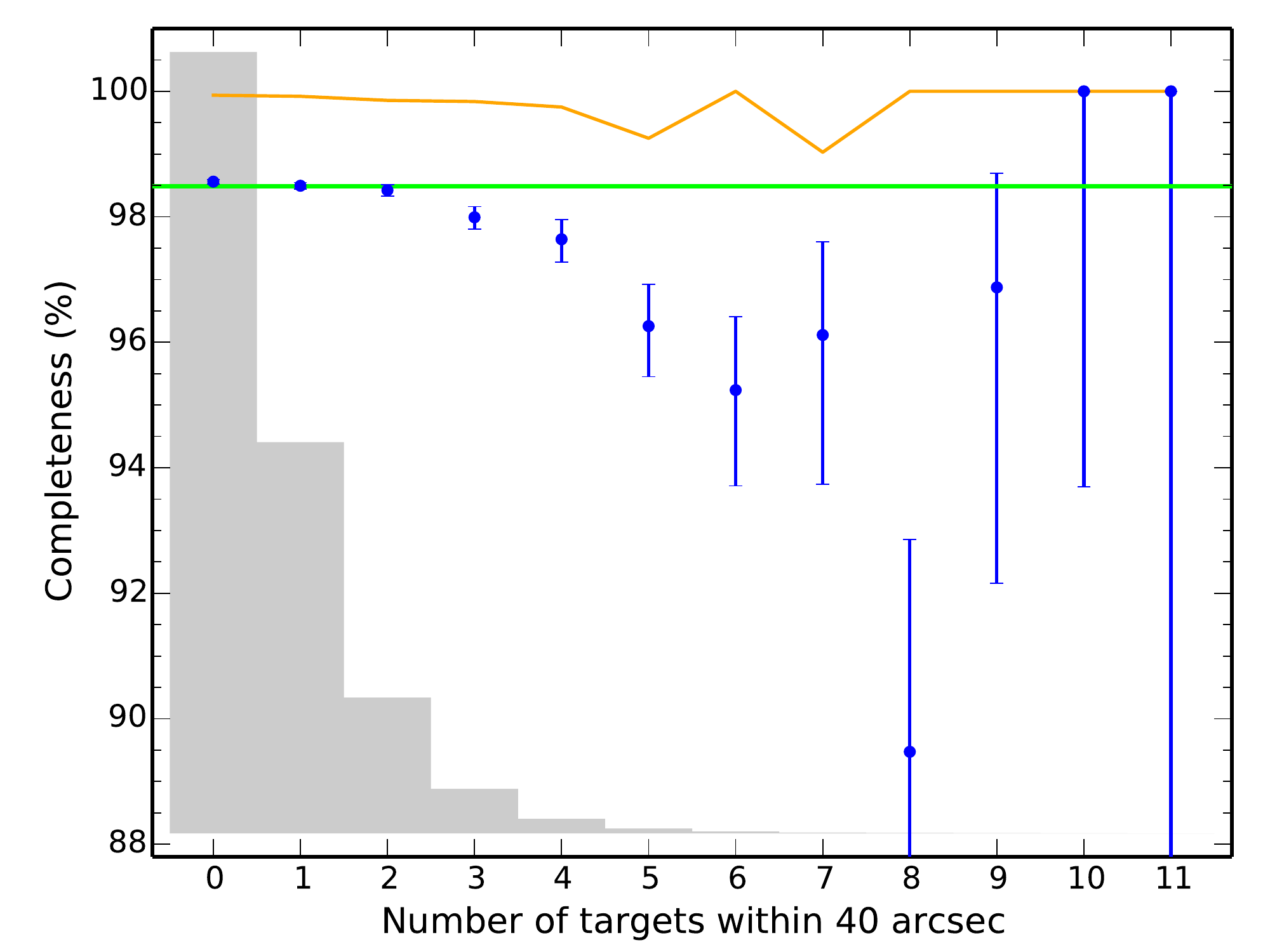}
\caption{The blue points and orange line show the redshift and
  targeting completeness of the equatorial survey regions as a
  function of the number of other main survey targets within
  $40$~arcsec. The horizontal green line shows the overall redshift
  completeness in the equatorial survey regions. The grey shaded
  histogram shows the distribution of the number of main survey
  targets within $40$~arcsec (on an arbitrary linear scale).}
\label{zcomp_nc}
\end{figure}

The cause of this dip can be found in Fig.~\ref{zcomp_nc}, where we
show the redshift completeness as a function of the number of main
survey targets within a distance of $40$~arcsec, $N_{40}$. For $3 \le
N_{40} \le 8$ there is a clear trend of decreasing redshift
completeness with increasing $N_{40}$. Since $N_{40}$ is
anti-correlated with nearest neighbour distance, it is this trend that
is responsible for the dip in Fig.~\ref{zcomp_nndist}. But what in
turn is the cause of this trend? Our fibre placement strategy has
evidently succeeded in maintaining the targeting completeness at near
$100$~per~cent for all $N_{40}$ (cf.\ orange line in
Fig.~\ref{zcomp_nc}). The reduced redshift completeness at high
$N_{40}$ must therefore be caused either by a reduced redshift success
or by a smaller duplication rate for $N_{40} \ge 3$.

In fact, we find that both are to blame. Although unsuccessfully
observed targets remain on the target list, they do so with a lower
priority than unobserved targets. This means that targets in dense
regions are less likely to receive a second observation than isolated
targets, thereby reducing the duplication rate for targets with high
$N_{40}$. The reason for the reduced redshift success is more
subtle. Targets with $N_{40} \ge 3$ are on average brighter, redder
and of higher surface brightness than targets with $N_{40} < 3$. Given
the completeness trends shown in Figs.~\ref{zcomp_r_mu} and
\ref{zcomp_col_mu} we would thus expect the redshift success to {\em
  increase} with $N_{40}$. However, we find that for larger $N_{40}$
values the completeness trends change, in the sense that faint, low
surface brightness galaxies in dense environments are even less likely
to yield a redshift that their isolated counterparts. In other words,
even for fixed target properties the redshift success depends on the
target's environment. Based on the visual inspection of targets with
failed observations and $N_{40} \ge 3$ we believe that this is due to
the fact that many of these faint targets lie in the extended halo of
a much brighter, nearby galaxy, so that the spectra of the faint
targets are frequently `polluted' with light from a bright
neighbour. Since the background is only measured globally for an
entire field, but not locally for each target, this `pollution' will
affect {\sc Autoz}'s ability to determine an unambiguous redshift.

We point out that most of the discussion in this section was focused
on the equatorial survey regions. The results are qualitatively
similar for the g23 and high-priority G02 regions, but all of the
effects are somewhat larger due to the lower overall redshift
completeness of these regions. We conclude this section by remarking
that for the equatorial regions at least, all of the completeness
issues described above are relatively minor compared to previous large
spectroscopic surveys of low-redshift galaxies.

\subsection{Redshift precision and incorrectness fraction}
\label{zqc}

In this section we briefly consider the quality of our redshifts, both
in terms of the redshift error and in terms of the incorrectness
fraction, i.e.\ the fraction of $nQ \ge 3$ redshifts that are
wrong. While we only used the {\sc Autoz} redshifts in the previous
two sections, we will now consider both the {\sc runz} and the {\sc
  Autoz} redshifts, and thus compare the performance of the two
redshift codes.

\begin{figure}
\includegraphics[width=\columnwidth]{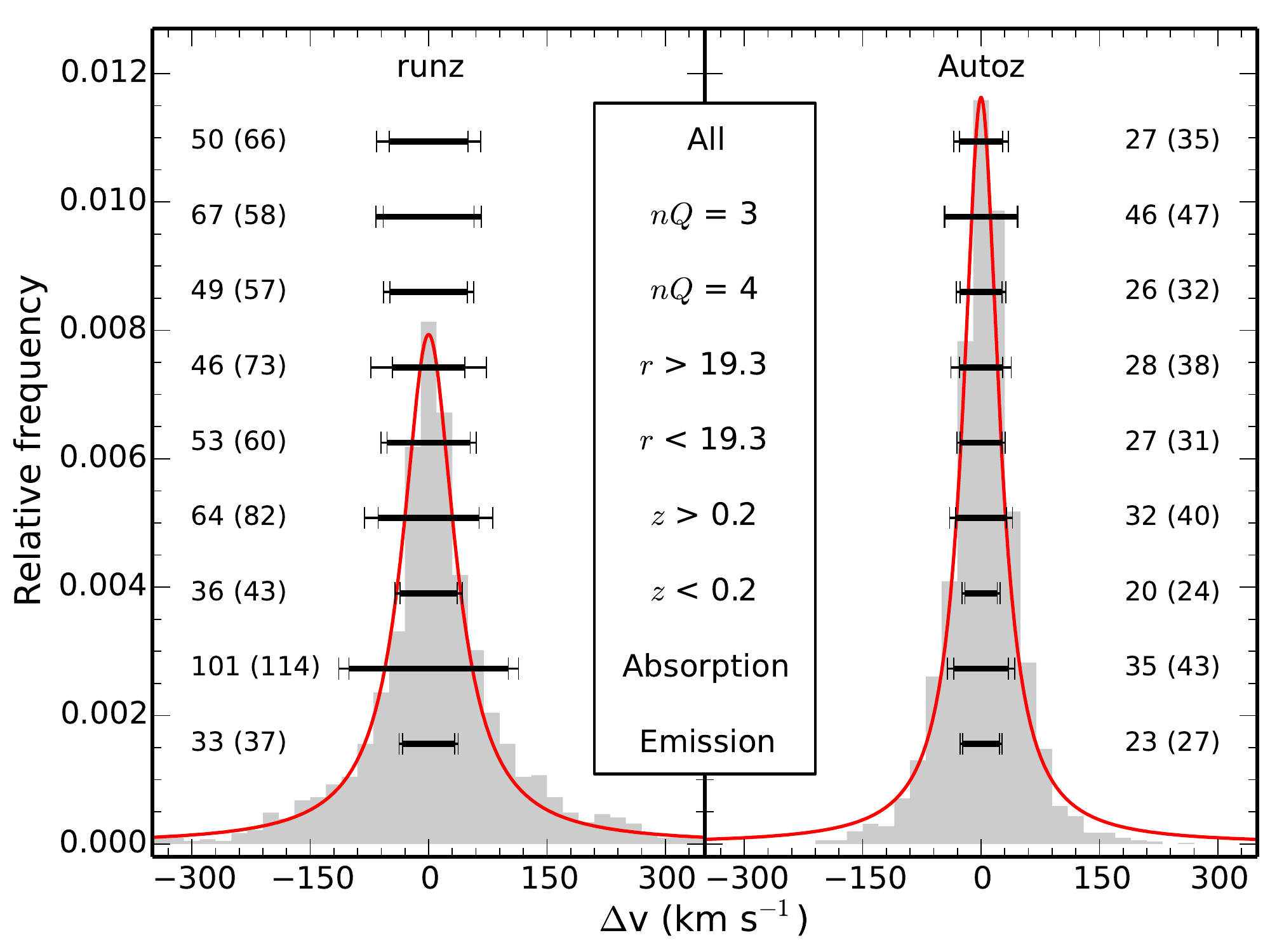}
\caption{The grey shaded histogram in the left panel shows the
  distribution of differences between the {\sc runz} redshifts
  measured from independent GAMA\,II spectra of the same main survey
  targets, where all redshifts have $nQ \ge 3$ and lie in the range
  $0.002 < z < 0.9$ ($2132$ pairs from $4096$ unique spectra of $2020$
  unique objects). These objects were selected for duplicate
  observations at random, independently of whether the first
  observation yielded a redshift or not. The red line shows a
  Lorentzian with $\gamma = 40$\kms\ for comparison. The thick,
  top-most horizontal errorbar shows the redshift error derived from
  the $68$ percentile range of this distribution. The thinner, more
  extended errorbar shows the redshift error derived from the
  distribution of redshift differences using {\em all} available
  duplicate observation, not just those of the randomly selected
  objects. The other errorbars show the same, but for various
  sub-samples as indicated in the middle box. The labels `$nQ=3$' and
  `$nQ=4$' refer to pairs where both redshifts have the respective
  quality. The labels `Absorption' and `Emission' refer to pairs where
  both redshifts were determined from {\sc Autoz} templates $40$--$42$
  or $43$--$47$, respectively \citep{Baldry14}. The numbers to the
  left are the values of the $1$$\sigma$ redshift errors in \kms\ for
  each sub-sample, those in parentheses refer to the errors derived
  from {\em all} duplicate observation. The right panel shows the same
  as the left, but now using {\sc Autoz} redshifts ($2540$ pairs from
  $4807$ unique spectra of $2358$ unique objects). In this case the
  Lorentzian is characterised by $\gamma = 27$\kms.}
\label{zqc_dv}
\end{figure}

\citet{Driver11} already estimated the GAMA error for {\sc runz}
redshifts by considering duplicate observations of the same objects,
using both intra-survey and inter-survey comparisons. In the former
case we compared all available duplicate redshifts with $nQ\ge3$ from
GAMA\,I only, in the latter we compared GAMA redshifts to those
from previous surveys (see also \citealp{Baldry14}). However,
\citet{Driver11} surmised that both of these samples likely yielded
biased results.

A large fraction of the objects in the intra-survey sample were
re-observed because the initial observation only yielded a low-quality
redshift (i.e. $Q=2$). These objects nevertheless ended up with {\em
  two} $nQ \ge 3$ redshifts because subsequent re-redshifting of the
initial spectra (after the re-observation) confirmed the initial
redshifts, which bumped them to $nQ\ge3$. Hence this sample was biased
towards lower quality spectra. Its median S/N was indeed found to be
$20$~per~cent lower than that of the full sample.

Due to the spectroscopic limit of the other surveys used in the
inter-survey comparison being brighter than that of GAMA, this sample
was also biased, but this time towards {\em higher} quality spectra:
the median S/N of the GAMA spectra in this sample was $70$~per~cent
higher than that of the full sample.

To avoid having to rely on these biased samples we subsequently
selected a random sample of main survey targets for duplicate
observations, irrespective of the quality of any existing
redshifts.\footnote{For their second observation these targets were
  treated as filler targets (cf.\ Section~\ref{inputcat}).} As a
result, $2020$ randomly selected main survey targets have more than
one $nQ \ge 3$ {\sc runz} redshift, yielding $2132$ redshift pairs
(from $4096$ unique spectra; some targets were observed more than
twice). Here, we only consider redshifts in the range $0.002 < z <
0.9$ in order to exclude both stars and QSOs. Using the {\sc Autoz}
redshifts we have $2540$ pairs from $4807$ unique spectra of $2358$
unique objects.

The distributions of the redshift differences of these pairs are shown
in Fig.~\ref{zqc_dv} in velocity units (left: {\sc runz}, right: {\sc
  Autoz}). Neither of these distributions is well described by a
Gaussian. Instead, they are approximately Lorentzian in velocity space
(red lines), indicating a Gaussian distribution in redshift space.  We
find $68$-percentile ranges of $141$ and $76$\kms\ for the {\sc runz}
and {\sc Autoz} distributions, respectively, indicating redshift
errors of $\sigma_z = 50$ and $27$\kms.

We first of all note that our value for the {\sc runz} error is
significantly lower than the value of $65$\kms\ found by
\citet{Driver11}. This is due to only using the duplicate observations
of the {\em random} sample here, as opposed to using {\em all}
available duplicate observations of main survey targets. Indeed, if we
use all duplicates ($12\,821$ pairs from $24\,920$ unique spectra of
$12\,340$ unique objects) we again find the same {\sc runz} redshift
error as \citeauthor{Driver11} (but with a sample larger by a factor
of $15$). Second, we note that the {\sc Autoz} redshifts are about
twice as precise than the {\sc runz} redshifts, demonstrating the
superiority of the {\sc Autoz} methods and templates. We also point
out that our overall redshift error for {\sc Autoz} is in reasonable
agreement with the median redshift error of $33$\kms\ identified by
\citet{Baldry14}.

Finally, the series of errorbars in Fig.~\ref{zqc_dv} illustrate how
the redshift precision varies as a function of a few selected spectral
and target properties. The $r$-band magnitude and redshift values at
which we have chosen to split our sample into faint/bright and
high-$z$/low-$z$ sub-samples are approximately the median values of the
sample. Qualitatively, the redshift error behaves as expected for all
sub-samples: a lower quality of the redshift, a fainter target, a
higher redshift or a spectrum dominated by absorption features all
have the effect of increasing the redshift error. We also note that,
for each sub-sample investigated, the use of {\em all} available
duplicate observations of main survey targets (instead of just those
of the random sample) always leads to a larger redshift error,
confirming the bias of the full sample, relative to the random sample.

We now turn to the redshift incorrectness fraction. Each {\sc runz}
and {\sc Autoz} redshift is accompanied by an estimate of the
probability, $p(z)$, that this redshift is correct. For any collection
of redshifts we can therefore compute which fraction of these should
be expected to be incorrect. We will now compare this expected
fraction with the actual fraction, which we again derive from
duplicate observations of the same objects.

\begin{figure}
\includegraphics[width=\columnwidth]{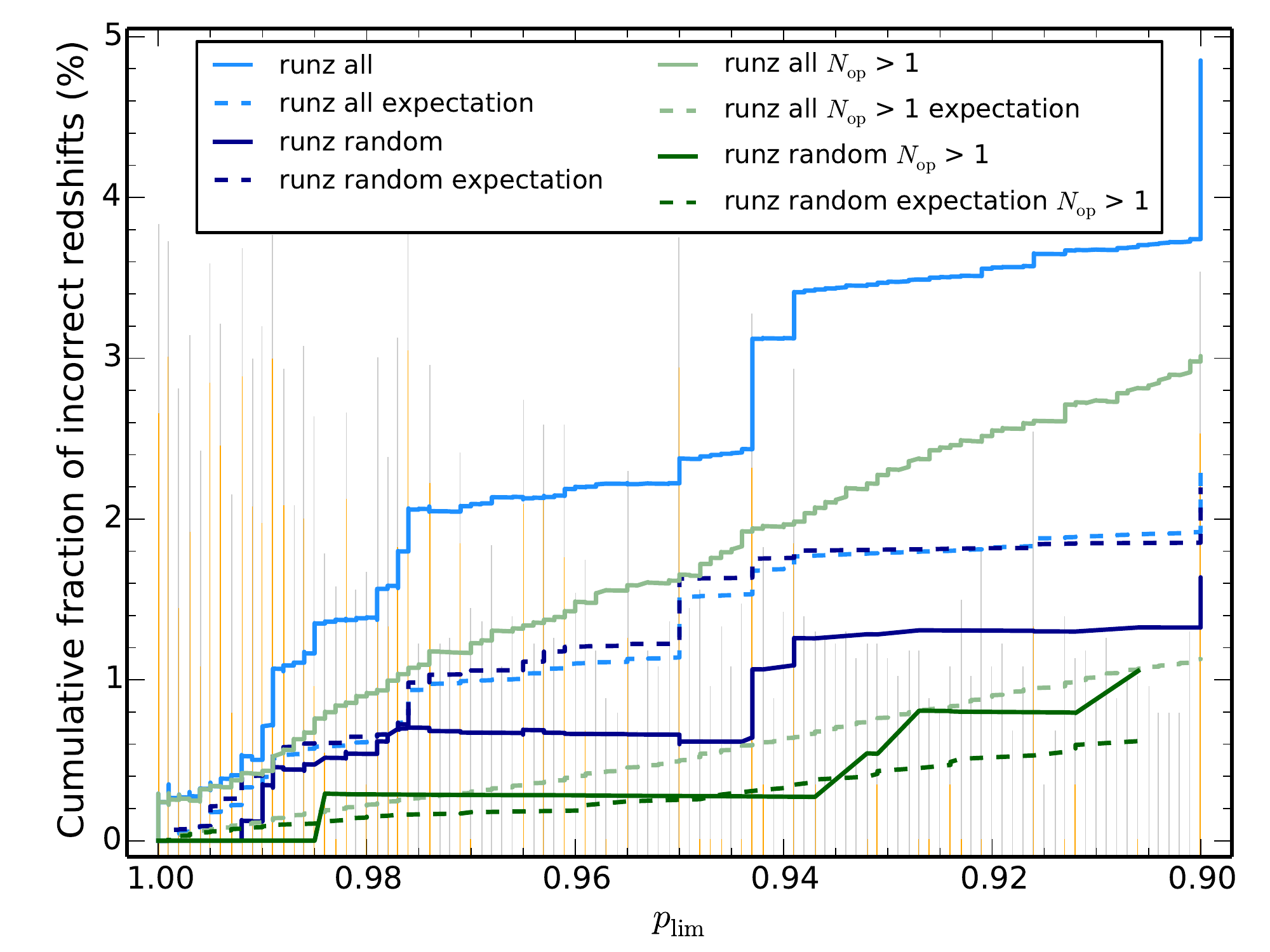}
\caption{The solid light blue line shows the cumulative fraction of
  incorrect {\sc runz} redshifts, i.e.\ the fraction of all {\sc runz}
  redshifts with $p(z) > p_{\rm lim}$ that are incorrect, using all
  available duplicate GAMA\,II redshifts of main survey targets. The
  dashed light blue line shows the cumulative incorrectness fraction
  that is expected from the $p(z)$ distribution of this sample, which
  is shown as the grey histogram in the background using a logarithmic
  scale. The dark blue lines show the same, but now only using the
  duplicate redshifts of the random sample. The orange histogram shows
  the $p(z)$ distributions of this sample. The green lines show the
  same as the blue ones, but now restricting both samples to those
  spectra that have been re-redshifted, i.e.\ that have $N_{\rm op} >
  1$.}
\label{zqc_blunder_runz}
\end{figure}

In the following we will consider any two redshifts of the same object
to disagree if they differ by more than $|\Delta v|_{\rm max} =
750$\kms\ ({\sc runz}) or $350$\kms\ ({\sc Autoz}). These values are
not simply multiples of the overall redshift errors, but were instead
chosen by carefully evaluating where the $|\Delta v|$ distributions
approach the `background' of random pairs. However, in practice the
exact values adopted for $|\Delta v|_{\rm max}$ make almost no
difference to the results. For any redshift pair found to disagree we
then assume that one (and only one) of the two redshifts is
wrong,\footnote{Note that we disregard the second order possibilities
  of both redshifts being incorrect and of both being correct. The
  latter may occur in cases where the target consists of two
  unresolved objects at different redshifts, and where the spectra
  were obtained at slightly different positions on the sky, resulting
  in different objects dominating the flux in the two spectra.} and we
mark the redshift with the lower $p(z)$ as being incorrect. The
spectra of objects with more than two observations are treated
analogously.

\begin{figure}
\includegraphics[width=\columnwidth]{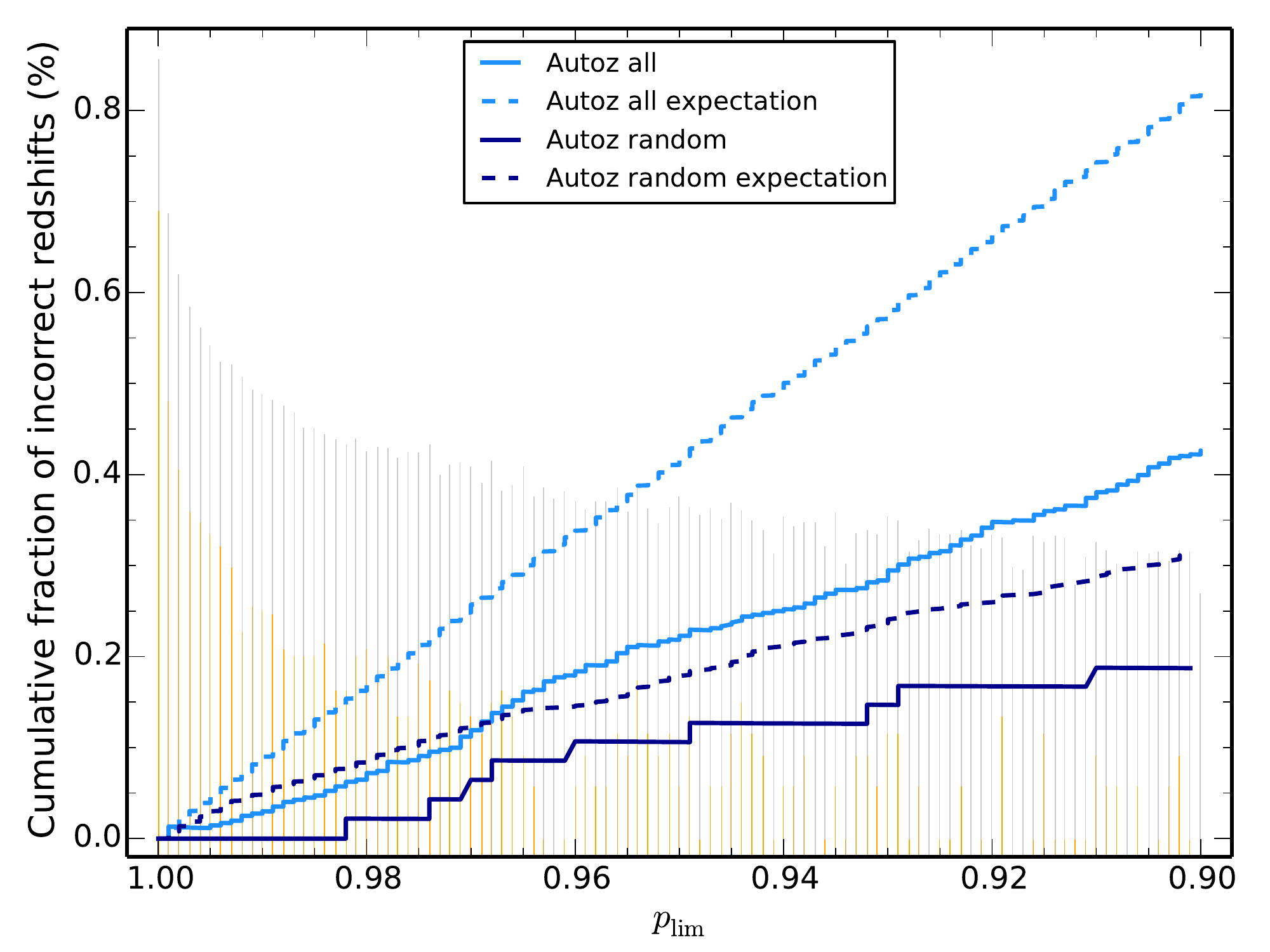}
\caption{Same as Fig.~\ref{zqc_blunder_runz} for {\sc Autoz}. The
  solid light and dark blue lines show the cumulative incorrectness
  fractions of {\sc Autoz} redshifts, using all available
  duplicate redshifts, and only those of the random sample,
  respectively. The dashed lines show the incorrectness fractions
  expected from the $p(z)$ distributions of the two samples, which are
  shown as the grey and orange histograms in the background using a
  logarithmic scale.}
\label{zqc_blunder_autoz}
\end{figure}

The solid lines in Figs.~\ref{zqc_blunder_runz} and
\ref{zqc_blunder_autoz} show the cumulative incompleteness fractions,
i.e.\ the fractions of redshifts with $p(z) > p_{\rm lim}$ that are
incorrect, for both the {\sc runz} and {\sc Autoz} redshifts,
respectively. In each case we show the incorrectness fractions using
all available duplicate redshifts of main survey targets (light blue),
and only those of the random sample (dark blue). The dashed lines show
the corresponding expected fractions computed from the $p(z)$
distributions of the various samples.

Recalling the connection between $p(z)$ and $nQ$ [{\sc runz}: see
  equation~\eref{nQ}; {\sc Autoz}: see Section~\ref{autoz}] we first
of all note that Figs.~\ref{zqc_blunder_runz} and
\ref{zqc_blunder_autoz} only contain redshifts with $nQ \ge 3$,
i.e.\ only those we consider of high enough quality to be accepted for
scientific analyses.

Let us now consider the {\sc runz} results. The step-like features in
the blue curves in Fig.~\ref{zqc_blunder_runz} are due to isolated
peaks in the $p(z)$ distributions of both the random and the full
samples (shown as the histograms in the background). These peaks are
of course due to those spectra in the sample with $N_{\rm op}=1$,
i.e.\ spectra that have not been re-redshifted
(cf.\ Section~\ref{final_z} and Fig.~\ref{rez_probs}). For these
redshifts we have $p(z) = p(i,Q)$ [cf.\ equations~\eref{p_x} and
  \eref{q_x}], meaning that the peaks simply reflect the probabilities
of individual redshifters to `correctly' identify a redshift.

The observed incorrectness fractions of the random and full samples
are clearly very different for all $p_{\rm lim}$ (solid dark and light
blue lines in Fig.~\ref{zqc_blunder_runz}, respectively). For the
random sample we find an incorrectness fraction among all $nQ\ge3$
(i.e.\ $p(z) \ge 0.9$) redshifts of $1.6$~per~cent, whereas for the
full sample we find $4.9$~per~cent. This again confirms the biased
nature of the full sample compared to the random sample.

\begin{table*}
\begin{minipage}{15.2cm}
\caption{Independent surveys from which GAMA is using imaging and/or
photometric data, sorted by wavelength.}
\label{gamasurveys}
\centerline{\begin{tabular}{lllll}
\hline
Survey & Facility & Wavelength / band & Type of data used & Reference\\
\hline
XXL             & \textit{XMM-Newton}$^a$   & $0.5$--$2$~keV                  & Proprietary data & \citet{Pierre11}\\
GALEX-GAMA      & \textit{GALEX}$^b$        & $0.15$, $0.22$~$\mu$m           & Public (MIS$^c$) and own data & this work (Section~\ref{galexphot})\\
SDSS            & Sloan telescope & $u$, $g$, $r$, $i$, $z$         & Public data       & DR7: \citet{Abazajian09}\\
KiDS$^d$        & VST$^e$          & $u$, $g$, $r$, $i$              & Proprietary data  & \citet{deJong13}\\
CFHTLenS$^f$    & CFHT$^g$         & $u$, $g$, $r$, $i$, $z$         & Public data  & \citet{Heymans12}\\
UKIDSS LAS$^h$  & UKIRT$^i$        & $Y$, $J$, $H$, $K$              & Public data       & \citet{Lawrence07}\\
VIKING$^j$      & VISTA$^k$        & $Z$, $Y$, $J$, $H$, $K_s$       & Proprietary data & \citet{Edge13}\\
WISE All-Sky DR & \textit{WISE}$^l$         & $3.4$, $4.6$, $12$, $22$~$\mu$m & Public data       & \citet{Wright10}\\
H-ATLAS$^m$     & \textit{Herschel}         & 100, 160, 250, 350, 500~$\mu$m  & Proprietary data  & \citet{Eales10}\\
DINGO$^n$       & ASKAP$^o$        & $21$~cm                         & In planning       & see \citet{Duffy12}\\
GMRT-GAMA       & GMRT$^p$         & $92$~cm                         & Own data          & \citet{Mauch13}\\
\hline
\end{tabular}}
$^a$\textit{X-ray Multi-Mirror Mission};
~$^b$\textit{Galaxy Evolution Explorer};
~$^c$Medium Imaging Survey;
~$^d$Kilo Degree Survey;
~$^e$VLT Survey Telescope;
~$^f$CFHT Lensing Survey;
~$^g$Canada-France-Hawaii Telescope;
~$^h$UKIRT Infrared Deep Sky Survey -- Large Area Survey;
~$^i$United Kingdom Infrared Telescope;
~$^j$VISTA Kilo-Degree Infrared Galaxy Survey;
~$^k$Visible and Infrared Survey Telescope for Astronomy;
~$^l$\textit{Wide-Field Infrared Survey Explorer};
~$^m$\textit{Herschel} Astrophysical Terahertz Large Area Survey;
~$^n$Deep Investigation of Neutral Gas Origins;
~$^o$Australian Square Kilometre Array Pathfinder;
~$^p$Giant Metrewave Radio Telescope;
\end{minipage}
\end{table*}

Comparing the observed incorrectness fractions with the expectations
from the $p(z)$ distributions (blue dashed lines), we find that they
do not agree for either of the two samples, with the prediction being
too low for the full sample and too high for the random sample. Note
also the similarity of the predictions for the two samples, which
implies a very similar shape of the $p(z)$ distributions. This is
somewhat puzzling at first. After all, we know that the full sample is
`worse' than the random one. Hence one would expect the $p(z)$
distribution of the full sample to be skewed towards lower values,
causing a steeper expected incompleteness fraction relative to the
random sample. The reason the two $p(z)$ distributions are
nevertheless so similar is the fact that both the full and the random
samples are dominated by spectra with $N_{\rm op}=1$, i.e.\ spectra
that have not been re-redshifted ($\sim$$90$ and $\sim$$75$~per~cent,
respectively). That means that the $p(z)$ distributions of both
samples essentially reflect their original $Q$ distributions, although
corrected for the biases of individual redshifters. Nevertheless,
these distributions are too `coarse' to capture the differences
between the two samples. What we are seeing here is a fundamental
limitation of the {\sc runz} dataset, which we already highlighted at
the end of Section~\ref{re_z_analysis}, namely that we are forced to
measure a redshifter's probability of `correctly' identifying a
redshift, $p(i,Q)$, as a function of the very coarse quality parameter
$Q$, and that we cannot capture any variation of $p(i,Q)$ within
$Q$. This is a clear limitation of the predictive power and usefulness
of our {\sc runz} $p(z)$ values when $N_{\rm op}=1$, i.e.\ in the
absence of any re-redshifting.

For re-redshifted data with $N_{\rm op}>1$, however, the situation is
different. The green lines in Fig.~\ref{zqc_blunder_runz} show the
result of restricting both the full and the random sample to only
those spectra with $N_{\rm op}>1$. First, we note that the expected
incorrectness fractions (green dashed lines) are now much lower than
before for both samples, as they should be, since redshifts with
independent confirmation should have a lower probability of being
incorrect. Second, we note that the expected incorrectness fractions
are now different for the two samples, in the sense one would expect,
i.e.\ a lower fraction for the random sample. Furthermore, the
observed incorrectness fraction for the random sample (solid dark
green line) now largely agrees with the expectation, although the
observations are plagued by low-number statistics. For the full
sample, however, the observed incorrectness fraction is still much
larger than the expectation. We attribute this to another fundamental
limitation of the {\sc runz} data: as we have pointed out repeatedly
throughout Section~\ref{re-redshifting}, $p(z)$ does {\em not}
represent the probability of a redshift being correct in any absolute
sense. Instead, it is the probability that multiple redshifters, given
the same data and code, will identify the same redshift. To see the
difference, consider a low-S/N spectrum that shows only a single,
marginally significant redshift. Since there are no other redshift
candidates, it is likely that two or even three redshifters will agree
that this is the best redshift. Even if each redshifter individually
only assigns a low confidence (i.e.\ $Q=2$), the agreement will
nevertheless result in a reasonably high value of $p(z)$, correctly
indicating the likelihood that this is the `best' redshift. However,
that does not change the fact that the redshift is of only marginal
significance and hence may well be wrong. In other words, $p(z)$ does
not incorporate any measure of the absolute significance of a
redshift.\footnote{This would best be done by comparing a measure of
  the significance of a redshift to those of other possible redshifts
  in the same spectrum. Indeed, this is the figure of merit used by
  {\sc Autoz}, see \citep{Baldry14}. However, such information is not
  available in the {\sc runz} data.} While the reasonable agreement
between the observed and expected incorrectness fractions for the
random sample indicates that this shortcoming does not affect the {\sc
  runz} redshifts on average, it does appear to affect the full sample
which is biased towards spectra that are more difficult to redshift.

We now turn to the {\sc Autoz} results in
Fig.~\ref{zqc_blunder_autoz}. For the random sample the incorrectness
fraction of all $nQ\ge3$ redshifts is just $0.2$~per~cent, a
remarkably low value. This is vastly superior to the performance of
{\sc runz}, even when restricting ourselves to the re-redshifted
data. For the full sample, the incorrectness fraction is higher by a
factor of $\sim$$2$, again confirming the biased nature of this
sample. Note that for {\sc Autoz} the comparison between the observed
and expected incorrectness fractions does not represent an independent
test of the reliability of the $p(z)$ values (as was the case for {\sc
  runz}), because the duplicate redshifts were already used in
establishing the relation between {\sc Autoz}'s figure of merit and
$p(z)$ \citep{Baldry14}. The result that the observed incorrectness
fractions are somewhat smaller than the expected ones for both samples
simply confirms that this relation was calibrated quite
conservatively.

In conclusion, we find that {\sc Autoz} significantly outperforms {\sc
  runz} (including re-redshifting) both in terms of the precision of
the redshifts as well as in terms of producing a higher confidence in
the redshifts. In addition, {\sc Autoz} finds more $nQ\ge3$ redshifts
than {\sc runz} (by $11$~per~cent). Thus, there is no trade-off:
{\sc Autoz} is unequivocally superior to {\sc runz}.

\section{Photometric procedure updates}
\label{photometry}

\begin{figure}
\includegraphics[width=\columnwidth]{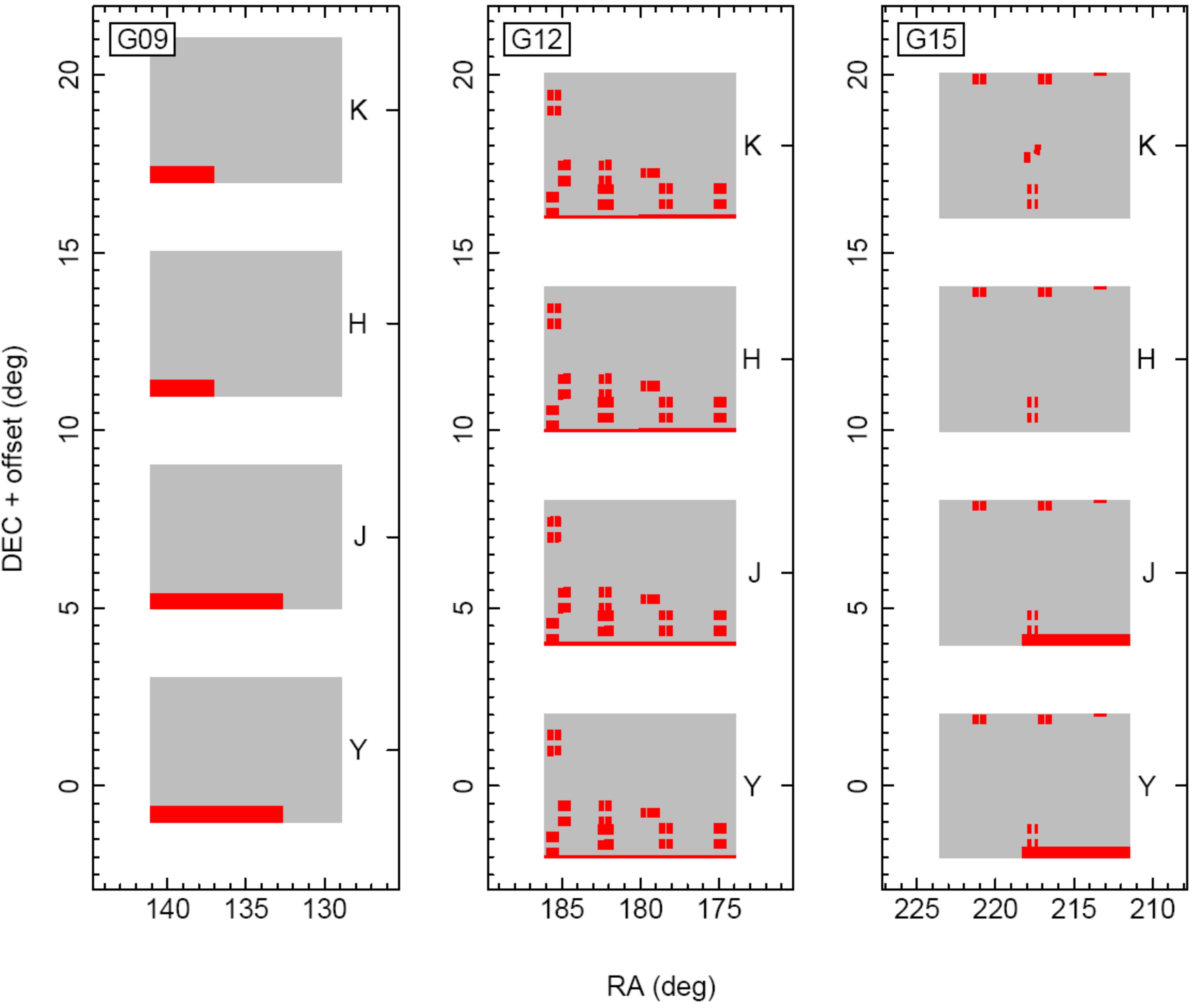}
\caption{UKIDSS LAS coverage of the GAMA\,I survey regions in the
  $YJHK$ bands, where each band is plotted with a declination
  offset. Red areas indicate missing data.}
\label{ukidss_coverage}
\end{figure}

Having discussed the GAMA spectroscopic survey in the previous two
sections, we now turn towards the photometric side of GAMA. Our
overall aim is to obtain photometric measurements of all GAMA main
survey objects across the entire accessible wavelength range, from the
X-ray to the radio regimes, in order to probe as wide a range of
galaxy properties and processes as possible. To this end we
collaborate with several independent imaging surveys, conducted our
own observing campaigns, and draw on publicly available data from a
number of sources. Table~\ref{gamasurveys} provides an overview of the
datasets that are currently being processed within GAMA. A
comprehensive data release of GAMA photometry will be presented by
Driver et al.\ (in preparation).

In this section we focus on the optical, NIR and UV data. We provide
an update of our aperture-matched optical and NIR photometry and
describe for the first time our procedure of deriving UV photometry
from \textit{GALEX} data.

\subsection{Aperture-matched optical and NIR photometry}
\label{onirphot}

\citet{Hill11} first described our procedure to derive
aperture-matched multi-band photometry from SDSS and UKIDSS LAS
imaging data, i.e.\ in the $u$, $g$, $r$, $i$, $z$, $Y$, $J$, $H$ and
$K$ bands, for the GAMA\,I survey regions. We will shortly
update these imaging data with deeper data from the ongoing VST KiDS
and VISTA VIKING surveys. In the meantime, however, we have updated
our photometric methods and procedures, which we describe in this
section.

In brief, the new v02 photometry improves on the original v01
photometry of \citet{Hill11} in the following ways: (i)~Visual
inspection and validation of all UKIDSS LAS images used in the
construction of the mosaics (see below) to overcome the previous
inclusion of poor quality frames (including strongly defocused and
trailed data). (ii)~Consistent modelling of the point-spread function
(PSF) across all data frames in all bands. Previously we had used the
PSF information provided by the SDSS and UKIDSS LAS image
headers. However, the two surveys employ different methods for
measuring the PSF.

\begin{figure}
\includegraphics[width=\columnwidth]{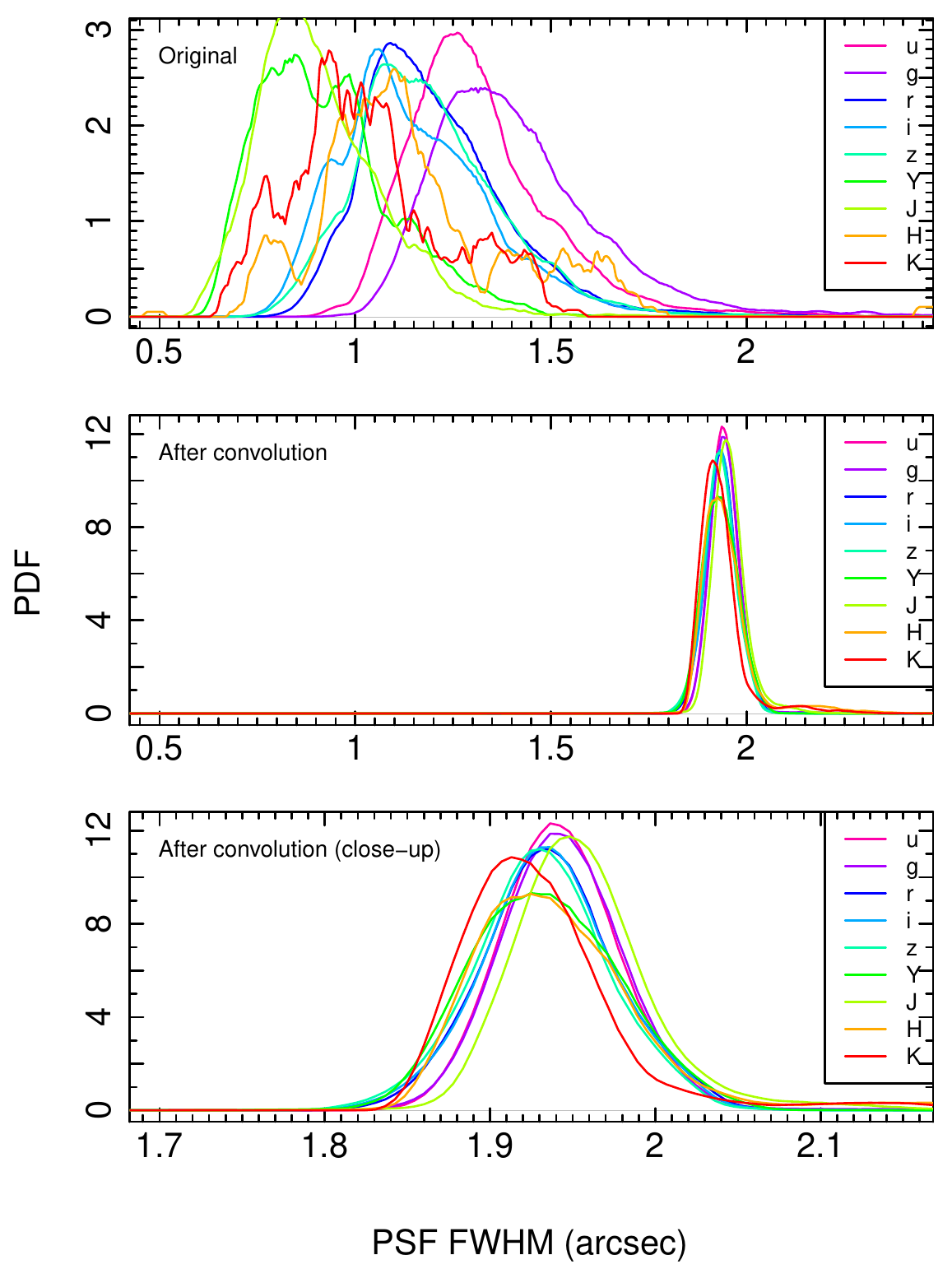}
\caption{Each line shows the distribution of the seeing values of all
  images in a particular band, as indicated by the legend. The top
  panel shows the original seeing distributions, while the middle and
  bottom panels show the distributions after the convolution process.
  All seeing values were measured using {\sc PSFEx} \citep{Bertin11}.}
\label{seeing_distr}
\end{figure}

\subsubsection{Data}

We have downloaded all fully reduced images that overlap with the
three GAMA\,I survey regions (cf.\ Table~\ref{gamaregions}) from the
SDSS DR7 and UKIDSS LAS DR6 and DR8 databases. For SDSS these were the
{\em fpC} frames, while for UKIDSS LAS we used the {\em stack} and
{\em leav-stack} frames. Given the extent and diversity of these data
it is clear that we need to homogenise them before we can obtain
reliable multi-band photometry. Following \citet{Hill11} we thus first
construct homogenised master mosaic images, one for each region and
band, and then use these mosaics to perform the photometry.

\subsubsection{Mosaic construction}

We begin by visually inspecting all images to check their quality. A
small number of frames in the NIR bands were discarded as a result of
these checks, mostly because they were either out of focus or
displayed a large amount of jitter. We discarded $33$, $13$, $49$, and
$48$ frames in the $Y$, $J$, $H$ and $K$ bands, respectively. Even
after removing these frames the coverage of the three GAMA\,I survey
regions remains high: $95.2$~per~cent in $Y$ and $J$, and
$97.5$~per~cent in $H$ and $K$. In Fig.~\ref{ukidss_coverage} we show
the coverage of our three survey regions in these bands in more
detail. The coverage in the SDSS $ugriz$ bands is essentially
$100$~per~cent, excluding only small regions that were masked because
of bright stars and artefacts. See \citet{Driver11} for details of the
GAMA mask.

\begin{figure}
\includegraphics[width=\columnwidth]{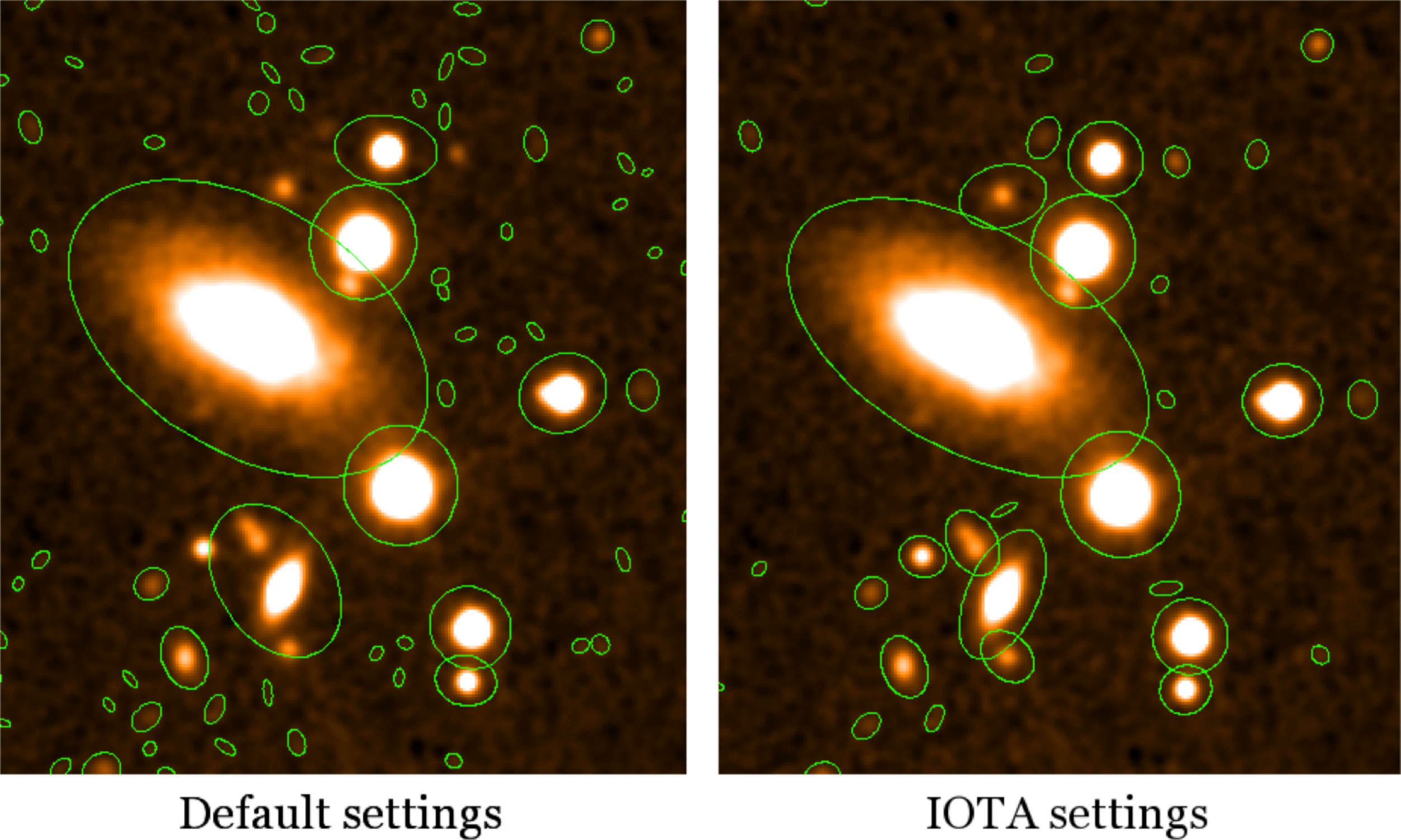}
\caption{Illustration of {\sc SExtractor}'s deblending choices for the
  default settings of the deblending parameters (left) and for the
  settings used by {\sc iota} (right). The image is a cutout from one
  of the PSF-homogenised $r$-band mosaics.}
\label{deblend}
\end{figure}

Next, we renormalise all frames to a common zero-point by multiplying
each frame with an appropriate factor derived from the frame's
original zero-point as given in its header. The common zero-point was
chosen as $30$~mag \citep{Hill11}.

Fig.~\ref{seeing_distr} (top) shows the distributions of the seeing in
all of our contributing frames for each band separately, as measured
using {\sc PSFEx} \citep{Bertin11}. Given the widths and offsets of
these distributions it is clear that performing aperture-matched
photometry on these images would yield poor-quality colour
measurements. To ensure uniformity we therefore elect to degrade all
of the imaging data to a uniform PSF FWHM of $2$~arcsec, which is
larger than the native PSF FWHM of essentially all of the NIR data and
of $\sim$$95$~per~cent of the SDSS data. To degrade a given image we
convolve it with a Gaussian kernel of FWHM $\Gamma_{\rm con}^2 =
(2~\mbox{arcsec})^2 - \Gamma_{\rm orig}^2$, where $\Gamma_{\rm orig}$
is the original PSF FWHM of the image. The middle panel of
Fig.~\ref{seeing_distr} shows the measured PSF FWHM distributions,
again using {\sc PSFEx}, after the convolution. The bottom panel shows
a close-up version which highlights the residual widths of the final
seeing distributions and their offsets from the target value of
$2$~arcsec. We believe these residual variations and offsets to be due
to the non-Gaussian nature of the original PSFs. We elect not to
refine the process further in anticipation of the higher quality data
from the VST KiDS and VISTA VIKING surveys.

At this point we have two sets of renormalised frames: those at native
seeing and those convolved to a common PSF. While we require the
PSF-homogenised data for our aperture-matched photometry, many other
scientific applications, such as e.g.\ structural decomposition,
require the data at their original resolution. Hence we now create two
large format mosaics for each survey region and for each band, one
using the convolved data, and one using the original data. To create
these mosaics we use the code {\sc SWarp} \citep{Bertin02}. The
mosaics are $\sim$$15 \times 5$~deg$^2$ in size (i.e.\ substantially
larger than the actual GAMA\,I survey regions) and have a pixel size
of $0.339$~arcsec (which is the pixel scale of VISTA). The mosaic
creation process is essentially identical to that described by
\citet{Hill11}.

\subsubsection{Aperture-matched photometry}

Aperture-matched photometry is performed on the convolved mosaics
using the code {\sc iota}, which is a wrapper around {\sc SExtractor}
\citep{Bertin96}. {\sc iota} takes as an input a list of positions at
which to perform flux measurements. For the v02 photometry presented
here, this list was generated by selecting all objects with
SURVEY\_CLASS $\geq 3$ from {\tt TilingCatv16}, resulting in
$152\,742$ galaxies. {\tt TilingCatv16} is the final GAMA\,I
targeting catalogue and is entirely based on SDSS DR6 (see
\citealp{Baldry10} and Section~\ref{dr2_ic}). For each object in this
list and for each band, {\sc iota} creates a $400 \times 400$ pixel
image cutout from the PSF-homogenised mosaics around the object's
position. It then runs {\sc SExtractor} eight times in dual-image
mode, each time using the $r$-band image as the detection image and
one of the remaining images as the measurement image. In this way the
Kron aperture used for flux measurements is defined in the $r$-band
and is consistently applied to all other bands. {\sc iota} then extracts
the relevant information from the {\sc SExtractor} output and
associates them with the input object.

We note that the above procedure of running {\sc SExtractor} only over
small image segments at pre-specified positions is significantly
faster than running it over the entire mosaics.

An important aspect of running {\sc SExtractor} is the setting of its
deblending parameters DEBLEND\_NTHRESH and DEBLEND\_MINCONT. After
some trial and error we now use the extreme values of $32$ and
$0.00005$, respectively. These extreme values are required because the
images have been low-pass filtered. In Fig.~\ref{deblend} we show an
example of {\sc SExtractor}'s deblending choices for a reasonably
complex region with the default settings of the deblending parameters
(left) and with our settings (right). The improvement is evident.

\begin{figure}
\includegraphics[width=\columnwidth]{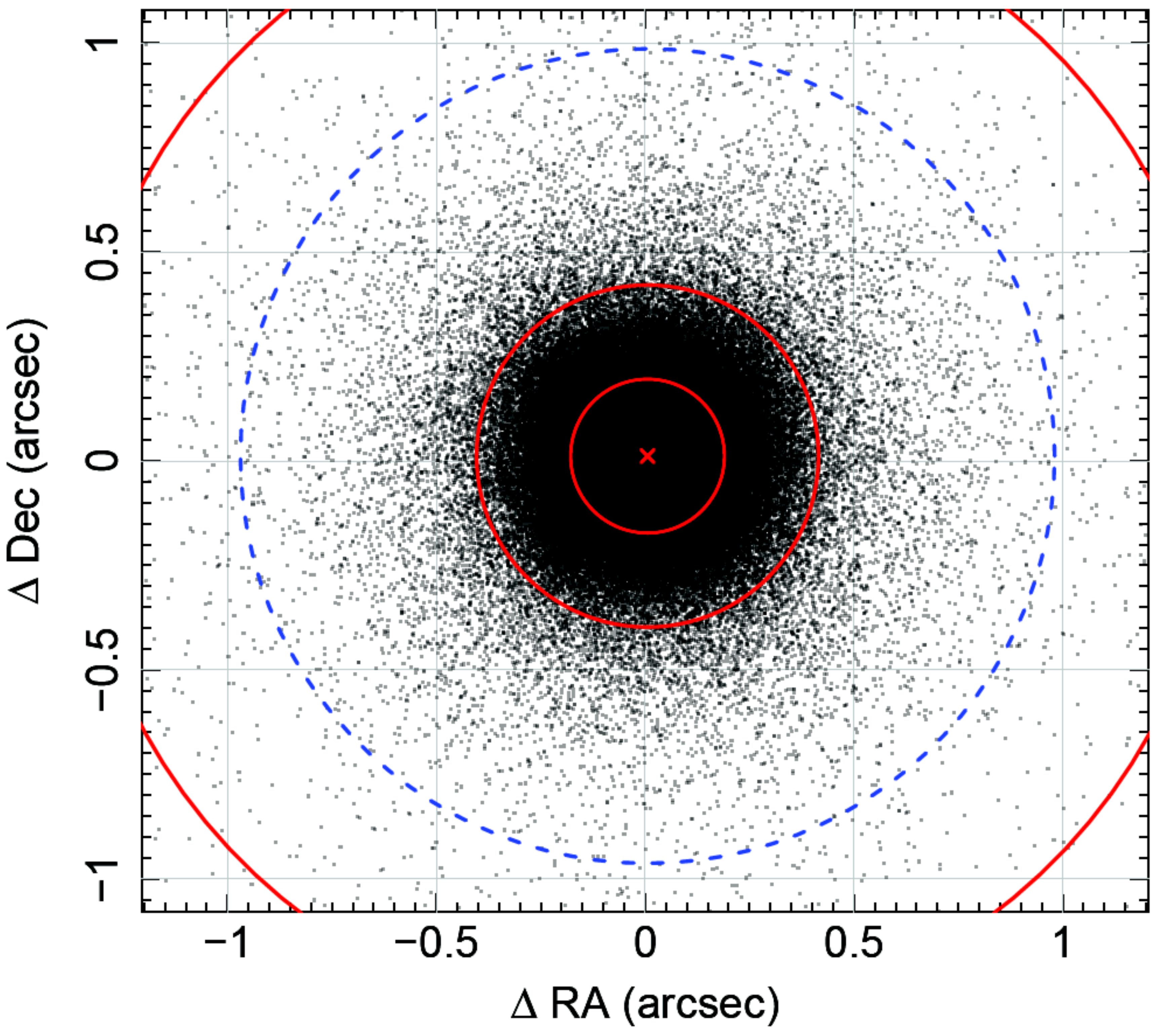}
\caption{Offsets between the SDSS and {\sc SExtractor} positions in
  the $r$-band. The red circles enclose $50$, $90$ and $99$~per~cent
  of the data. The blue dashed circle shows the PSF FWHM of the
  convolved mosaics.}
\label{pos_offsets}
\end{figure}

Nevertheless, given the different codes and resolution of the data,
one may ask to what extent the SDSS-defined objects of the input
catalogue correspond to the {\sc SExtractor}-defined objects derived
here. Fig.~\ref{pos_offsets} shows the offsets between the $r$-band
input positions from the SDSS and the $r$-band positions found by {\sc
  SExtractor}. The red circles enclose $50$, $90$ and $99$~per~cent of
the data. According to this diagnostic, at least, the correspondence
is good.

\begin{figure*}
\includegraphics[width=0.85\textwidth]{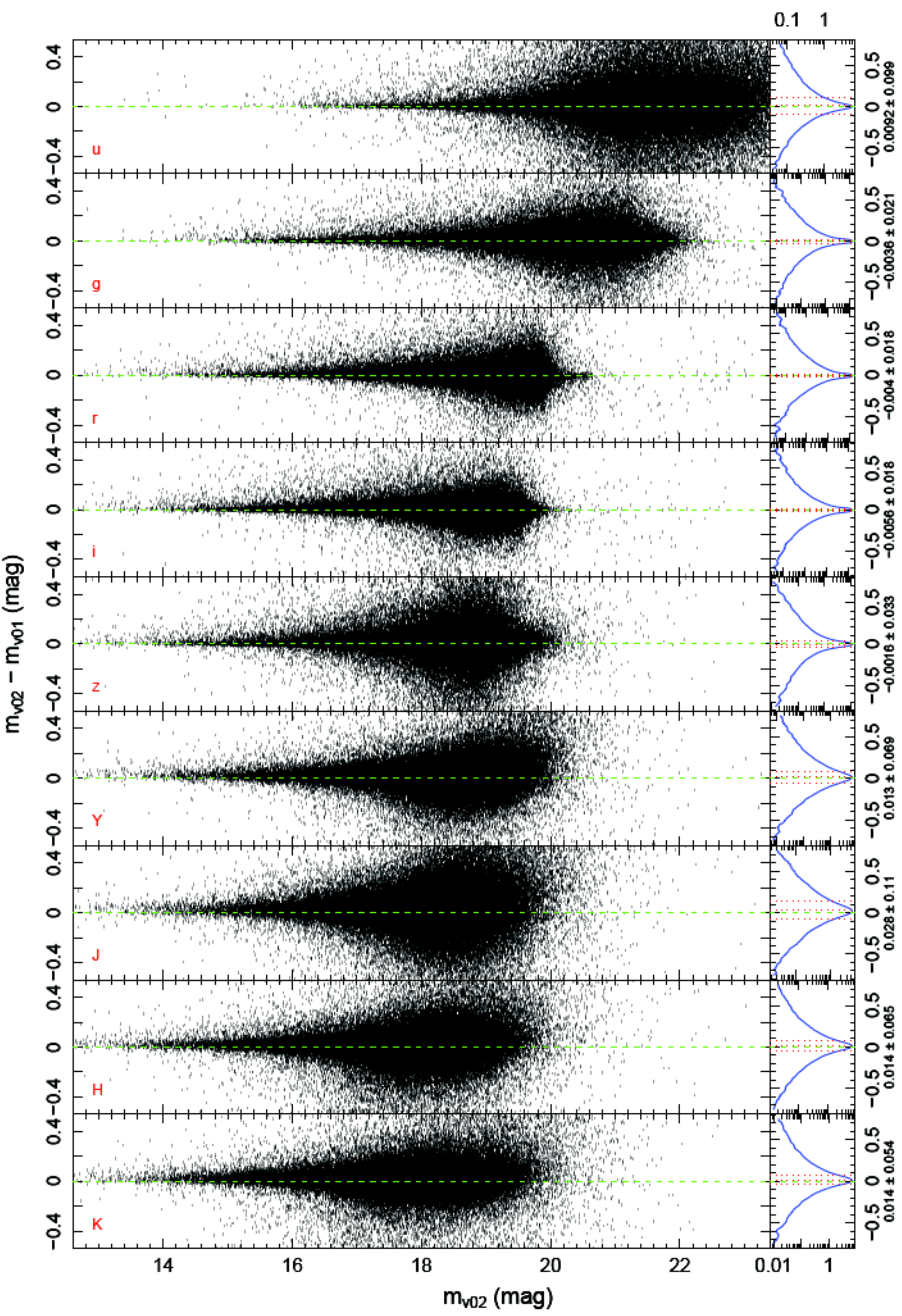}
\caption{Comparison between the Kron magnitudes of v01 \citep{Hill11}
  and v02 (this work) of the aperture-matched photometry. Each panel
  shows the magnitude differences in a different band, as
  indicated. The right-hand panels show the distributions of the
  magnitude differences. The numbers to the right of these panels are
  the means and standard deviations of these distributions. These are
  also marked by the red dotted lines.}
\label{trumpets}
\end{figure*}

\subsubsection{Comparison between v01 and v02 photometry}

Fig.~\ref{trumpets} provides a direct comparison between the v01
photometry of \citet{Hill11} and the v02 photometry presented here. We
note that both the mean offsets from zero and the standard deviations
are all rather minor for the optical bands (as shown by the
distributions of the magnitude differences in the right-hand panels),
with the $g$, $r$, and $i$ bands showing particularly small
dispersions. However, in the NIR bands the differences between the v01
and the v02 photometry are significantly larger, both in terms of the
mean offset and the dispersion. The primary reason for these
differences is the change in the way we determine the width of the
Gaussian kernel with which the images are convolved.

\subsubsection{Photometric errors}

The convolution of the mosaics with a Gaussian kernel obviously
changes the error properties of the images dramatically. The resulting
correlations among the errors on the pixel data is not taken into
account by {\sc SExtractor}, and hence the photometric errors
calculated by {\sc SExtractor} cannot be relied upon in any absolute
sense. We will, however, assume that the {\sc SExtractor} errors are
meaningful in a relative sense. We thus rescale these errors in the
following way:
\begin{equation}
\sigma_{\rm f} = \sqrt{k_1 \; \sigma_{\rm \sc SEx}^2 + k_2},
\end{equation}
where $\sigma_{\rm \sc SEx}$ and $\sigma_{\rm f}$ are an object's {\sc
  SExtractor} and final rescaled errors, respectively, and $k_1$ and
$k_2$ are band-specific positive constants. In the $u$$g$$r$$i$$z$
bands, $k_1$ and $k_2$ are derived by minimising the differences
between the mean values of $\sigma_{\rm f}$ and of the SDSS DR6
Petrosian magnitude errors as a function of SDSS magnitude and
aperture size. For the NIR data, $k_1$ and $k_2$ are derived by
matching the UKIDSS LAS photometric errors as a function of magnitude
only (as the UKIDSS LAS does not provide any aperture size
information).

\subsection{\textit{GALEX} photometry}
\label{galexphot}

Photometry in the rest-frame, non-ionizing UV wavelength regime is a
sensitive probe of the star-formation activity of galaxies, and as
such it plays an important role within GAMA's multi-wavelength
campaign, enabling a wide range of studies of the connections between
star formation activity and other galaxy properties. Moreover, in
conjunction with measurements of the dust emission in the far-infrared
and sub-mm regimes (provided by the \textit{Herschel}-ATLAS data in
the GAMA regions) and measurements of the size, inclination and
morphology of galaxies, UV photometry provides the observational basis
for a quantitative description of the transport of starlight in the
dusty disks of spiral galaxies, allowing the relative contributions to
the heating of dust by optical and UV photons to be separated. This,
in turn, allows us to break the age/reddening degeneracy, to quantify
the {\em intrinsic} emission of stars in galaxies throughout the
UV-optical-NIR range, and to robustly determine the star formation
histories of GAMA galaxies.

In this section we describe our methods of deriving UV photometry for the
GAMA survey regions from imaging data obtained with \textit{GALEX}
\citep{Martin05}.

\subsubsection{The GALEX-GAMA survey}

Archival and newly obtained data from \textit{GALEX} have been used to
construct an imaging survey -- the GALEX-GAMA survey -- of
$92$~per~cent of the area of the five GAMA\,II survey regions
(cf.\ Table~\ref{gamaregions}) to a detection limit for galaxies of at
least $m_{\rm AB} = 24.5$~mag ($0.59$~$\mu$Jy) in the \textit{GALEX}
near-ultraviolet (NUV) band ($1750$--$2750$~\AA). In addition,
$69$~per~cent of the \textit{GALEX} NUV footprint are also covered in
the \textit{GALEX} far-ultraviolet (FUV) band ($1350$--$1750$~\AA) to a
limit of at least $m_{\rm AB} = 24.3$~mag ($0.72$~$\mu$Jy). These
limits correspond to the typical depth reached in the GAMA\,II
regions\footnote{The quoted limits are the mean $2.5$$\sigma$ upper
  limits in integrated emission from the optically emitting regions of
  all undetected GAMA galaxies with total \textit{GALEX} exposure
  times in the range $1400$ to $1600$~s. These limits therefore take
  into account the photon statistics integrated over the angular
  extent of the galaxies for the actual background levels encountered
  towards the GAMA regions at the epoch of the observations. Foreground
  extinction by dust in the Milky Way is not taken into account in
  these limits; over the GAMA\,II regions this dims galaxies by a
  median of $0.26$ and $0.25$~mag in the NUV and FUV, respectively.}
if an area of sky covered by the circular \textit{GALEX} field of
view, of diameter $1.2$~deg, is continuously observed for the typical
$\sim$$1500$~s duration spent by \textit{GALEX} in eclipse in each
orbit. This depth is commonly referred to as Medium Imaging Survey
(MIS) depth, after the \textit{GALEX} survey of selected regions of
the sky observed in the same manner \citep{Martin05,Bianchi14}. As
illustrated below, MIS-depth coverage has proved to be well matched to
the spectroscopic depth of GAMA\,II, and is capable of detecting a galaxy
with the present-day emergent NUV luminosity of the Milky Way out to 
a redshift of $0.53$.

\begin{figure}
\includegraphics[width=\columnwidth]{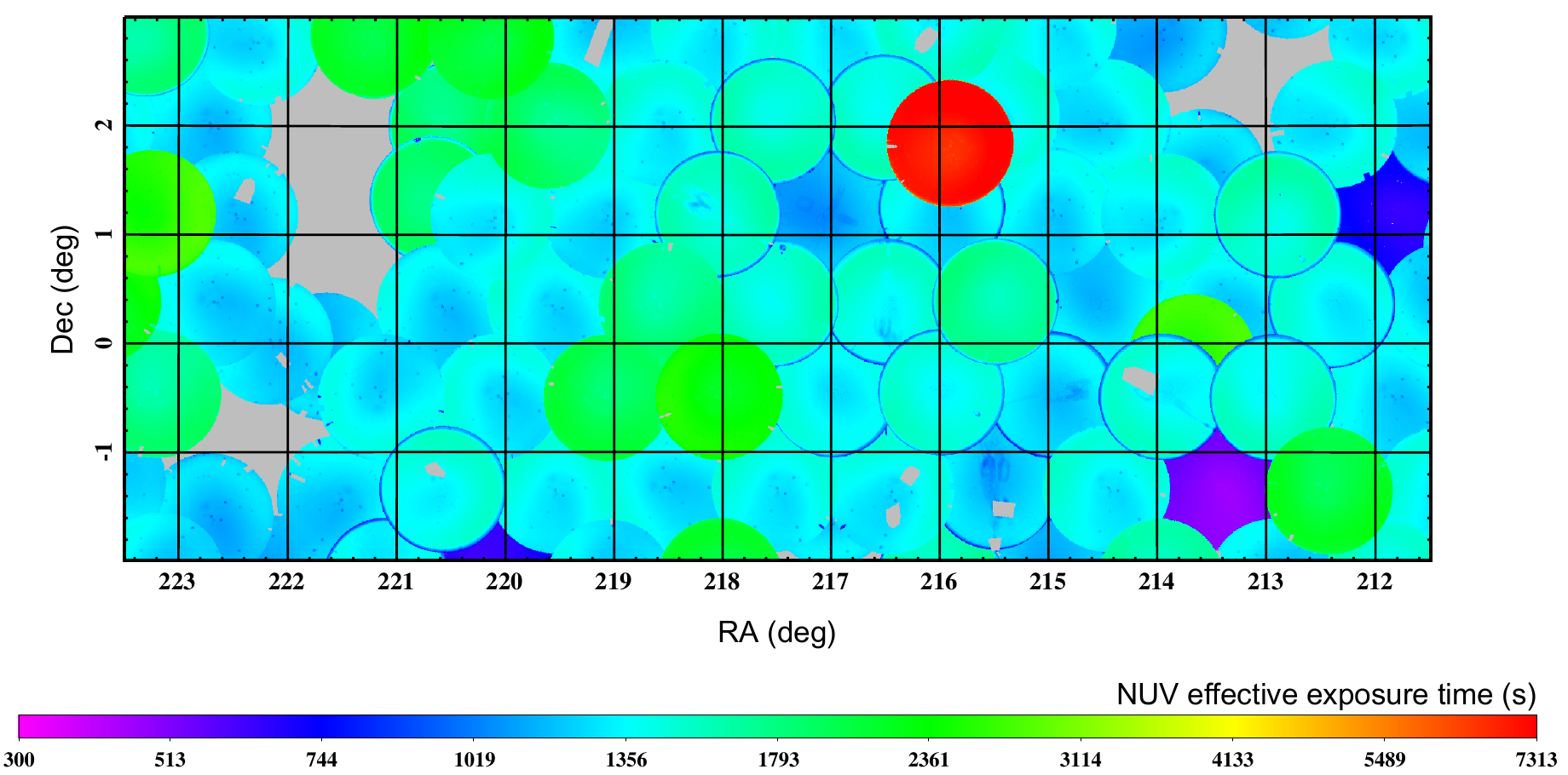}
\caption{Effective exposure time in the NUV in the G15 survey region
  using only pointings with an exposure time of at least
  $400$~s. Holes in the coverage are due to bright stars and
  reflection artefacts caused by bright stars on neighbouring tiles.}
\label{nuv_cov}
\end{figure}

Fig.~\ref{nuv_cov} shows an exposure map of the MIS-depth coverage of
the G15 region. This illustrates the closely packed, overlapping,
hexagonal tiling pattern used to cover all of the GAMA\,II survey
regions, which is only broken to avoid bright stars. This almost
complete NUV coverage of the GAMA regions at MIS depth was achieved by
combining archival data from previous MIS-depth programmes with those
from two new programmes dedicated to GAMA. The latter were the
\textit{GALEX} guest observer programme GI5-0048, designed to complete
the MIS-depth NUV coverage of GAMA's equatorial survey regions, and a
program performed in the final year of \textit{GALEX} operations
(during its extended mission) to map the G23 region. Furthermore, the
G02 region lies within the area of the multi-wavelength extension of
the Cosmological Evolution Survey \citep[COSMOS;][]{Scoville07}. Hence
it is completely covered to the $20$~ks depth of the \textit{GALEX}
Deep Imaging Survey \citep[DIS;][]{Martin05,Zamojski07} in both the
NUV and FUV bands. We point out that, due to the failure of the
\textit{GALEX} FUV detector in 2009, FUV coverage at MIS or greater
depth is incomplete. FUV coverage is therefore primarily confined to
G15 and G02, and partially extends to G09 and G12. Except in the
vicinity of bright stars, all regions not covered to MIS depth are
nevertheless covered in both bands by the \textit{GALEX} All-Sky
Imaging Survey \citep[AIS;][]{Martin05,Bianchi14} with a typical
exposure time of $100$~s.

\subsubsection{Extraction of photometry}

The starting point for our data analysis is the set of standard
\textit{GALEX} pipeline products described by \citet{Morrissey07}.
Both the initial processing, done using the standard Caltech
\textit{GALEX} pipeline, as well as the subsequent analysis by the
GAMA team, differ substantially according to the depth of the data,
due to differing noise and blending characteristics. This results in
three different sets of GALEX-GAMA products, one for each survey depth
(AIS, MIS or DIS-depth). Since in DR2 we only release photometry
derived from the MIS-depth data (see Section~\ref{dr2_galex}), we
restrict our description below to the analysis of these data.  A more
complete description of this analysis, as well as of the analysis of
the AIS and DIS-depth data, will be provided by Andrae et al.\ (in
preparation).

The resolution of the \textit{GALEX} images is significantly lower
than that of the SDSS data used to define the GAMA\,II sample: the
FWHM of the PSF is $4.2$ and $5.3$~arcsec in the FUV and NUV bands,
respectively \citep{Morrissey07}. Given the faint flux levels and
corresponding high source densities of the GAMA sample, we must
therefore expect that a significant fraction of \textit{GALEX}
detections consist of the blended UV emission from multiple GAMA
galaxies, and that the assignment of UV flux to GAMA objects is
non-trivial.

To address this issue we have employed three different methods to
derive the NUV and FUV fluxes of each GAMA galaxy. We label these
methods `simple match photometry', `advanced match photometry' and
`curve-of-growth (CoG) photometry', and we describe each of these in
detail below. Briefly, the first method simply associates each GAMA
object with its nearest neighbour \textit{GALEX} source, as detected
by the standard \textit{GALEX} pipeline, within a maximum distance of
$4$~arcsec. The second method extends the first by identifying those
cases where multiple GAMA and/or \textit{GALEX} objects are associated
with each other, and attempting to distribute the UV flux correctly
among the GAMA objects involved. Finally, in our third method we go
back to the \textit{GALEX} imaging data and perform our own surface
photometry at the known positions of GAMA objects. The UV fluxes found
for a GAMA object by these three different methods are affected
differently by blending, allowing the definition of objective criteria
to decide which method should be used under which circumstances in
order to minimise systematic errors in the photometry.

The first two of the above methods use the catalogue of blind UV
detections produced by the \textit{GALEX} pipeline as an input. We
therefore describe these data first.

\paragraph {Blind UV photometry}
\label{galex_blind}

A catalogue of blind UV photometry for each GAMA\,II survey region was
constructed by concatenating the catalogues of UV fluxes and UV
structural parameters of discrete sources output by version 7.0.2 of
the \textit{GALEX} pipeline for each tile (generally corresponding to
a single \textit{GALEX} pointing in eclipse for MIS-depth coverage).
As described by \citet{Morrissey07}, the source identification,
background removal and shape fitting was done using a modified version
of {\sc SExtractor} \citep{Bertin96}, adapted to handle the transition
from Poisson-dominated backgrounds (as is generally the case for
MIS-depth FUV images) to Gaussian-dominated backgrounds (as is
generally the case for MIS-depth NUV images).

In the present analysis we made no attempt to combine the data on
sources that lie in the overlap region of two or more tiles. In order
to prevent multiple detections of the same source (on different tiles)
entering our catalogue, we first had to associate each position in the
GAMA\,II survey regions with a `primary' tile. For regions of sky covered
by more than one tile the primary was chosen firstly according to
whether or not a tile has unmasked MIS-depth FUV coverage at the
position under consideration, and secondly according to the effective
exposure time in the NUV.

A key characteristic of the blind catalogue is that the source detection
and the definition of the aperture for photometry are performed
exclusively in the NUV band. The FUV flux of each NUV-detected source
was then measured using the NUV-defined aperture. This procedure was
adopted because of: (i)~the more complete sky coverage in the NUV
compared to the FUV; (ii)~the improved precision of FUV-NUV colours;
(iii)~the smoother background in the NUV, where it is dominated by
zodiacal light, compared to the FUV, where it is more highly
structured due to a larger fractional contribution from cirrus
structure in the interstellar medium of the Milky Way. Since the
measurement error depends on the local brightness of the background,
the contents of the blind catalogue more closely approximate a
flux-limited sample when selected according to detectability in the
NUV rather than in the FUV. Because at MIS-depth NUV sensitivity is
very similar to FUV sensitivity for typical galaxies, and because
almost all stars are more easily detected in the NUV, relatively few
sources are missed due to the choice not to consider sources that
might be detected in the FUV but not in the NUV.

We note that the \textit{GALEX} blind catalogue does not include {\em
  all} NUV detections. Instead it is limited to those sources that are
detected at a significance of at least $2.5$$\sigma$ in the NUV.

\paragraph{Simple match photometry}
\label{galex_simple}

In the simple match photometry method we positionally match the
NUV-detected sources from the blind catalogue above to the optically
detected GAMA objects in the GAMA\,II input catalogue. In this process the
match to a GAMA object is considered to be the nearest \textit{GALEX}
source in the blind catalogue within a maximum distance of
$4$~arcsec. Matches can of course involve any type of object contained
in the input catalogue, including galaxies that are spectroscopic
targets, fainter galaxies, and stars (down to the input catalogue's
limit of $r=20$~mag).

The main parameters of the matched \textit{GALEX} source, such as its
NUV position, ellipticity, size, and NUV and FUV fluxes,
are included in the simple match catalogue. We point out that, as a
consequence of the blind catalogue construction, the NUV flux of a
GAMA object in the simple match catalogue is guaranteed to have a
statistical significance of at least $2.5$$\sigma$, whereas the
significance of the corresponding FUV flux measurement may often fall
below this level. Indeed, the flux may even be negative.

\begin{figure*}
\includegraphics[width=\textwidth]{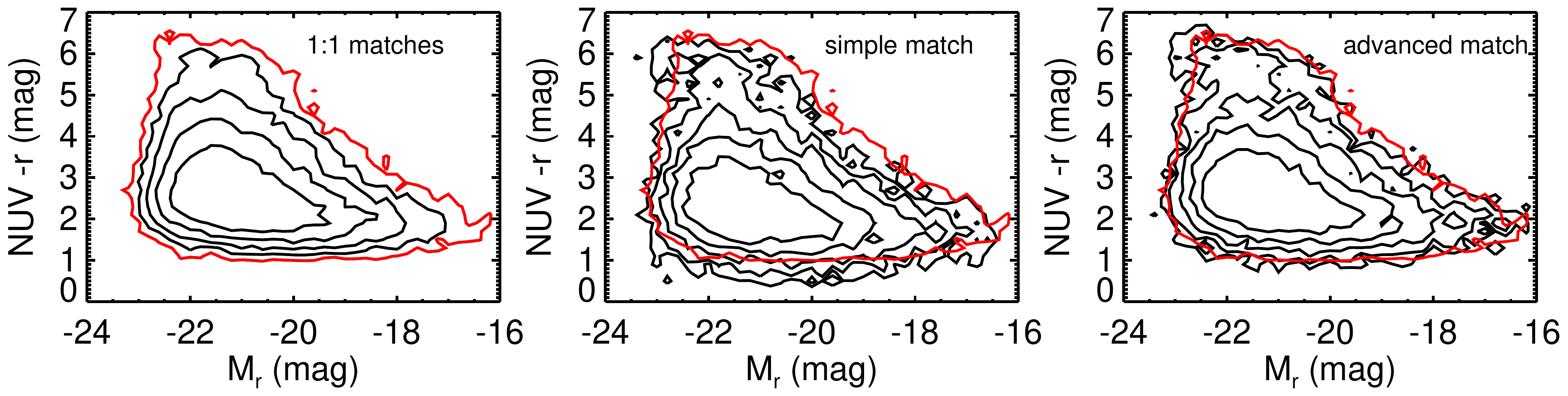}
\caption{Object number densities of various samples of GAMA galaxies
  in the colour-magnitude plane spanned by NUV$-r$ colour and absolute
  $r$-band magnitude. The contours are drawn at $0.5$, $1$, $2$, $4$
  and $8$~per~cent of the total source number density. In the left
  panel we only use GAMA objects that have an unambiguous
  (i.e.\ one-to-one) match with a single \textit{GALEX} source. The
  outermost contour in this panel (marked in red) is reproduced in the
  other panels for comparison. In the middle panel we only use those
  objects that are {\em not} unambiguously matched to a single
  \textit{GALEX} source, and we use the NUV flux from its nearest
  \textit{GALEX} neighbour, i.e.\ as returned by the simple match
  procedure. In the right panel we use the same objects as in the
  middle panel, but we now use the redistributed NUV flux as returned
  by the advanced match procedure.}
\label{nuv-r}
\end{figure*}

GAMA objects without a nearest neighbour in the blind catalogue within
$4$~arcsec are considered to be unmatched, and were not included in
the simple match catalogue. To be able to distinguish between GAMA
objects that were not detected in the NUV and those that were not
covered by \textit{GALEX} (at MIS-depth) we also constructed a
catalogue containing basic observational information for each GAMA
object. This includes the effective exposure time and background
level, the corresponding point source detection limit, as well as any
map flags influencing the object's detectability, for both the NUV and
FUV bands. We note that the detection threshold for the integrated
flux of UV extended sources will be higher than the given threshold
for a point source, and can in principle be calculated for any
hypothesised size and shape of the UV source using the effective
exposure time and background level quoted in this catalogue.

A useful indicator of objects potentially affected by blending is
provided in the simple match catalogue by two columns specifying the
total number of sources in the \textit{GALEX} blind catalogue within
the maximum matching radius of $4$~arcsec, and the total number of
(other) GAMA objects to which the UV source has (also) been
matched. If one or both of these numbers is $>1$, then it is possible
that blending may affect the UV fluxes assigned to the GAMA object. In
this circumstance either the advanced match or the CoG technique is to
be preferred. Conversely, both numbers being unity indicates a
one-to-one match ($57$~per~cent of cases). In these cases we consider
the \textit{GALEX} pipeline flux measurements the most accurate
(unless the galaxy is very extended, see below).

\paragraph{Advanced match photometry}
\label{galex_adv}

The advanced match photometry method addresses the issue of blending
by carefully identifying cases where multiple GAMA and \textit{GALEX}
objects are associated with each other, and then distributing the UV
flux from the \textit{GALEX} sources among the GAMA objects based on
our knowledge of the NUV and $r$-band positions and sizes of all of
the involved objects. This is a further development of the method
introduced by \citet{Robotham11a}.

The GAMA and \textit{GALEX} objects considered in the advanced match
are the same as in the previous section. In a first step, optical
shape information for each GAMA object, taken from the
single-component S{\'e}rsic model fits of \citet{Kelvin12}, is used to
define a target area within which any UV sources listed in the blind
catalogue are deemed to be at least in part associated with the GAMA
object, and therefore contributing UV flux to the GAMA object. A
circular area with a radius of $4$~arcsec (the maximum matching radius
used for the simple matching) is adopted for unresolved or compact
GAMA objects.

In a second step, for each UV source within the target area we make a
list of any other GAMA objects in the input catalogue which lie within
the NUV elliptical footprint of the \textit{GALEX} source. For
one-to-one matches, all of the UV flux of the \textit{GALEX} source is
allocated to the GAMA object (in which case the advanced match
procedure returns the same NUV and FUV fluxes for a given GAMA object
as the simple match procedure). If, however, there are more than one
\textit{GALEX} objects in the target area, or if more than one
potential optical counterpart to one or more of the \textit{GALEX}
sources in the target area is found, then the NUV and FUV fluxes of
each of the \textit{GALEX} objects are split among all optical
counterparts of that source, weighted inversely by angular distance
(using a minimum distance of $0.3$~arcsec to account for positional
uncertainties). This weighting is motivated by the expectation that
the position of a blended UV detection returned by the \textit{GALEX}
pipeline (i.e.\ by {\sc SExtractor}) is simply the flux-weighted mean
position of the individual UV emitters contributing to the blend.
Finally, the UV flux contributions to the target GAMA object from all
of the \textit{GALEX} sources in the target area are summed, to obtain
the total redistributed NUV and FUV fluxes of the object. The object
is then included in the advanced match catalogue if its total NUV flux
resulting from the redistribution has a statistical significance of
more than $2.5$$\sigma$.

We note that a GAMA object may be included in the advanced match
catalogue but not in the simple match catalogue, and vice versa. The
former happens when a GAMA object is offset from its nearest
\textit{GALEX} neighbour by more than the maximum matching radius of
$4$~arcsec, but still receives flux from one or more UV sources as a
result of the flux redistribution. This might for example happen when
the centroid of the resolved UV emission of an extended galaxy is
offset by more than the matching radius from the galaxy's $r$-band
position. The latter (more common) case happens when the redistributed
flux received by a GAMA object is less than $2.5$$\sigma$. This
commonly happens when the flux of a single UV source is shared among
multiple GAMA objects. Indeed, in general, the main effect of the flux
redistribution is to lower the UV fluxes assigned to GAMA objects.

A demonstration that this flux redistribution actually improves the
measurement of the UV flux of GAMA galaxies in a statistical sense is
shown in Fig.~\ref{nuv-r}, which shows the distribution of GAMA
objects in the plane spanned by NUV$-r$ colour and $r$-band absolute
magnitude. In the left panel we only use objects with unambiguous
(i.e.\ one-to-one) matches with \textit{GALEX} objects. This sample
provides a benchmark for the true colour-magnitude distribution. In
the other two panels we use those objects that are {\em not}
unambiguously matched to a single \textit{GALEX} object, but instead
are involved in a one-to-many, many-to-one or many-to-many match. In
the middle panel we use the NUV flux returned by the simple match
procedure, in the right panel we use that returned by the advanced
match procedure. We can see that the multiple matches, if using the
NUV flux returned by the simple match, are biased towards bluer
NUV$-r$ colours by about $0.2$~mag, due to the effect of blending
boosting the NUV fluxes. This bias is, however, not present when using
the redistributed NUV flux returned by the advanced match technique,
which recovers a very similar colour-magnitude distribution as that of
the one-to-one matches. Also apparent in the middle and right panels
is a slight shift of the distribution towards brighter absolute
magnitudes compared to the one-to-one matched sample. This arises
because luminous sources are more extended, and are therefore more
likely to have multiple matches than fainter, unresolved sources.

\paragraph{Curve-of-growth photometry}
\label{galex_cog}

Our final photometric method involves performing surface photometry on
the \textit{GALEX} images at the (optically defined) location of each
GAMA galaxy, using a CoG technique with an automated edge detection
algorithm. To this end we reprocessed all MIS-depth data using version
7.0.2 of the \textit{GALEX} pipeline, resulting in various maps for
each tile and for each band, of which we use count maps, background
maps, effective exposure maps and flag maps for the CoG analysis. In
addition, all images were visually inspected to flag reflection
artefacts from bright stars on neighbouring tiles, which escape
automatic flagging in the \textit{GALEX} pipeline.

Unlike the simple and advanced match photometry, CoG photometry is
only performed for galaxies that are spectroscopic targets, as defined
by the GAMA\,II tiling catalogue. For each target galaxy, a cutout is
made from the pipeline map with the longest exposure time. The maps
are masked over the areas covered by all known unrelated sources in
both the $r$-band (as listed in the GAMA\,II input catalogue and using
the shape and size information from the single-component S{\'e}rsic
catalogue of \citealp{Kelvin12}) and in the NUV (as listed in the
blind catalogue). In addition, all pixels marked in the flag map as
being affected by window and dichroic reflections are masked.

Radial profiles in NUV and FUV brightness are then constructed by
measuring the mean brightness of all unmasked pixels in elliptical
annuli. The ellipticity of these annuli is determined by the
convolution of the \textit{GALEX} PSF with the footprint of the galaxy
as returned by the single-S{\'e}rsic fits of \citet{Kelvin12}. An edge
detection algorithm is then employed to identify the elliptical
aperture which, on the one hand, encloses all flux from a source as
completely as possible without imposing any preconception on the
extent or shape of the radial profile of the source, while on the
other hand minimising the aperture area and hence the noise. This
algorithm, described fully by Andrae et al.\ (in preparation),
compares measurements of the brightness interior and exterior of a
hypothesised edge of the source, averaged over radial extents
optimised for the noise and structural characteristics of the
underlying background. Because the size of galaxies may be
systematically different in the NUV and FUV (e.g., extended UV disks
around galaxies often have very blue FUV$-$NUV colours;
\citealp{GildePaz05}) we determine the source's edge separately in
both bands. Once the aperture is defined, the background is determined
in the optimised region exterior to the aperture, and subtracted from
the flux inside the aperture. The integrated flux of the source is then
taken as the sum of the remaining flux inside the aperture. The
uncertainty on this flux is computed taking into account the measured
fluctuations on the background, thus incorporating the contribution of
background structure to the uncertainty.

\begin{figure}
\includegraphics[width=\columnwidth]{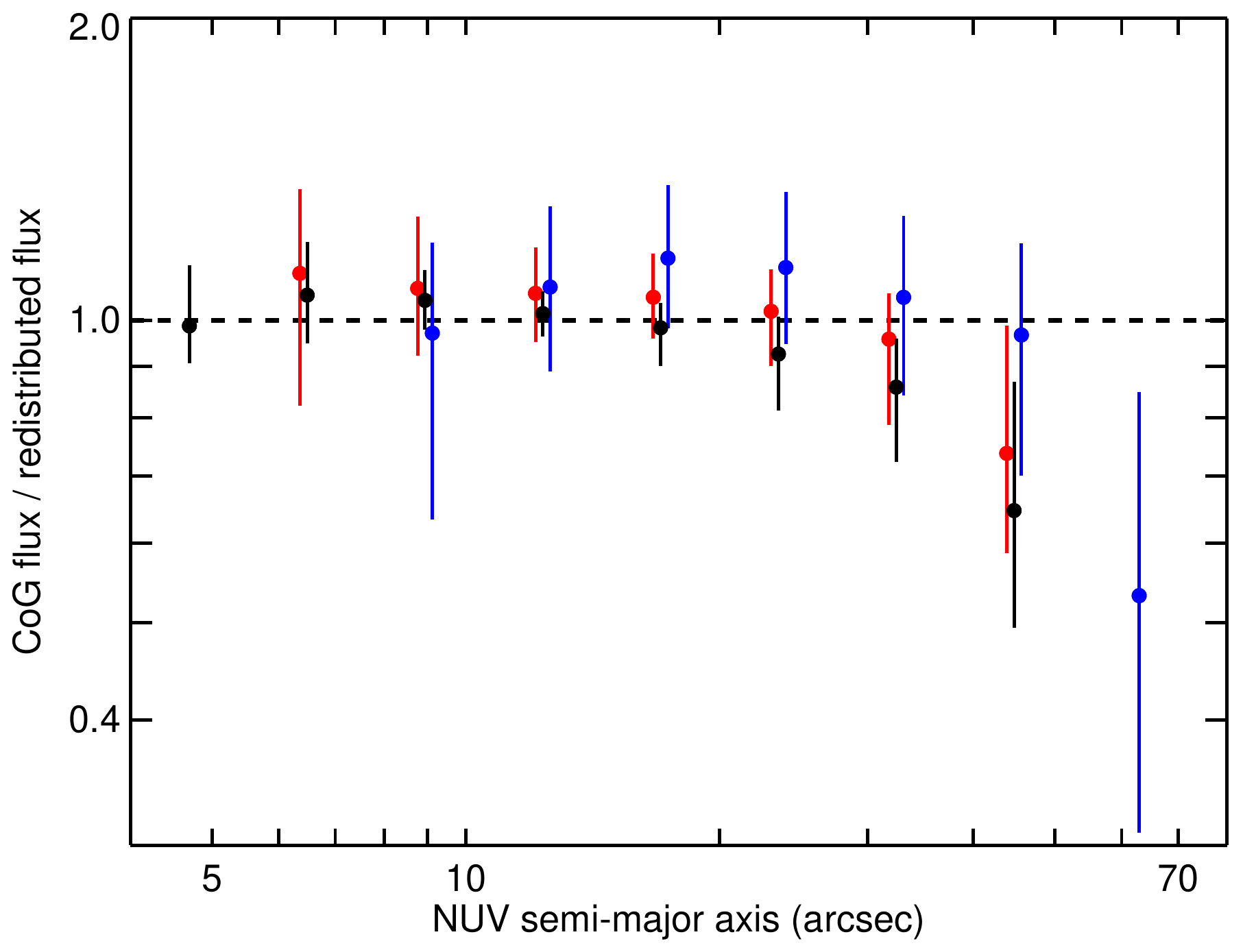}
\caption{The ratio of the NUV flux returned by the CoG method to that
  returned by the advanced match method for all GAMA galaxies detected
  by the CoG and advanced match techniques, as a function of the
  semi-major axis as measured by the \textit{GALEX} pipeline. The
  vertical errorbars indicate the $1$$\sigma$ spread in flux ratios in
  each bin. Black symbols show the ratios for galaxies with one-to-one
  matches, red symbols for galaxies with one-to-two or two-to-one
  matches, and blue symbols for galaxies involved in multiple
  matches.}
\label{cog_size}
\end{figure}

In cases where no clear edge can be detected, due to the galaxy not
being sufficiently bright, the semi-major axis of the elliptical
aperture is defined as being three times the effective radius of the
galaxy as measured in the single-S{\'e}rsic fits in $r$-band,
after convolution with the \textit{GALEX} PSF. Even though in such
cases the returned integrated fluxes can be lower than the
$2.5$$\sigma$ threshold adopted for inclusion of sources in the simple
and advanced match catalogues, or even negative, such sources are
retained in the CoG catalogue. This is due to the inherently different
approach of measuring the UV flux at the known positions of GAMA
galaxies, which will allow statistical analyses of populations of
individually non-detected galaxies (e.g.\ stacking analyses).

\paragraph{`Best' photometry}
\label{galex_best}

We have performed a number of tests of, and comparisons among, the
three different photometry methods described above, including the
insertion of artificial galaxies into the data and comparing the input
and recovered fluxes. These will be discussed in detail by Andrae et
al.\ (in preparation). In summary, we find the \textit{GALEX} pipeline
photometry to be reliable for objects that are not affected by
blending and are not too large. The CoG method, on the other hand, has
proved to be the most robust method to measure the UV fluxes of GAMA
objects that are blended with other objects in the UV, and of very
extended objects.

For blended objects the CoG method of measuring the UV flux in an
aperture whose position and shape is determined by the higher
resolution optical data, while masking out other nearby objects,
turned out to be more accurate than the advanced match method of
indiscriminately sharing the UV flux among nearby GAMA objects.

For very extended objects we also believe our CoG photometry to be the
most robust. In Fig.~\ref{cog_size} we show the ratio of CoG flux to
that returned by the advanced match method, as a function of the
\textit{GALEX} pipeline NUV semi-major axis of the the nearest
neighbour \textit{GALEX} object. We can see that this ratio
systematically drops below one for sizes larger than $20$~arcsec, even
for those objects that are not affected by blending (shown in
black). In these cases the UV flux returned by the advanced match
method is simply the \textit{GALEX} pipeline flux of the nearest
neighbour \textit{GALEX} object. Since we have tested the accuracy of
our CoG photometry even for large galaxies using simulations, we
believe the \textit{GALEX} pipeline photometry to be flawed for these
objects.

On the other hand, for smaller objects unaffected by blending we
consider the \textit{GALEX} pipeline photometry to be superior to our
CoG photometry. The reason is that the \textit{GALEX} pipeline's
procedure of fitting a simple parametric model to the source, and then
integrating over this model to obtain the total flux, results in lower
random noise than that accumulated by integrating over an aperture.
This decrease in random noise, however, comes at the expense of an
increased systematic error when the source morphology is too complex
to be adequately represented by the simple models used by the
\textit{GALEX} pipeline, which is the case both for very well-resolved
and for blended sources.

We thus define the `best' UV photometry to be that returned by the CoG
method when the NUV semi-major axis is larger than $20$~arcsec or when
the GAMA object does not have an unambiguous counterpart in the
\textit{GALEX} blind catalogue, in which cases systematic errors
dominate. In all other cases, where random errors dominate, we use the
fluxes returned by the simple match technique as the `best'
photometry.

\begin{figure}
\includegraphics[width=\columnwidth]{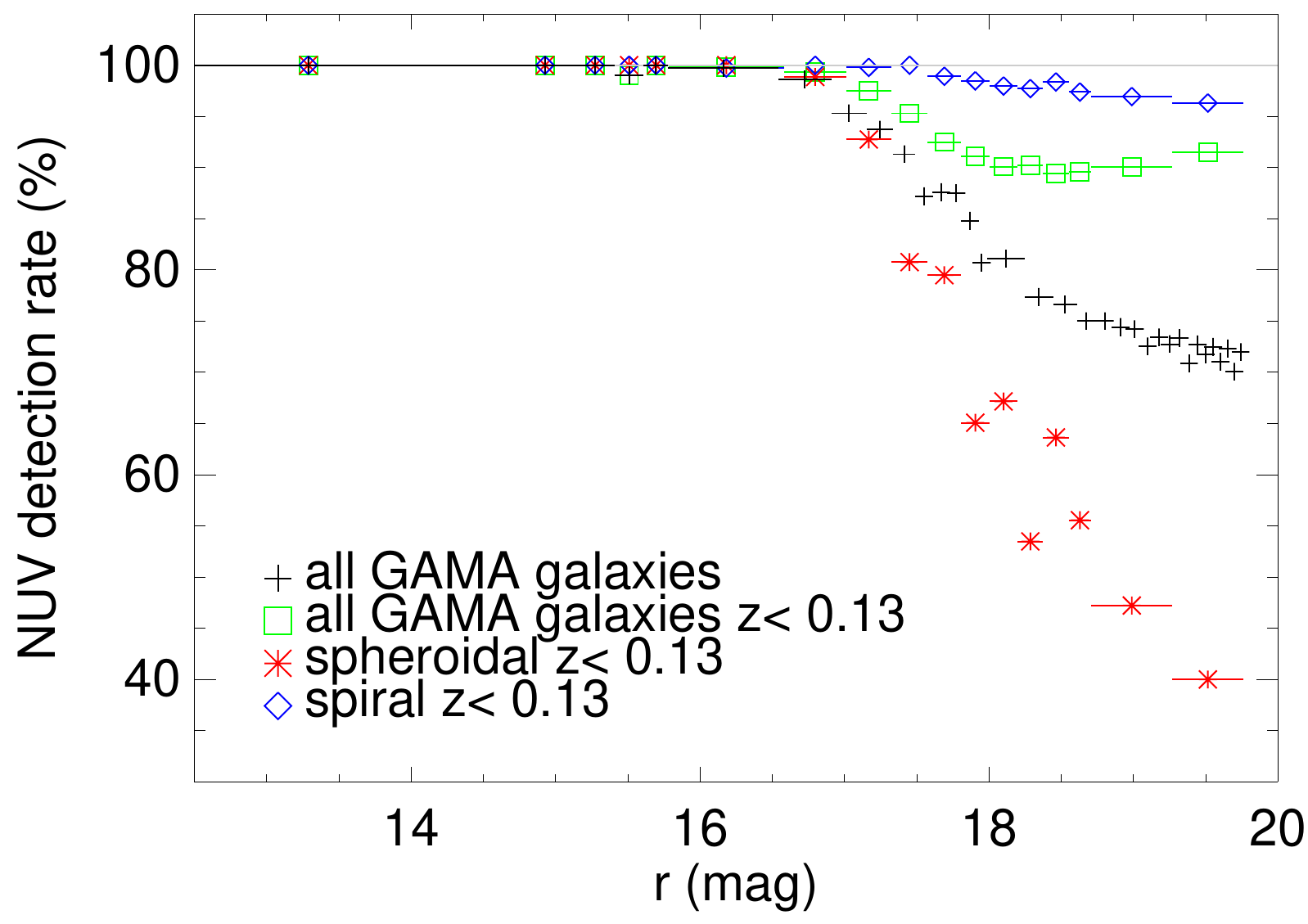}
\caption{NUV detection rate as a function of $r$-band magnitude. The
  black crosses show the detection rate for all spectroscopic
  targets. The green squares show the detection rate for galaxies with
  $z < 0.13$ where reliable morphological classification using the
  method of \citet{Grootes14} is possible. The red crosses and blue
  diamonds show the detection rates for spheroidal and spiral galaxies
  at $z < 0.13$, respectively.}
\label{nuv_detect_rate}
\end{figure}

\subsubsection{Detection statistics}

In Fig.~\ref{nuv_detect_rate} we show as black crosses the NUV
detection rate of all GAMA galaxies that are spectroscopic targets as
a function of their $r$-band magnitude. This confirms that the MIS
depth of the \textit{GALEX} data is reasonably well matched to the
depth of the GAMA\,II spectroscopic survey, providing a detection rate
of $72$~per~cent at the survey's limit of $r = 19.8$~mag.

\citet{Grootes14} showed that at least at low redshift ($z < 0.13$) it
is possible to morphologically classify galaxies using a proxy that
only involves the photometric quantities $i$-band magnitude,
S{\'e}rsic index and $r$-band effective radius. Applying this
classification to our sample, we also show in
Fig.~\ref{nuv_detect_rate} the NUV detection rates separately for
spiral and spheroidal galaxies (blue diamonds and red crosses,
respectively). We can see that at $z < 0.13$ the detection rate of
spirals stays at a level of at least $90$~per~cent for all
magnitudes. In contrast, the NUV detection rate of spheroids falls
continuously from $100$~per~cent at $r \approx 17$~mag to
$\sim$$40$~per~cent at $r = 19.8$~mag.

\section{Data Release 2}
\label{dr2}

Following the first public data release (DR1) described by
\citet{Driver11}, we now present the second public release of GAMA
data (DR2), which is available at {\tt http://www.gama-} {\tt
  survey.org/dr2/}, in this final part of the paper.

In summary, DR2 provides AAT/AAOmega spectra, redshifts and a wealth
of ancillary information for $72\,225$ objects from GAMA\,I. These
data are served by the GAMA DR2 database, which consists of a MySQL
database and a data file server. The MySQL database contains all of
the catalogues that are part of DR2, as well as the accompanying
meta-data. The file server hosts the actual data files, i.e.\ all
spectra and catalogues. Public access to the DR2 database is provided
by a web interface at the above URL.

DR2 represents a significant extension of DR1. In DR1 we released
spectra and redshifts only from the first year of observations, and
only for targets with $r < 19.0$~mag (except for a very narrow strip
in G12). In contrast, DR2 includes data from all of GAMA\,I
(i.e.\ from the first three years of observations), and extends the
limiting magnitude in one of the survey regions to $r=19.4$~mag. We
also provide additional information such as SFRs, stellar masses and
group data, which was not present in DR1.

Overall, DR2 differs significantly from DR1, not only in terms of the
data being released, but also in the way in which the data are
served. In this section we thus describe the various aspects of DR2 in
more detail.

\begin{figure*}
\includegraphics[width=\textwidth]{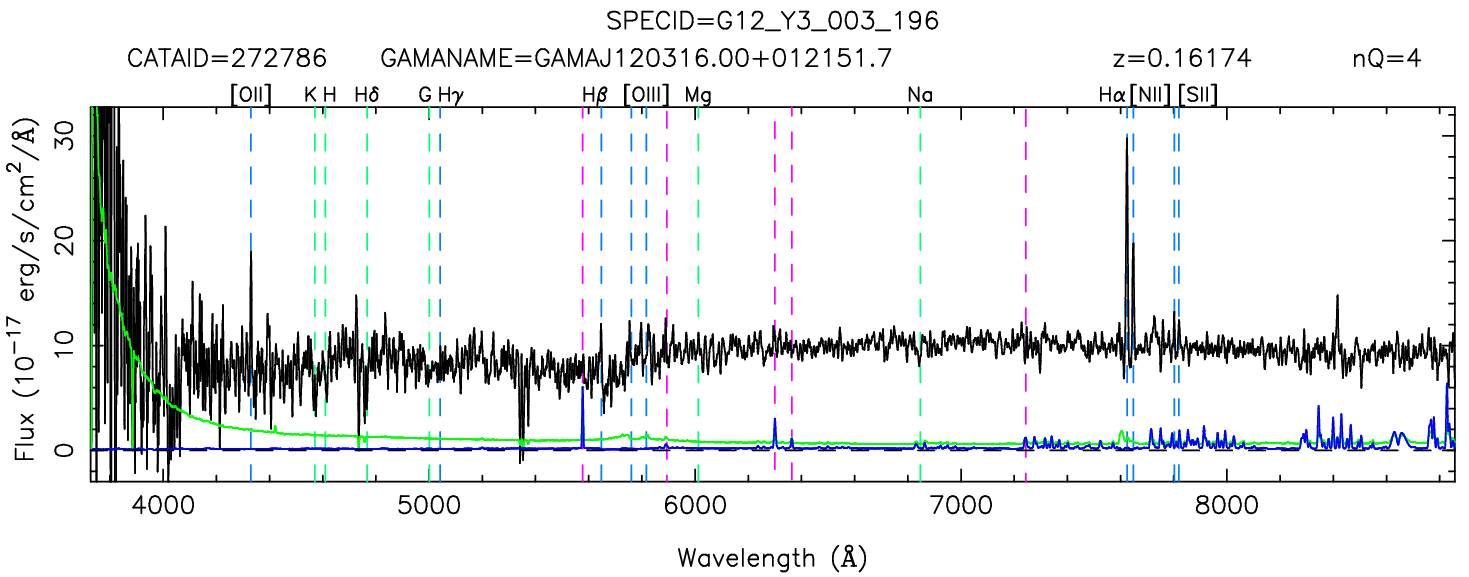}
\caption{Typical GAMA AAT spectrum. The flux-calibrated spectrum is
  shown in black, the $1\sigma$ error spectrum in green, and the
  field's mean sky spectrum (with arbitrary scaling) in blue. The
  vertical dashed lines mark the positions of common nebular emission
  (blue) and stellar absorption lines (green) at the redshift of the
  galaxy, and of strong telluric lines (purple). The spectrum was
  smoothed with a boxcar of width 5 pixels.}
\label{example_spec}
\end{figure*}

\subsection{Data description}
\label{data_description}

We begin this section by explaining the selection of the objects
included in DR2. In DR2 we are releasing data for all GAMA\,I main
survey objects with $r < 19.0$~mag in survey regions G09 and G12, and
for all objects with $r < 19.4$~mag in region G15. Refer to
Table~\ref{gamaregions} for the definition of the GAMA\,I survey
regions. Note that for G15 we are essentially releasing all GAMA\,I
data. The total number of objects included in DR2 is $72\,225$. Of
these, $70\,726$ objects have secure redshifts. The overall redshift
completeness of the DR2 sample is thus $97.9$~per~cent. Split by
survey regions the completeness is $97.7$, $98.8$ and $97.5$~per~cent
in G09, G12 and G15, respectively.

As described in Section~\ref{inputcat}, the qualifier `GAMA\,I' above
refers to the fact that the objects for DR2 were selected from the
input catalogue for the first phase of the GAMA survey ({\tt
  InputCatAv05}). DR2 only contains data for main survey targets, data
for filler targets (cf.\ Section~\ref{inputcat}) are not included. The
$r$-band selection magnitude above is the Petrosian $r$-band magnitude
from SDSS DR6 \citep{AdelmanMcCarthy08}, corrected for Galactic
extinction.

For the above objects DR2 provides all spectra obtained in GAMA\,I,
publicly available spectra from previous surveys in the GAMA\,I
regions, input catalogue and targeting information,
redshifts,\footnote{DR2 only includes {\sc runz} redshifts. {\sc
    Autoz} redshifts will be made available in the next data release.}
optical and NIR ($u$ to $K$) aperture-matched photometry derived from
SDSS and UKIDSS LAS imaging data, photometry and structural parameters
from single-component S{\'e}rsic fits in the same bands, FUV and NUV
photometry from \textit{GALEX}, $k$-corrections, stellar masses,
spectral line measurements and H$\alpha$-derived SFRs, three different
environment measures (only for G15), and last, but by no means least,
the GAMA Galaxy Group Catalogue (G$^3$C, again for G15 only).

In the following we will describe each of these datasets in
turn. Before we can continue, however, we must introduce another piece
of GAMA vocabulary. The GAMA data flow has been broken up into
individual tasks which are performed by what we refer to as Data
Management Units (DMUs). Each DMU performs a specific data reduction
or data analysis step on some input data, and as a result produces
some output, which is stored in the GAMA database. A DMU's output
(also referred to as the DMU's products) may consist of pixel data,
one or more tables, or other advanced data products, and always
includes the meta-data required to use these data in a scientific
context. The modular structure of the data flow provides a convenient
and natural structure for the database, and we will follow this
structure in the description of the DR2 data that follows.

Finally, we point out that more detailed, exhaustive descriptions are
available from the DR2 web pages as part of the meta-data accompanying
the DMU products. Appropriate references to the GAMA literature or to
previous sections of the present paper are also provided in each
section below.

\subsubsection{Spectra}
\label{spectra}

DR2 provides all $59\,345$ spectra of DR2 objects that were obtained
at the AAT as part of the GAMA\,I survey, including all duplicate
observations. These data were obtained, reduced and calibrated using
the procedures described by \citet{Robotham10}, \citet{Driver11} and
\citet{Hopkins13a}. The spectra cover the wavelength range
$3740$--$8850$~\AA\ at a resolution of $R = \lambda/\Delta \lambda
\approx 1000$ at the blue end increasing to $R\approx 1600$ at the red
end, and with a pixel size of $1.04$~\AA.

The spectra are provided as FITS files. Each FITS file contains the
fully reduced, sky-subtracted, wavelength-calibrated, telluric
absorption-corrected and flux-calibrated spectrum, the reduced
spectrum without flux calibration, the corresponding $1\sigma$ error
arrays, and the mean sky spectrum of the field from which this
spectrum was taken. We show a typical example spectrum in
Fig.~\ref{example_spec}.

DR2 also includes $19$ spectra of fibre-bright DR2 objects obtained at
the Liverpool Telescope (LT). The observing and data reduction procedures
for these spectra are described in Section~\ref{LTObs}. Again, the
fully reduced spectra are provided as FITS files. Note that these
spectra are not flux-calibrated.

Beyond these spectra obtained by the GAMA team, DR2 also provides
publicly available spectra from previous surveys covering the GAMA\,I
survey regions, as listed in Table~\ref{dr2spec}. In total, we
have obtained $30\,828$ spectra (including all duplicate observations
for completeness) from the databases of the various surveys. The FITS
files containing these spectra provided by DR2 are essentially those
of the originating surveys, except that multiple extensions (sometimes
used to store duplicate observations) were extracted to individual
files. For each spectrum we have also added a number of GAMA standard
keywords to the FITS header in order to provide some homogeneity
across all spectra. Note that only the spectra from the SDSS are
flux-calibrated.

In total we thus provide $90\,192$ spectra of the $72\,225$ unique
objects included in DR2.

\subsubsection{Input catalogues}
\label{dr2_ic}

The {\tt InputCat} DMU provides various input catalogues for the
spectroscopic survey. \citet{Baldry10} described the construction of
these catalogues in detail, and so we only provide a brief summary
here. 

{\tt InputCatA} is the master input catalogue. It was constructed from
various queries to the SDSS DR6 {\tt PhotoObj} table and contains only
information from that table. In addition, {\tt InputCatA} introduces a
unique numeric GAMA object identifier, CATAID, which is always used in
GAMA tables when referring to objects.

{\tt TilingCat} is, as the name suggests, the catalogue from which we
actually select objects for observation. As described in more detail
by \citet{Baldry10}, it is derived from {\tt InputCatA} by applying
magnitude limits, star-galaxy separation criteria, surface-brightness
limits, and our mask. It also contains information on the best
available redshift (if any) for each object, and it is this catalogue
that we use to keep track of the state of the survey. During survey
operations it is updated as soon as new redshifts are available
(whereas {\tt InputCatA} remains static). The version of this
catalogue released in DR2 is the final GAMA\,I version (i.e.\ after
the completion of all GAMA\,I observations).

Note that this table contains the {\em entire} tiling catalogue, not
just the objects for which redshifts are released in DR2. Objects
included in DR2 may be identified using the column DR2\_FLAG. Those
objects for which redshifts are not released in DR2 have their
redshift column Z set to $-9.99999$. However, the redshift quality
column NQ (see Section~\ref{final_z} for a definition) has not been
modified, so this column informs users whether a good quality redshift
for this object exists (but is not yet released).

\begin{table}
\caption{Breakdown of the origin of the spectra included in GAMA DR2
  and served by the DR2 database.}
\label{dr2spec}
\begin{tabular}{llrl}
\hline
Survey & Source / & No.\ of spectra & Reference\\
& Data release & in DR2 &\\
\hline
GAMA\,I         & AAT         & $59\,345$ & \citet{Hopkins13a}\\
                & LT          & $      19$ & Section~\ref{LTObs}\\
\hline
SDSS		& DR7         & $16\,267$ & \citet{Abazajian09}\\
2dFGRS	        & Final DR    & $11\,906$ & \citet{Colless01}\\
MGC		& Final DR    & $ 2\,154$ & \citet{Driver05}\\
6dFGS	        & Final DR    & $  248$    & \citet{Jones09}\\
2QZ		& Final DR    & $  150$    & \citet{Croom04b}\\
2SLAQ-LRG	& Final DR    & $   44$    & \citet{Cannon06}\\
2SLAQ-QSO	& Final DR    & $   43$    & \citet{Croom09}\\
WiggleZ	        & DR1	      & $   16$    & \citet{Drinkwater10}\\
\hline
Total           &             & $90\,192$ &\\
\hline
\end{tabular}
\end{table}

{\tt SpStandards} is a table of standard stars, again selected from
SDSS DR6. In each 2dF/AAOmega field that we have observed so far, we
have assigned a small number of fibres (typically $3$) to calibration
stars picked from this table. These standard star spectra have been
used to tie all of our AAOmega spectra to the SDSS spectrophotometric
calibration, at least in an average sense, as described in detail by
\citet{Hopkins13a}.

Finally, this DMU includes the table {\tt Galactic}-{\tt Extinction}, which
provides the Galactic foreground extinction in all \textit{GALEX},
MGC, SDSS and UKIDSS bands for every object in {\tt InputCatA} and
{\tt SpStandards}, using the dust maps of \citet{Schlegel98} and the
relative extinction values listed in Table~\ref{rel_extinction}.

\subsubsection{Spectra and redshift catalogues}
\label{speccat}

There are two DMUs that provide spectra and redshift catalogues: while
the {\tt ExternalSpec} DMU is only concerned with spectra and
redshifts from previous surveys, the {\tt SpecCat} DMU provides all
catalogues related to GAMA's own spectroscopic data, as well as the
final catalogues that combine all available GAMA and external data.

We already mentioned in Section~\ref{spectra} above that DR2 includes
publicly available spectra from previous surveys
(cf.\ Table~\ref{dr2spec}). These spectra are tabulated in the
catalogue {\tt ExternalSpecAll} of the {\tt ExternalSpec} DMU. This
table identifies the spectra by their unique GAMA SPECID, provides
their locations on the DR2 file server, and lists, among other
properties, their redshifts. Note that these are the redshifts
published by the originating surveys; we have not attempted to
re-measure them. We have, however, translated the various redshift
quality parameters provided by the originating surveys to our $nQ$
system (see Section~\ref{final_z}) for ease of use.

For each spectrum this table also identifies the GAMA object the
spectrum was matched to. The matching GAMA object is defined as the
object closest to the position at which the spectrum was recorded
(within a maximum of $2$~arcsec) in the catalogue resulting from the
union of {\tt TilingCat} and {\tt SpStandards} from the {\tt InputCat}
DMU (see previous section).

\begin{table}
\caption{Relative extinction values in \textit{GALEX}, MGC, SDSS and
  UKIDSS bands, as used by the table {\tt GalacticExtinction}. See
  also \citet{Schlafly11}.}
\label{rel_extinction}
\begin{tabular}{lll}
\hline
Filter & $A/E(B-V)$ & Reference\\
\hline
FUV &   $8.376$ & \citet{Wyder05}\\
NUV &   $8.741$ & \\
$B_{\rm MGC}$   & $4.23$  & \citet{Liske03b}\\
$u$   & $5.155$ & \citet{Schlegel98}\\
$g$   & $3.793$ & \\
$r$   & $2.751$ & \\
$i$   & $2.086$ & \\
$z$   & $1.479$ & \\
$Y$   & $1.211$ & WFCAM Science Archive$^a$\\
$J$   & $0.889$ & \\
$H$   & $0.578$ & \\
$K$   & $0.360$ & \\
\hline
\end{tabular}
$^a${\tt http://surveys.roe.ac.uk/wsa/}
\end{table}

Frequently, multiple spectra from the same survey are matched to the
same object (because we have included all duplicate observations). For
each spectrum in {\tt ExternalSpecAll} we thus ask (and flag the
spectrum accordingly) whether it is the one from its originating
survey that provides the most reliable redshift of its matched
object.\footnote{Note that this is not necessarily the same as the
  highest S/N spectrum.} The set of spectra thus flagged is provided
as the table {\tt ExternalSpec} for convenience. This table has all
{\em intra}-survey duplications removed, but still retains the {\em
  inter}-survey ones.

Finally, the {\tt ExternalSpec} DMU also provides the table {\tt
  ExternalzAll} which contains a small number of redshifts for DR2
objects from NED and the UZC \citep{Falco99}. The original spectra
from which these redshifts were measured are not available to GAMA,
and are hence not included in DR2.

Moving on to the {\tt SpecCat} DMU, the table {\tt AATFields} lists
all $392$ 2dF/AAOmega observations (fields) obtained at the AAT as
part of the GAMA\,I survey. Each of these observations delivered on
average $345$ spectra of galaxy targets. {\tt AATFields} provides
information pertaining to an entire field, such as its date and time
of observation, total exposure time, number of galaxy targets and
calibration stars observed, and rudimentary redshift success
statistics.

As described extensively in Section~\ref{redshifting}, all spectra
collected for GAMA at the AAT were redshifted at the telescope using
the code {\sc runz}, and many were redshifted again subsequently, in
part multiple times. The table {\tt AATRunzResults} contains the
complete redshifting results (i.e.\ essentially the {\sc runz} output)
for all GAMA AAT spectra that are part of DR2.

{\tt AATSpecAllzAll} is a table containing one line for each GAMA AAT
spectrum in DR2, summarising all of the (re-)red-shifting results
for this spectrum, as well as listing the results of the analysis to
determine the `best' redshift based on the reliabilities of the
redshifters (see Section~\ref{final_z}).

The table {\tt AATSpecAll} again contains one line for each GAMA AAT
spectrum included in DR2, giving its `best' redshift as well as
listing a number of other properties of the spectrum, including its
location on the DR2 file server. It also identifies the object that
was targeted. Note that duplicate observations of the same object are
retained in this table. As in table {\tt ExternalSpecAll}, we again
flag the spectrum that provides the most reliable redshift for a given
object.

As described in Section~\ref{LTObs}, a small number of fibre-bright
targets were not observed at the AAT but rather at the LT. These
spectra are tabulated in {\tt LTSpecAll}, along with their redshifts
and their location on the DR2 file server.

The table {\tt SpecAll} then synthesises much of the information
above. It combines tables {\tt AATSpecAll} and {\tt LTSpecAll} with
tables {\tt ExternalSpecAll} and {\tt ExternalzAll} from the {\tt
  ExternalSpec} DMU, thus providing a complete list of all spectra and
redshifts that are available for the objects included in DR2,
including GAMA spectra and those from previous spectroscopic
surveys. Note that all duplicate observations of the same
object are still retained in this table.

Finally, the table {\tt SpecObj} contains one line for each object
named as a target in {\tt SpecAll}, giving details of the spectrum
that provides the most reliable redshift (from GAMA or otherwise, thus
purging all intra and inter-survey duplications), including of course
the redshift and its quality. Note that this table contains $72\,213$
objects, which is $12$ fewer than the number of objects nominally
included in DR2. For these $12$ objects DR2 simply contains no
spectroscopic or redshift data.

We expect that table {\tt SpecObj} is the table most users will be
most interested in, along with the table {\tt TilingCat} in the {\tt
  InputCat} DMU (which also contains the best redshifts, see
Section~\ref{dr2_ic} above). We point out that all GAMA redshifts
provided in DR2 are {\sc runz} redshifts. The {\sc Autoz} redshifts
will be included in the next data release.

\begin{figure*}
\includegraphics[width=0.49\textwidth]{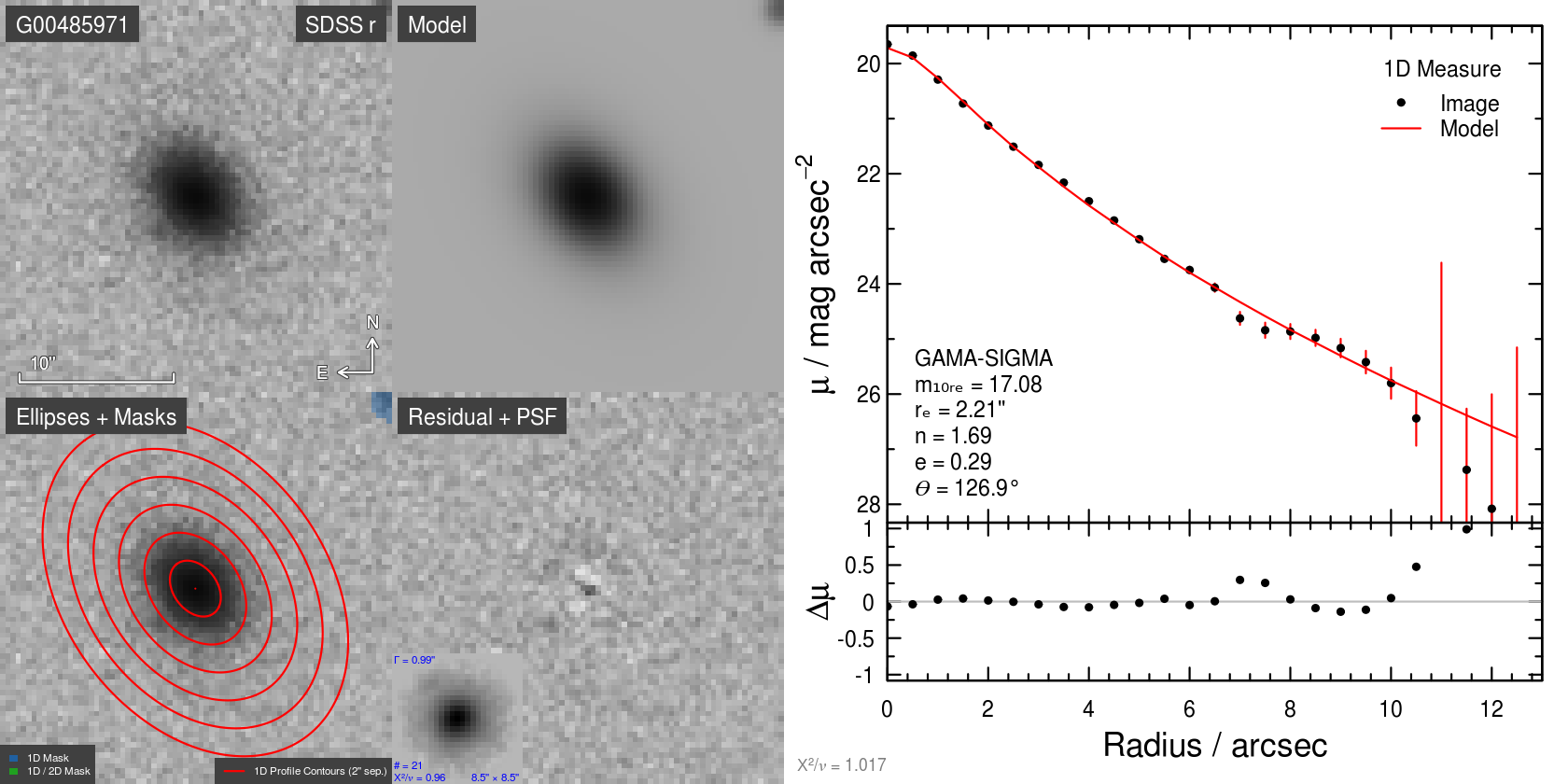}
\includegraphics[width=0.49\textwidth]{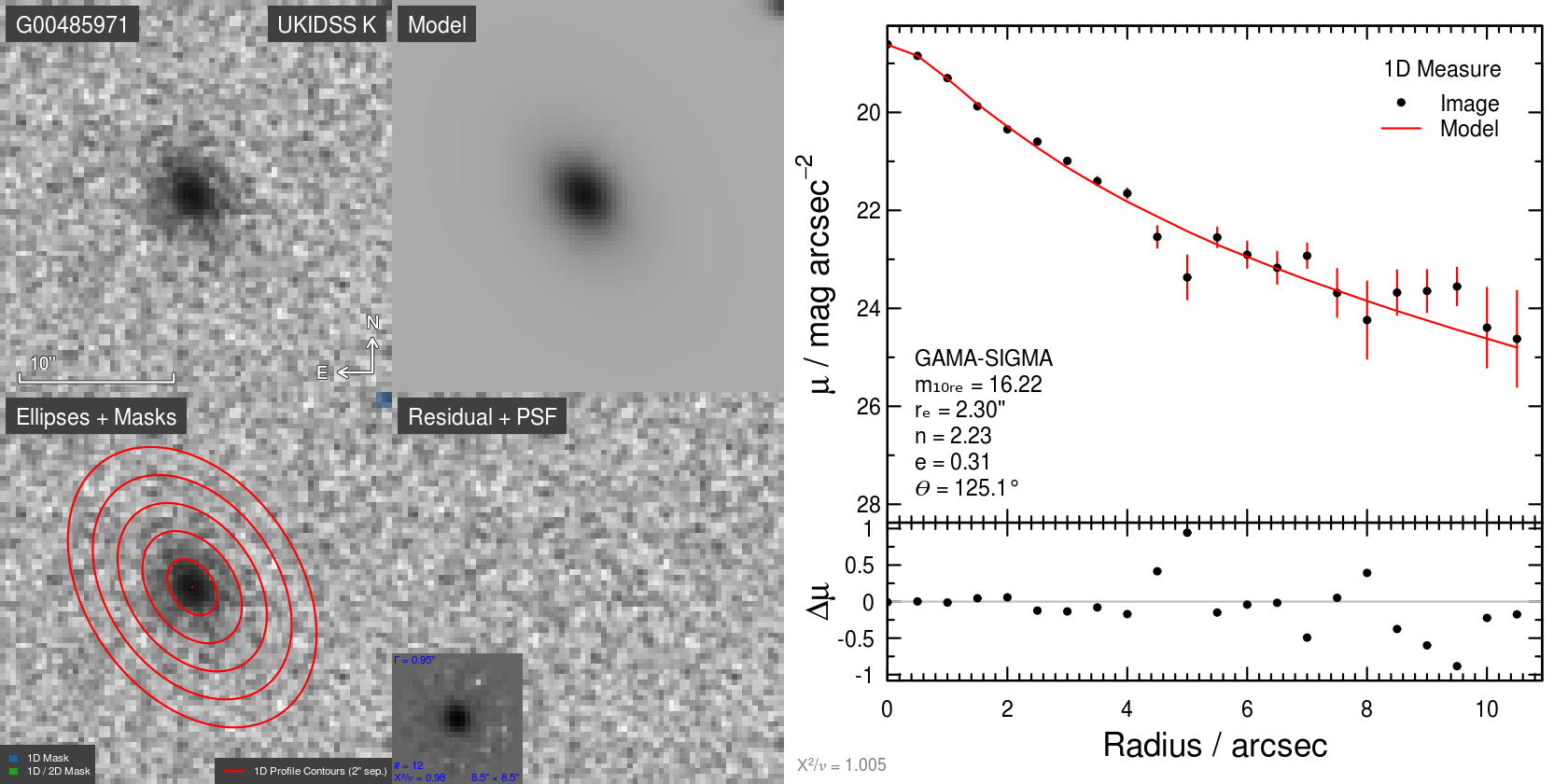}
\vspace{1mm}\\
\includegraphics[width=0.49\textwidth]{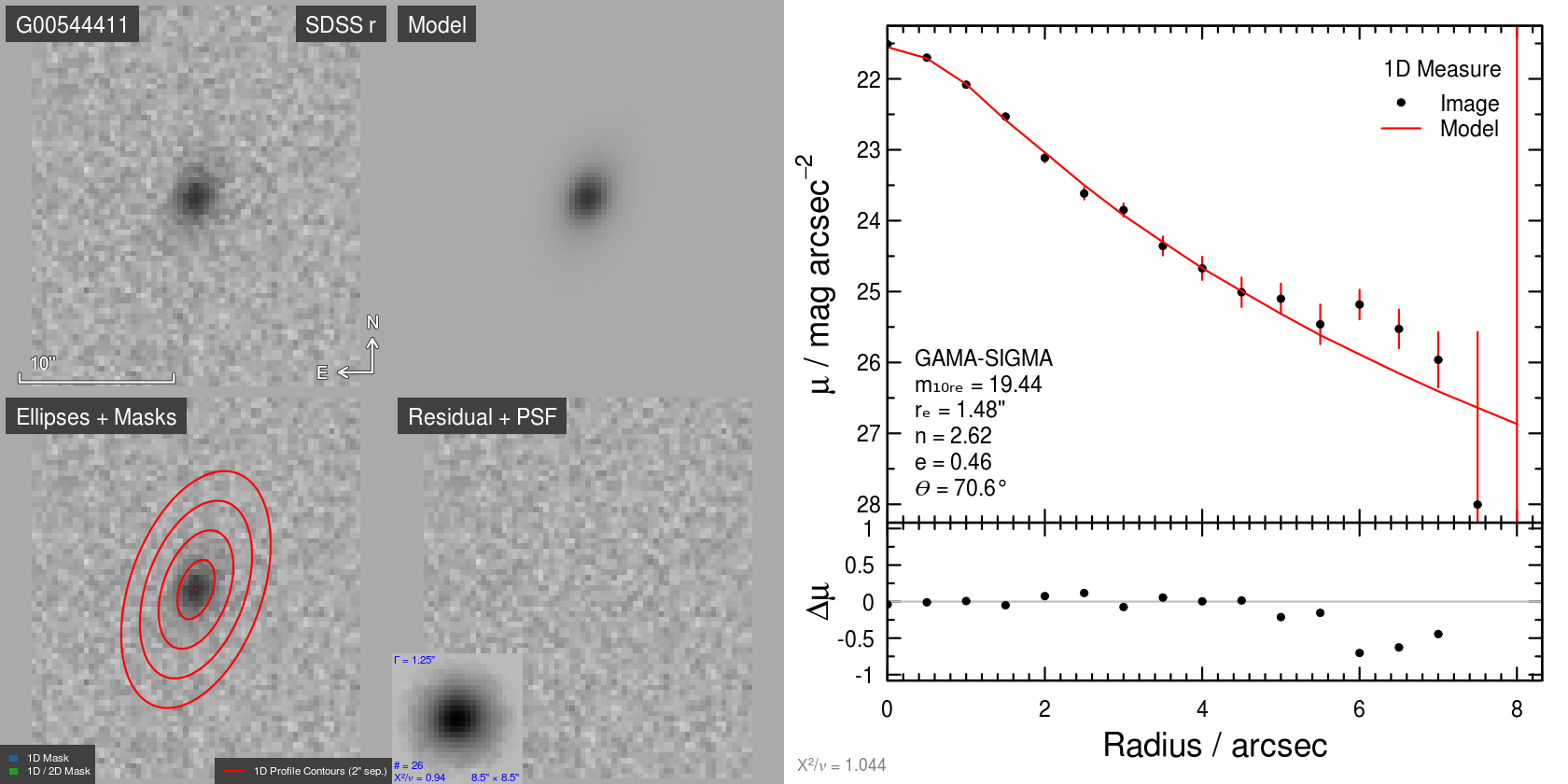}
\includegraphics[width=0.49\textwidth]{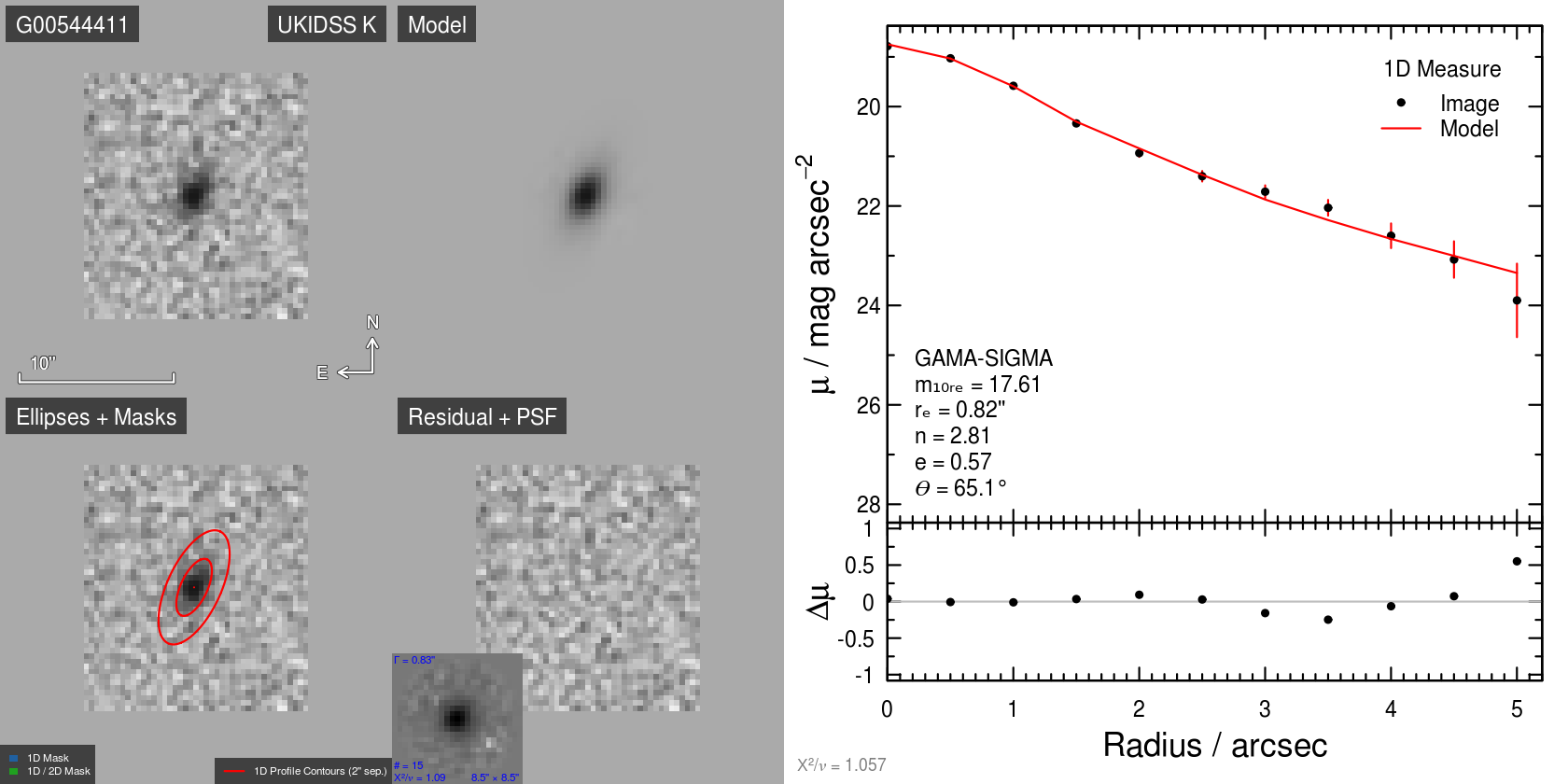}
\caption{Single-component S{\'e}rsic fits to a bright ($r_{\rm Petro}
  = 17$~mag, top row) and a faint ($r_{\rm Petro} = 19.4$~mag, bottom
  row) example galaxy in the $r$ (left) and $K$ bands (right). The
  sub-panels of the image panels show (from top left to bottom right)
  the data, the final S{\'e}rsic model, the detailed isophotes, and
  the residual image, respectively. The insets show the PSFs. The
  other panels show the corresponding azimuthally averaged surface
  brightness profiles of the galaxies (black dots with red errorbars)
  along with the profiles of the final models (continuous red lines)
  as a function of semi-major axis. The parameters of the models are
  also given. The residuals of the fit are shown below these panels.}
\label{sersic}
\end{figure*}

\subsubsection{Local flow correction of redshifts}
\label{local_flow}

The {\tt LocalFlowCorrection} DMU transforms our redshifts to various
reference frames, and provides redshifts and distance moduli corrected
for the local flow. Specifically, the heliocentric redshifts provided
by the {\tt SpecAll} table in the {\tt SpecCat} DMU (with $z >
-0.01$ and $nQ \ge 2$) are transformed to the reference frame of the
Cosmic Microwave Background (CMB) using the dipole of
\citet{Lineweaver96}, and to that of the Local Group using the
prescription of \citet{Courteau99}. The local flow correction uses the
flow model of \citet{Tonry00}. The details of these transformations
are described in section~2.3 of \citet{Baldry12}.

\subsubsection{Aperture-matched optical and NIR photometry}
\label{apmatchedphotom}

The {\tt ApMatchedPhotom} DMU provides Kron and Petrosian
aperture-matched photometry covering the $ugrizYJHK$ bands for all DR2
objects. As described in detail in Section~\ref{onirphot}, this
photometry was derived from processed SDSS and UKIDSS LAS imaging data
using {\sc SExtractor}. The original images were renormalised to a
common zero-point of $30$~mag, convolved with Gaussians to a common
PSF, resampled to a common grid with a pixel size of $0.339$~arcsec,
and then combined into very large mosaics, one for each of the above
bands and for each of the three GAMA\,I survey regions. {\sc
  SExtractor} was then run eight times in dual-image mode on small
regions of these mosaics corresponding to the positions of the objects
in {\tt TilingCat}. Each time the $r$-band image was used as the
detection image while the image in one of the other bands was used as
the measurement image, thus ensuring identical, $r$-defined apertures
for the flux measurements in all bands. 

Table {\tt ApMatchedCat} provides the above photometry (along with
various other {\sc SExtractor} outputs) for all DR2 objects. We also
release the $27$ mosaics from which the photometry was derived.

We point out that we have recently discovered, from comparisons with
VIKING and 2MASS data, an apparent zero-point offset in our photometry
derived from the UKIDSS LAS data. The cause of this offset is at
present not fully understood. This issue will be described in more
detail in a future paper presenting the GAMA\,II panchromatic
photometry (Driver et al., in preparation). In the meantime, users may
wish to consider applying the following zero-point offsets to bring
the photometry presented in table {\tt ApMatchedCat} into agreement
with VIKING: $m_{\rm corr} = m_{\rm DR2} - (0.13, 0.12, 0.07,
0.11)$ for $Y$, $J$, $H$, $K$, respectively.

\subsubsection{Optical and NIR S{\'e}rsic photometry}
\label{sersicphotom}

The {\tt SersicPhotometry} DMU provides the results of fitting a
single-component S{\'e}rsic model \citep{Sersic68,Graham05} to the
two-dimensional surface brightness distribution of every GAMA DR2
object in each of the bands $ugrizYJHK$ independently. This is
achieved by using the code Structural Investigation of Galaxies via
Model Analysis ({\sc sigma} v0.9-0) on processed SDSS and UKIDSS LAS
imaging data. {\sc sigma} is a wrapper around {\sc SExtractor}
\citep{Bertin96}, {\sc PSFEx} \citep{Bertin11} and {\sc galfit} v3
\citep{Peng10}. The code, the fitting procedure and the results are
described in detail by \citet{Kelvin12}, and so we only provide a
brief summary here.

For a given input galaxy and band {\sc sigma} proceeds as follows:
(1)~An image of appropriate size is cut out from the appropriate
mosaic (see Section~\ref{onirphot}; here we use the mosaics that
were constructed from the renormalised images at their original
resolution); (2)~{\sc SExtractor} is run over the image using
parameters optimised for the detection of unresolved sources;
(3)~detected objects originating from the same original imaging data as
the target galaxy are fed into {\sc PSFEx} in order to determine the
PSF at the location of the target galaxy; (4)~{\sc SExtractor} is
re-run over the image, this time with parameters optimised for the
detection of extended sources; (5)~{\sc galfit} is used to fit a
single-component S{\'e}rsic model to the target galaxy; neighbouring
objects are either included in the fit or masked, as appropriate, and
the initial values of the various fitting parameters are based on the
{\sc SExtractor} output; (6)~several sanity checks are conducted to
assess whether a catastrophic error has occurred, and, if necessary,
the object is re-fit with alternative constraints (e.g.\ additional
background smoothing, different masking vs.\ modelling choices);
(8)~all output information from the entire process is collated before
moving on to the next band or object.

The above process results in the table {\tt SersicCatAll} which
comprises no fewer than $531$ columns. In addition to {\sc galfit}
output for each band, this includes all of the {\sc PSFEx} and {\sc
  SExtractor} output for completeness. For ease of use, we thus also
provide the table {\tt SersicCat}, which only contains the most useful
subset of these columns.

In addition to these two tables we also make available the image
cutouts and PSF images used as inputs to the modelling process (the
location of a galaxy's data on the file server is given in {\tt
  SersicCat}), as well as the full original-resolution
mosaics. Finally, for each galaxy and band we provide convenient
summary plots showing the result of the fit. We show four examples of
these plots in Fig.~\ref{sersic}, for a bright (top) and faint
(bottom) galaxy, in the $r$ (left) and $K$ bands (right).

One frequent use of modelling the surface brightness distribution of a
galaxy is to obtain an estimate of its {\em total} flux (as opposed to
the flux measured in an aperture) by integrating the model to
infinity. However, given the varied behaviour of the surface
brightness profiles of especially late-type spiral galaxies at large
radii, which frequently show both downturns and upturns
\citep[e.g.][]{Pohlen06}, it is not clear that integration to infinity
is justified. On the other hand, it is not {\em a priori} clear where
to truncate the integration either. We refer the reader to section
4.3.3 of \citet{Kelvin12} for a discussion of this issue, and simply
point out here that {\tt SersicCat} provides S{\'e}rsic magnitudes
integrated both to infinity and to $10$ effective radii, and that we
recommend using the latter.

\subsubsection{\textit{GALEX} photometry}
\label{dr2_galex}

The {\tt GalexPhotometry} DMU provides \textit{GALEX} FUV and NUV
photometry for all DR2 objects that were detected by \textit{GALEX}.
The data, photometric procedures and matching were detailed in
Section~\ref{galexphot}, and so here we only describe the structure of
the {\tt GalexPhotometry} DMU.

The table {\tt GalexPhot} provides \textit{GALEX} NUV and FUV
photometry in the GAMA\,I survey regions. The data in this table were
derived using the \textit{GALEX} {\sc SExtractor}-based pipeline
without any reference to any GAMA data (see
Section~\ref{galex_blind}). In other words, these are the `blind'
\textit{GALEX} detections in the GAMA\,I survey regions. The table was
restricted to primary sources with S/N $\ge 2.5$ in the NUV band.

The table {\tt GalexSimpleMatch} contains the result of a simple
nearest neighbour match (see Section~\ref{galex_simple}) between {\tt
  GalexPhot} and the table {\tt InputCatA} from the {\tt InputCat} DMU
(see Section~\ref{dr2_ic} above), listing only the objects included in
DR2. GAMA DR2 objects without a \textit{GALEX} nearest neighbour
within a distance of $4$~arcsec are considered unmatched and are not
included in this table.

The table {\tt GalexAdvancedMatch} contains the result of an advanced
match between {\tt GalexPhot} and {\tt InputCatA}, again only listing
objects included in DR2. The advanced matching procedure attempts to
reconstruct the true UV flux of a given GAMA object in cases where
multiple GAMA and \textit{GALEX} objects are associated with each
other (see Section~\ref{galex_adv}). GAMA DR2 objects without a
\textit{GALEX} match are not included in this catalogue.

In addition to the \textit{GALEX}-pipeline-generated photometry of
`blind' \textit{GALEX} detections presented in table {\tt GalexPhot},
the table {\tt GalexCoGPhot} provides NUV and FUV photometric
measurements of all GAMA DR2 objects at their {\em a priori} known
optical positions using a curve-of-growth method (see
Section~\ref{galex_cog}). We deem this photometry to be superior to
that generated by the \textit{GALEX} pipeline in some circumstances.

The table {\tt GalexMain} is, as the name suggests, the main catalogue
of this DMU. It should cover the needs of most users. It duplicates
the most important information from the other tables in this DMU, and
provides estimates of the `best' NUV and FUV fluxes for all GAMA DR2
objects that were detected by \textit{GALEX}, i.e.\ an appropriate
choice is made between the \textit{GALEX} pipeline photometry and the
curve-of-growth photometry (see Section~\ref{galex_best}).

Finally, the table {\tt GalexObsInfo} provides basic \textit{GALEX}
observational information, i.e.\ exposure times, background levels and
\textit{GALEX} pipeline detection limits, for all GAMA DR2
objects. Note that this table includes all DR2 objects that are
currently not covered by \textit{GALEX} data (the rows for these
objects are `empty'). Including these objects here enables users to
discriminate between objects that were covered by \textit{GALEX} but
not detected, and those that were not covered by \textit{GALEX}.

\subsubsection{k-corrections}

The {\tt kCorrections} DMU provides $k$-corrections in the
\textit{GALEX}, SDSS and UKIDSS bands for all DR2 objects with $nQ \ge
2$. The $k$-corrections were calculated with {\sc Kcorrect} v4\_2
\citep{Blanton07} using SDSS DR6 model magnitudes and the local
flow-corrected redshifts provided by the {\tt LocalFlowCorrection} DMU
(see Section~\ref{local_flow} above). Note that, strictly speaking,
geocentric redshifts should be used to calculate $k$-corrections, but
here we have used the flow-corrected redshifts for consistency with
calculations of the maximum distance at which a given survey object
would still be included in the survey.

We provide $k$-corrections to both redshift $0$ (table {\tt
  kcorr\_z00}) and to redshift $0.1$ (table {\tt kcorr\_z01}). These
tables also include the coefficients of a polynomial fit to the
$k$-corrections in each band, as detailed by \citet{Loveday12}.

\subsubsection{Stellar masses}

The {\tt StellarMasses} DMU provides stellar masses, restframe
photometry, and other ancillary stellar population parameters from
stellar population fits to $ugriz$ spectral energy distributions
(SEDs) for all galaxies with $0 < z < 0.65$ and $nQ \ge 2$ from the
GAMA DR2 sample. The details of the derivation of the stellar masses
were described by \citet{Taylor11}, and so we only provide a brief
summary here.

The data provided by the {\tt StellarMasses} table have been derived
through stellar population synthesis (SPS) modelling of broadband
optical ($ugriz$) photometry. The modelling is done using the
\citet{Bruzual03} stellar evolution models, assuming a
\citet{Chabrier03} stellar initial mass function and the
\citet{Calzetti00} dust curve. The SPS models used in the fitting are
defined by four parameters: e-folding time for the (exponentially
declining) star formation history, time since formation (i.e.\ age),
stellar metallicity, and dust attenuation (see section 3.1 of
\citealp{Taylor11}). The SPS grid spans the range $0 < z < 0.65$;
objects with $z > 0.65$ have not been fit.

For each galaxy {\tt StellarMasses} contains the values of various
stellar population parameters that have been inferred from the SPS
fits in a Bayesian way. These include the luminosity-weighted mean
age, metallicity, and total mass of the stars, as well as restframe
photometry and colours. For the restframe luminosities we provide both
intrinsic and observed values (i.e.\ before and after internal dust
attenuation).

Note that the NIR photometry available for GAMA galaxies have not been
used at this stage, for the simple reason that the models do not
provide a good description of the full optical-to-NIR SED shapes
(section 4 of \citealp{Taylor11}).

Note further that the results contained in {\tt StellarMasses} were
derived from the aperture-matched (i.e.\ {\tt SExtractor} AUTO)
photometry provided by the {\tt ApMatchedPhotom} DMU (see
Section~\ref{apmatchedphotom} above). An aperture correction is
therefore required for integrated quantities such as stellar mass or
luminosity in order to account for flux/mass that falls beyond the
finite AUTO aperture used for the SEDs. For this purpose we provide
the quantity FLUXSCALE, which is the ratio between the $r$-band
aperture flux and the total S{\'e}rsic flux integrated to $10$
effective radii (taken from the {\tt SersicPhotometry} DMU, see
Section~\ref{sersicphotom} above). This correction has {\em not} been
applied to the values in {\tt StellarMasses}. This step is instead
left to the user.

\subsubsection{Spectral line measurements and star formation rates}

The {\tt SpecLineSFR} DMU provides emission and absorption line
measurements for all GAMA DR2 AAT spectra, as well as derived physical
properties, including the SFR, for all DR2 AAT spectra and for all
SDSS spectra of DR2 objects.

This DMU provides four catalogues. The table {\tt SpecLines} provides
emission and absorption line measurements for all GAMA DR2 AAT spectra
with a redshift measurement, i.e.\ all spectra with $nQ \ge 2$ listed
in {\tt AATSpecAll} (see Section~\ref{speccat} above). As described in
more detail by \citet{Hopkins13a}, common emission lines were fit with
single Gaussians, assuming a common redshift and a common line width
for adjacent line groups, while simultaneously fitting the local
continuum.

Table {\tt EmLinesPhysGAMA} translates these raw measurements to
physical properties, as detailed by \citet{Gunawardhana13}. In
particular, this table provides Balmer decrements, H$\alpha$
luminosities, H$\alpha$-derived SFRs, and emission line
classifications for all spectra in {\tt SpecLines} with any measured
H$\alpha$ emission and $z > 0.001$. Note that the H$\alpha$ selection
effectively limits this catalogue to $z \la 0.36$. The H$\alpha$
luminosities (and hence the SFRs) are corrected for stellar
absorption, dust obscuration and aperture effects
\citep{Gunawardhana13}.

Similarly, table {\tt EmLinesPhysSDSS} provides the same quantities for
all SDSS spectra of DR2 objects (see Section~\ref{spectra} above) with
any measured H$\alpha$ emission and $z > 0.001$. The values in this
table were derived from the line measurements provided by the MPA/JHU
SDSS line database,\footnote{\tt
  http://www.mpa-garching.mpg.de/SDSS/DR7/} which were originally
performed by \citet{Tremonti04} and \citet{Brinchmann04}.

Finally, table {\tt EmLinesPhys} is this DMU's main catalogue, which
we expect to cover the needs of most users. It combines tables {\tt
  EmLinesPhysGAMA} and {\tt EmLinesPhysSDSS} to provide (where
available) Balmer decrements, H$\alpha$ luminosities,
H$\alpha$-derived SFRs, and emission line classifications for all DR2
galaxies with a redshift measurement (i.e.\ $nQ \ge 2$ and $z >
-0.01$, the same selection as that of the {\tt LocalFlowCorrection}
DMU).

An important limitation of the current version of this DMU is the fact
that it only covers the GAMA AAT and SDSS spectra. The GAMA observing
campaign on the AAT did not systematically include objects that had
previously already been observed by other surveys (see
Sections~\ref{spectra} and \ref{speccat} above). Although the spectra
from these other surveys are available in the GAMA DR2 database, no
spectral line measurements for these spectra are included in this
DMU. For the SDSS spectra such measurements are provided to the public
by the MPA/JHU database, and we make use of these measurements in {\tt
  EmLinesPhysSDSS}. However, the spectra from the other (non-SDSS)
surveys could not be used in this DMU because they are not
flux-calibrated. Hence, the selection functions of these other surveys
will be imprinted on this DMU. This needs to be corrected for when
using table {\tt EmLinesPhys}, using e.g.\ the method of
\citet{Gunawardhana13}.

\subsubsection{Environment measures}

The {\tt EnvironmentMeasures} DMU provides several different metrics
of the local environment of GAMA DR2 galaxies: a surface density, the
number of galaxies within a cylinder, and the density of galaxies
within an adaptive Gaussian ellipsoid. Note that this release only
covers the G15 survey region, because only in this region are we
releasing redshifts down to the GAMA\,I survey limit of $r <
19.4$~mag.

All three environment measurements are performed on a density-defining
pseudo-volume-limited population of galaxies. This population is
defined as all galaxies with $M_r(z_{\rm ref}=0, Q_e=0.78) <
-20.4$~mag, where $Q_e$ defines the expected evolution of the absolute
Petrosian magnitude $M_r$ as a function of redshift, and is taken from
\citet{Loveday12}. Given the depth of the GAMA\,I survey ($r <
19.4$~mag), the above absolute magnitude limit implies a redshift
(i.e.\ volume) limit of $z=0.18333$. However, in order to account for
the upper edge of the velocity range employed when searching for
nearby galaxies (see below), the environment measurements are only
provided for galaxies out to $z=0.18$. The exact sample included in
this DMU is: all GAMA DR2 galaxies in G15 with redshift quality $nQ
\ge 3$ (i.e.\ reliable redshifts) and within the redshift limits of
$0.002 < z \le 0.18$, where $z$ is the local flow-corrected redshift
provided by the {\tt LocalFlowCorrection} DMU (see
Section~\ref{local_flow} above). All three environment measures are
corrected for redshift incompleteness where necessary.

The first environment measure provided by the table {\tt
  EnvironmentMeasures} is the surface density 
\begin{equation}
\Sigma_5 = \frac{5}{\pi \, d_5^2}
\end{equation}
at the position of a given galaxy. $d_5$ is the distance (in Mpc) in
the plane of the sky from the galaxy in question to its fifth nearest
neighbour among that part of the density-defining population that lies
within $\pm 1000$\kms\ of the redshift of the galaxy \citep{Brough13}.

The second measure, $N_{\rm cyl}$, is the number of (other) galaxies
from the density-defining population within a cylinder centred on the
galaxy in question and of co-moving radius $1$~Mpc and thickness
$\pm 1000$\kms.

Finally, the third measure is the density of galaxies from the
density-defining population in an adaptive Gaussian ellipsoid defined
by
\begin{equation}
\left(\frac{r_a}{3 \sigma}\right)^2 + \left(\frac{r_z}{3 c_z \sigma}\right)^2 
\le 1,
\end{equation}
where $r_a$ and $r_z$ are the distances from the centre in the plane
of the sky and along the line-of-sight in co-moving Mpc, respectively,
and $\sigma = 2$~Mpc. The adaptive scaling factor, $c_z = 1 + 0.2\,n$,
where $n$ is the number of galaxies from the density-defining
population within $2$~Mpc, is used to scale the value of $\sigma$
along the redshift axis by up to a factor of $3$ for the highest
density environments to compensate for the `finger-of-God' effect
\citep{Schawinski07,Thomas10}.

\subsubsection{Group catalogue}
\label{g3c}

The {\tt GroupFinding} DMU provides the GAMA Galaxy Group Catalogue
(G$^3$C), which was first introduced by \citet{Robotham11b}. The GAMA
spectroscopic survey was specifically designed to enable group science
\citep{Robotham10}, and the G$^3$C is hence one of the key data
products of the survey. In the present release the G$^3$C is
restricted to the G15 survey region, as this is the only region for
which DR2 includes data down to the GAMA\,I survey limit ($r <
19.4$~mag).

The G$^3$C is constructed using a friends-of-friends (FoF)
algorithm. The parameters of this algorithm were determined using a
set of GAMA-style mock galaxy catalogues [constructed from the
  Millennium dark matter simulation \citep{Springel05b} and the
  GALFORM semi-analytical model of galaxy formation \citep{Bower06}]
such that the medians of the most important properties of the groups
recovered by the FoF algorithm from the mock catalogues are unbiased
with respect to the `true' groups in the mocks (which are defined as
groups of galaxies inhabiting the same dark matter halo).

The number of groups included in this release is $4242$, of which
$466$ have five or more members. The multiplicity, velocity dispersion
and size distributions of these groups are quite similar to those
derived from the mock catalogues, except that we find fewer
high-multiplicity groups in the real data than in the mocks. The
details of the FoF algorithm, its application to the mock catalogues
and the real data, and the resulting group catalogue are described
extensively by \citet{Robotham11b}.\footnote{Despite the difference in
  the version labels used by \citet{Robotham11b} (v1) and in DR2 (v05),
  the version released here is in fact identical to the one described
  by \citeauthor{Robotham11b}, except for its restriction to G15.} Here we
only describe the structure of the {\tt GroupFinding} DMU's data
products.

Table {\tt G3CGal} contains the sample of galaxies on which the FoF
grouping algorithm was run. This sample was selected as all main
survey galaxies in the G15 survey region ($r < 19.4$~mag) with $nQ \ge
3$ and $0.01 < z < 0.5$. The purpose of the redshift limits is to
avoid luminosity function and distance uncertainties at very low
redshift.  For those galaxies that were identified as being a member
of a group the table also contains a reference to the appropriate
group.

\begin{figure}
\includegraphics[width=\columnwidth]{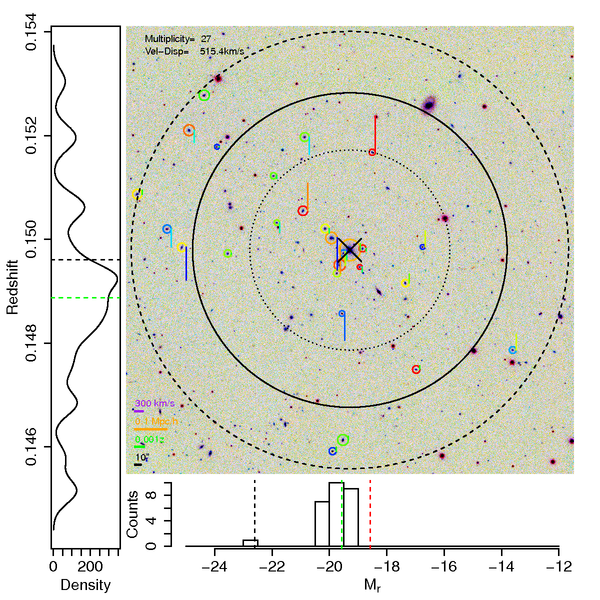}
\caption{Example group image. The background image is a $urK$
  composite. Group members are marked by circles. The size of a circle
  scales with the galaxy's $r$-band flux, while its colour reflects
  the galaxy's $u-r$ colour. A galaxy redshifted with respect to the
  group median redshift has a red upwards pointing line, the length of
  which scales with the velocity difference, while for a blueshifted
  one the line is blue and points downwards. The rings represent the
  $50$, $68$ and $100$ percentiles of the radial galaxy distributions
  relative to the iterative group centre. The velocity probability
  density function smoothed with a Gaussian kernel of width
  $50$\kms\ (the typical GAMA velocity error) is shown on the left,
  where the group median is shown with a green dashed line and the BGG
  with a black dashed line. The bottom panel presents the absolute
  $r$-band magnitude distribution of the group, with the effective
  GAMA survey limit shown with a red dashed line, the group median
  absolute magnitude with a green line and the BGG absolute magnitude
  with a black line.}
\label{group}
\end{figure}

Table {\tt G3CFoFGroup} lists a large number of properties of the
groups that were identified by running the grouping algorithm on {\tt
  G3CGal}. These include the group's multiplicity, position, redshift,
size, velocity dispersion, estimates of its total $r$-band luminosity
and halo mass, and identification of its Brightest Group Galaxy (BGG),
among others. For each group we also provide a summary plot, an
example of which is shown in Fig.~\ref{group}.

Table {\tt G3CLink} provides all of the galaxy-galaxy links found when
running the FoF algorithm. This table is useful for users who wish to
know which galaxies are linked the most with other galaxies within a
group, or to identify the most tenuously associated galaxies within a
group.

Table {\tt G3CGalsInPair} is a list of all galaxies that are paired
with another galaxy within a projected physical separation of $50$~kpc
and a velocity separation of $1000$\kms. Note that this is a list of
paired galaxies, not of galaxy pairs. The number of pairs is hence
half the number of galaxies in this table.

In addition to the above we also make available the equivalent tables
derived from the set of mock galaxy catalogues already mentioned
above. An additional table provides the properties of the `true'
groups in the mocks. Nine mock catalogues were created in
all. However, each of these covers not only the G15 region as is the
case for the real data, but represents a complete analogue of the full
GAMA\,I survey, i.e.\ of all three survey regions. Since the three
regions have the same size and shape this means that the mock
catalogues provide a total of $9 \times 3 = 27$ comparison
volumes. The tables derived from the mock catalogues contain the
results from all $27$ volumes.

\subsection{Data access}

Public access to all of the data described above, as well as to the
meta-data accompanying these, is provided by means of a MySQL
database, a file server, and a set of web pages which act as an
interface. These are available at {\tt http://www.gama-survey.}
{\tt org/dr2/}.

\subsubsection{MySQL database}

All of the tabular data described in Sections~\ref{dr2_ic}--\ref{g3c}
were ingested into a MySQL database. The contents of this database are
most conveniently explored using the schema browser we provide for
this precise purpose. The schema browser affords an overview of the
tables available for query (structured by DMUs), and provides access
to all of the meta-data provided by the DMUs, including DMU
descriptions, individual table descriptions, and the information
describing individual columns. These meta-data are required to be
complete and detailed enough to enable the use of the actual data in a
scientific context.

Having used the schema browser to identify the tables and columns that
contain the data of interest, a user may submit an appropriate
free-form SQL query. We provide a set of example queries for those
unfamiliar with MySQL. Alternatively, we offer an SQL Query Builder
which allows users to construct SQL queries largely by point and
click. This is a very powerful tool that is extremely helpful when
constructing complex queries across multiple tables, and we encourage
DR2 users to make use of it.

Query results may be returned in a variety of formats, including FITS
binary tables. The query results page also lists the first $100$ rows
of the query result, and provides links to upload individual or all
objects to the Single Object Viewer (see next section) or the SDSS
Image List tool, or to download data files from the file server, as
appropriate.

Finally, we point out that a Python interface to the DR2 MySQL
database is available in the {\sc Astroquery} package.\footnote{{\tt
    http://astroquery.readthedocs.org/}}

\subsubsection{Single Object Viewer}

The Single Object Viewer (SOV) provides a convenient way to access
{\em all} of the data that are available in the DR2 database for a
given object. 

The SOV can be queried with one or more CATAIDs or SPECIDs (the unique
object and spectrum identifiers used by GAMA). If multiple IDs are
given the SOV provides an effective way of moving along the list. For
a given object (or the object that is associated with a given
spectrum) the SOV displays the most important data from the {\tt
  InputCat} and {\tt SpecCat} DMUs, alongside an SDSS DR7 five-band
composite image of the object, the best or requested spectrum (see
Fig.~\ref{example_spec} for an example), and the summary plots from
the {\tt SersicPhotometry} and {\tt GroupFinding} DMUs (see
Figs.~\ref{sersic} and \ref{group}, respectively). The SOV is thus an
effective tool for visually inspecting the images, spectra, S{\'e}rsic
fits and group environments even of large samples of objects.

In addition to this overview, the SOV provides convenient links to
query {\em any} table in the DR2 database for the object or spectrum
under consideration, thus making it easy to explore individual objects
in complete detail.

\subsubsection{File server}
\label{file_server}

All of the data described in Section~\ref{data_description} (i.e.\ the
GAMA spectra, spectra from previous surveys, catalogues and
accompanying meta-data, mosaic images, summary plots, etc.) are made
available for download on the DR2 file server. The data are organised
in a directory tree structure that is intended to be
self-explanatory. The file server web page provides a convenient way
of browsing and accessing this directory structure. It also provides
information regarding the contents of directories, data formats and
file naming conventions.

Catalogues are organised into sub-directories by DMU. These DMU
directories contain the actual catalogue data, as well as all
accompanying meta-data.

Files with filename extension {\tt .notes} are plain text files
containing detailed descriptions, either of the DMU as a whole ({\tt
  DMUName.notes}) or of individual tables ({\tt
  TableName.notes}). Together these files provide a comprehensive
description of the DMU and all of its data products.

For convenience, all catalogues are provided in two formats: as a
space-delimited ASCII file, and as a binary FITS table. The former is
always accompanied by another file (with the same root filename, but
with filename extension{\tt .par}), which contains the basic table
meta-data, including the table's creation date, contact person, short
description, and the column meta-data such as column name, units, and
short column description. The FITS version is in the `FITS-plus'
format\footnote{{\tt http://www.starlink.ac.uk/topcat/}} used by the
popular table manipulation tools {\sc topcat} \citep{Taylor05} and
{\sc stilts} \citep{Taylor06}. This format allows us to conveniently
store the table data and basic meta-data (i.e.\ the contents of the
{\tt .par} file) together in the same file.

Note that the contents of a given DMU's {\tt .notes} and {\tt .par}
files are identical to the information on this DMU provided by the
schema browser described above.

Moving on from catalogues to spectra, these are organised on the file
server in sub-directories according to their originating surveys
(cf.\ Table~\ref{dr2spec}). For each spectrum we also provide a plot
in PNG format equivalent to that shown in Fig.~\ref{example_spec}.

The {\tt imaging} part of the file server contains the large-format
mosaics used by the {\tt ApMatchedPhotom} and {\tt SersicPhotometry}
DMUs, the complete input and output data used by the {\tt
  SersicPhotometry} DMU, the summary plots produced by the {\tt
  GroupFinding} DMU, as well as SDSS DR7 five-band composite postage
stamps for all objects in {\tt TilingCat} and {\tt SpStandards}.

\section{Summary}
\label{summary}

This rather technical paper essentially consists of four parts.
First, we report in Section~\ref{updates} on a number of aspects
concerning the implementation of the GAMA\,II spectroscopic survey.
Specifically, in Section~\ref{inputcat} we provide an overview of the
changes to the input catalogue and the target selection that were
implemented following the completion of the first phase of the GAMA
survey. In Section~\ref{redshifting} we discuss in depth the full
procedure by which we measure a spectrum's redshift using the
semi-automatic code {\sc runz}. We detail our motivation for
developing an extensive double-checking process (re-redshifting),
describe its implementation as well as the analysis of the resulting
data, and discuss its overall effect. Having briefly described our
new, fully automated redshift code {\sc Autoz} in Section~\ref{autoz},
we end this part of the paper in Section~\ref{LTObs} with a summary of
our observations with the Liverpool Telescope of a small number of
targets that were too bright to be observed during regular survey
operations at the AAT. 

We point out that this first part of the paper supplements the series
of earlier technical papers describing the implementation of the GAMA
spectroscopic survey
\citep{Baldry10,Robotham10,Driver11,Hopkins13a,Baldry14,Davies15}.

Second, following its recent completion, we present the end product of
the GAMA \,II spectroscopic survey in Section~\ref{progress_qc}. We
show and discuss a series of diagnostics to assess the final state of
the survey and the quality of the redshift data. Our final dataset
includes reliable redshifts for over $263\,000$ objects. In its three
equatorial survey regions GAMA has achieved an exceptionally high
overall redshift completeness of $98.48$~per~cent, while the two
southern regions G02 and G23 were completed to levels of $94.95$ and
$94.19$~per~cent, respectively.  Despite these high values, weak but
nevertheless significant completeness trends with brightness, surface
brightness and colour remain. In contrast, the spatial distribution of
the redshift completeness is extremely homogeneous, both on large and
small angular scales. The high redshift completeness even in densely
populated regions of the sky is a particular hallmark of the GAMA
survey, one that sets it apart from its predecessors. It is this
feature, in combination with its faint limit of $r < 19.8$~mag, that
makes the GAMA survey a unique resource for studies that rely on
accurate measurements of the properties of galaxy pairs and groups.
Section~\ref{progress_qc} concludes by comparing the redshift
precision and reliability of {\sc runz} with those of the newer {\sc
  Autoz} code. We find that the latter outperforms the former on both
accounts. The average $1$$\sigma$ error of {\sc Autoz} redshifts is
just $27$\kms, and only $0.2$~per~cent of {\sc Autoz} redshifts
classified as reliable turn out to be incorrect.

The third part of the paper is concerned with two aspects of GAMA's
photometric programme. Section~\ref{onirphot} provides an update on
our procedures to extract aperture-matched optical and NIR photometry
from processed SDSS and UKIDSS LAS imaging data, while in
Section~\ref{galexphot} we describe our methods to obtain FUV and NUV
photometry for GAMA galaxies from the data of the \textit{GALEX}-GAMA
survey. This part of the paper essentially continues the series of
technical papers on GAMA photometry \citep{Hill11,Kelvin12,Cluver14}.

Finally, in Section~\ref{dr2} we describe the second public release of
GAMA data. In DR2 we release GAMA\,I spectra, redshifts and a wealth
of additional information for all main survey objects with $r <
19.0$~mag in survey regions G09 and G12, and for all objects with $r <
19.4$~mag in region G15 ($72\,225$ objects in total). The additional
information is comprised of input catalogue and targeting information,
optical and NIR ($u$ to $K$) aperture-matched photometry, photometry
and structural parameters from single-component S{\'e}rsic fits in the
same bands, FUV and NUV photometry from \textit{GALEX},
$k$-corrections, stellar masses, spectral line measurements and
H$\alpha$-derived SFRs, three different environment measures (only for
G15), and the GAMA Galaxy Group Catalogue (G$^3$C, again for G15
only). Together these data represent a valuable resource for studies
of the low-redshift galaxy population.

In future data releases we will extend the publicly available spectra
and redshifts both to fainter limiting magnitudes and to the southern
survey regions G02 and G23. We will also release additional data
products not yet included in DR2, including the {\sc Autoz} redshifts,
mid and far-infrared photometry, photometry and bulge-disk
decompositions derived from KiDS and VIKING data, morphologies, and
additional environmental measures. In due course, all GAMA data and
data products will be made publicly available.

We conclude by encouraging interested readers to contact the GAMA team
if they already would like to use GAMA data that are currently still
proprietary. We actively support (and engage with) collaboration
projects, as long as there are no conflicts with already existing
projects. Details of the different collaboration possibilities are
available at the GAMA website.

\section*{Acknowledgements} 

GAMA is a joint European-Australasian project based around a
spectroscopic campaign using the Anglo-Australian Telescope. The GAMA
input catalogue is based on data taken from the Sloan Digital Sky
Survey and the UKIRT Infrared Deep Sky Survey. Complementary imaging
of the GAMA regions is being obtained by a number of independent
survey programs including \textit{GALEX} MIS, VST KiDS, VISTA VIKING,
\textit{WISE}, \textit{Herschel}-ATLAS, GMRT and ASKAP providing UV to
radio coverage. GAMA is funded by the STFC (UK), the ARC (Australia),
the AAO, and the participating institutions. The GAMA website is {\tt
  http://www.gama-survey.org/}.

Funding for the SDSS and SDSS-II has been provided by the Alfred
P.\ Sloan Foundation, the Participating Institutions, the National
Science Foundation, the U.S.\ Department of Energy, the National
Aeronautics and Space Administration, the Japanese Monbukagakusho, the
Max Planck Society, and the Higher Education Funding Council for
England. The SDSS website is {\tt http://www.sdss.org/}.

The SDSS is managed by the Astrophysical Research Consortium for the
Participating Institutions. The Participating Institutions are the
American Museum of Natural History, Astrophysical Institute Potsdam,
University of Basel, University of Cambridge, Case Western Reserve
University, University of Chicago, Drexel University, Fermilab, the
Institute for Advanced Study, the Japan Participation Group, Johns
Hopkins University, the Joint Institute for Nuclear Astrophysics, the
Kavli Institute for Particle Astrophysics and Cosmology, the Korean
Scientist Group, the Chinese Academy of Sciences (LAMOST), Los Alamos
National Laboratory, the Max-Planck-Institute for Astronomy (MPIA),
the Max-Planck-Institute for Astrophysics (MPA), New Mexico State
University, Ohio State University, University of Pittsburgh,
University of Portsmouth, Princeton University, the United States
Naval Observatory, and the University of Washington.

Based on observations made with the NASA Galaxy Evolution Explorer.

This research has made use of the NASA/IPAC Extragalactic Database
(NED) which is operated by the Jet Propulsion Laboratory, California
Institute of Technology, under contract with the National Aeronautics
and Space Administration.

\bibliographystyle{mn2e}
\setlength{\bibhang}{1.5em}
\setlength\labelwidth{0.0em}
\bibliography{}

\label{lastpage}

\end{document}